\documentclass[12pt,a4paper]{article}
\usepackage[OT2,T1]{fontenc}
\usepackage{verbatim}
\usepackage{amssymb}
\usepackage{amsmath}
\usepackage{mathrsfs}
\usepackage{mathtools}
\usepackage{bm}
\usepackage{graphicx}
\usepackage[leftcaption]{sidecap}
\usepackage{caption}
\usepackage{subcaption}
\usepackage{here}
\usepackage[margin=1in]{geometry}
\usepackage{cite}
\usepackage[most]{tcolorbox}
\newtcbox{\othermathbox}[1][]{nobeforeafter, math upper, tcbox raise base, 
          enhanced, rounded corners, colback=black!5, colframe=black}
\usepackage{empheq}
\DeclareMathAlphabet{\mathscrbf}{OMS}{mdugm}{b}{n}
\usepackage{tikz}
\usetikzlibrary{decorations.markings}
\usepackage{wrapfig}
\usepackage{parskip}
\usepackage{enumitem}

\usepackage{marvosym}

\usepackage{sectsty}
\allsectionsfont{\sffamily}
\usepackage{titlesec}
\titleformat*{\paragraph}{\sffamily\bfseries\boldmath}
\usepackage{tocloft}

\let\OLDthebibliography\thebibliography
\renewcommand\thebibliography[1]{
  \OLDthebibliography{#1}
  \setlength{\parskip}{0pt}
  \setlength{\itemsep}{3pt plus 0.3ex}
}

\usepackage[colorlinks,allcolors=blue]{hyperref}

\definecolor{MyRed}{RGB}{220,60,10}
\definecolor{MyOrange}{RGB}{240,155,0}
\definecolor{MyYellow}{RGB}{215,215,38}
\definecolor{MyGreen}{RGB}{10,220,60}
\definecolor{MyTurk}{RGB}{60,140,180}
\definecolor{MyBlue}{RGB}{60,10,220}
\definecolor{MyPurple}{RGB}{165,0,230}
\definecolor{LightYellow}{RGB}{255,245,108}
\definecolor{LightBrown}{RGB}{255,188,0}
\definecolor{MiddleBrown}{RGB}{199,146,0}
\definecolor{DarkBrown}{RGB}{143,104,0}
\definecolor{DarkerBrown}{RGB}{87,62,0}
\definecolor{Purple}{RGB}{255,0,188}

\newcommand{\be}{\begin{equation}}
\newcommand{\ee}{\end{equation}}
\newcommand{\ben}{\begin{enumerate}}
\newcommand{\een}{\end{enumerate}}
\newcommand{\bi}{\begin{itemize}}
\newcommand{\ei}{\end{itemize}}
\newcommand{\bmm}{\begin{pmatrix}}
\newcommand{\emm}{\end{pmatrix}}

\newcommand{\cyclotron}{B}
\newcommand{\dd}{\text{d}}

\newcommand{\der}{\partial}

\newcommand{\ds}{\displaystyle}

\newcommand{\eg}{{\it e.g.}\ }

\newcommand{\ie}{{\it i.e.}\ }

\newcommand{\nn}{\nonumber}
\newcommand{\phii}{\varphi}

\newcommand{\bA}{\textbf A}
\newcommand{\bB}{\textbf B}

\newcommand{\bp}{\textbf p}

\newcommand{\bx}{\textbf x}
\newcommand{\by}{\textbf y}

\newcommand{\cI}{{\cal I}}

\newcommand{\cK}{{\cal K}}

\newcommand{\cN}{N}
\newcommand{\cO}{{\cal O}}
\newcommand{\cR}{\Delta}

\renewcommand{\AA}{\mathbb{A}}
\newcommand{\CC}{K}

\newcommand{\NN}{\mathbb{N}}
\newcommand{\RR}{\mathbb{R}}

\newcommand{\ZZ}{\mathbb{Z}}

\newcommand{\upchi}{\text{X}}

\begin{document}

\hrule
\vspace{.3em}
\begin{center}
\Large{\bfseries{\textsf{Ergodic Edge Modes in the 4D Quantum Hall Effect}}}
\end{center}
\vspace{.7em}
\hrule
~\\

\begin{center}
\large{\textsf{Benoit Estienne,$^{\blacklozenge}$
Blagoje Oblak,$^{\blacklozenge\,\spadesuit}$ 
and Jean-Marie St\'ephan$^{\bigstar}$}}\\
~\\
\begin{minipage}{\textwidth}\small\it
\begin{center}
{\tt{estienne@lpthe.jussieu.fr}, \tt{boblak@lpthe.jussieu.fr}, \tt{stephan@math.univ-lyon1.fr}}\\
~\\
$^{\blacklozenge}$ Sorbonne Universit\'{e}, CNRS, Laboratoire de Physique Th\'{e}orique\\ et Hautes \'{E}nergies, LPTHE, F-75005 Paris, France; \\
$^{\spadesuit}$ Centre de Physique Th\'eorique, CNRS, Ecole Polytechnique,\\ UMR 7644, F-91128 Palaiseau, France;\\
$^{\bigstar}$ Univ Lyon, CNRS, Universit\'e Claude Bernard Lyon 1, UMR 5208,\\ Institut Camille Jordan, F-69622 Villeurbanne, France.
\end{center}
\end{minipage}
\end{center}

\vspace{5em}

\begin{center}
\begin{minipage}{.915\textwidth}
\begin{center}{\bfseries{\textsf{Abstract}}}\end{center}\vspace{.4em}
The gapless modes on the edge of four-dimensional (4D) quantum Hall droplets are known to be anisotropic: they only propagate in one direction, foliating the 3D boundary into independent 1D conduction channels. This foliation is extremely sensitive to the confining potential and generically yields chaotic flows. Here we study the quantum correlations and entanglement of such edge modes in 4D droplets confined by harmonic traps, whose boundary is a squashed three-sphere. Commensurable trapping frequencies lead to periodic trajectories of electronic guiding centers; the corresponding edge modes propagate independently along $S^1$ fibers, forming a bundle of 1D conformal field theories over a 2D base space. By contrast, incommensurable frequencies produce quasi-periodic, ergodic trajectories, each of which covers its invariant torus densely; the corresponding correlation function of edge modes has fractal features. This wealth of behaviors highlights the sharp differences between 4D Hall droplets and their 2D peers; it also exhibits the dependence of 4D edge modes on the choice of trap, suggesting the existence of observable bifurcations due to droplet deformations.
\end{minipage}
\end{center}

\newpage
\textsf{\tableofcontents}
~\\[-.3cm]

\newpage
\section{Introduction}

Spurred by the discovery of the quantum Hall effect (QHE) \cite{vonKlitzing,Tsui}, the study of topological phases has been one of the driving forces of condensed matter physics in recent decades \cite{Hasan,QiZhang}. By now, numerous such topological systems have indeed been found, both in genuine condensed matter \cite{Bernevig_2006,Konig_2007,Hsieh_2008} and in its various simulations \cite{Cooper,Dalibard,Goldman,GoldmanDal,WangPhoton,Ozawa}. The latter offer the exciting prospect of realizing exotic phases that do not, otherwise, occur spontaneously in Nature. One such exotic system is the QHE in dimensions higher than two.

Theoretical aspects of higher-dimensional Hall droplets have been investigated ever since the proposal of \cite{Zhang2001}, where electrons on a four-sphere were subjected to a non-Abelian background gauge field. It was quickly noticed, however, that the interesting phenomenology of this model --- Landau levels, non-commutative space, non-trivial bulk topological invariants and chiral edge modes --- also occurs in 4D systems that simply support an Abelian magnetic field \cite{Polchinski2002,Karabali2002,Karabali2003,Karabali2004,Polychronakos:2004es,Karabali:2010ss}, readily generalizing the 2D QHE. (In particular, as in 2D, the symplectic interpretation of non-commutative geometry \cite{karabali2005fuzzy,Karabali2006,Nair:2020lvk} then connects quantum Hall physics to geometric quantization \cite{Douglas,Klevtsov,Klevtsov:2016bos,Charles_2019,Nair_2020}.)

A staple of these works is the {\it anisotropic} nature of edge modes \cite{Polchinski2002,Karabali2002,Karabali2003}: while low-energy excitations of a 2D droplet propagate along the entire 1D boundary, their higher-dimensional peers are localized on 1D fibers embedded in a higher-dimensional manifold. The gapless edge modes of a 4D droplet thus span a collection of (1+1)-dimensional chiral conformal field theories (CFTs), rather than a (3+1) CFT. As a result, edge dynamics is exceedingly sensitive to the trapping potential: it can range from fully integrable to chaotic, depending on the trap. The present paper is therefore devoted to a detailed analysis of such boundary excitations in a microscopic model of trapped 4D `electrons' in a strong magnetic field.\footnote{The 4D restriction is a matter of convenience: our conclusions extend to higher-dimensional droplets.} Since we shall rely on quantum-mechanical methods similar to those used to describe ultracold atoms \cite{Cooper,Dalibard,Goldman,GoldmanDal} and topological photonics \cite{WangPhoton,Ozawa}, our hope is also to provide a bridge between field-theoretic considerations \cite{Polchinski2002,Karabali2002,Karabali2003,Karabali2004,Polychronakos:2004es,Karabali:2010ss} and the vast literature on synthetic dimensions \cite{Kraus2012,Kraus2013,Price2015,Ozawa2016,Zilberberg2017,Lohse2018,Yuan,PriceOzawa,OzawaCarusotto,Ozawa2019}.

\begin{figure}[t]
\centering
\includegraphics[width=0.4\textwidth]{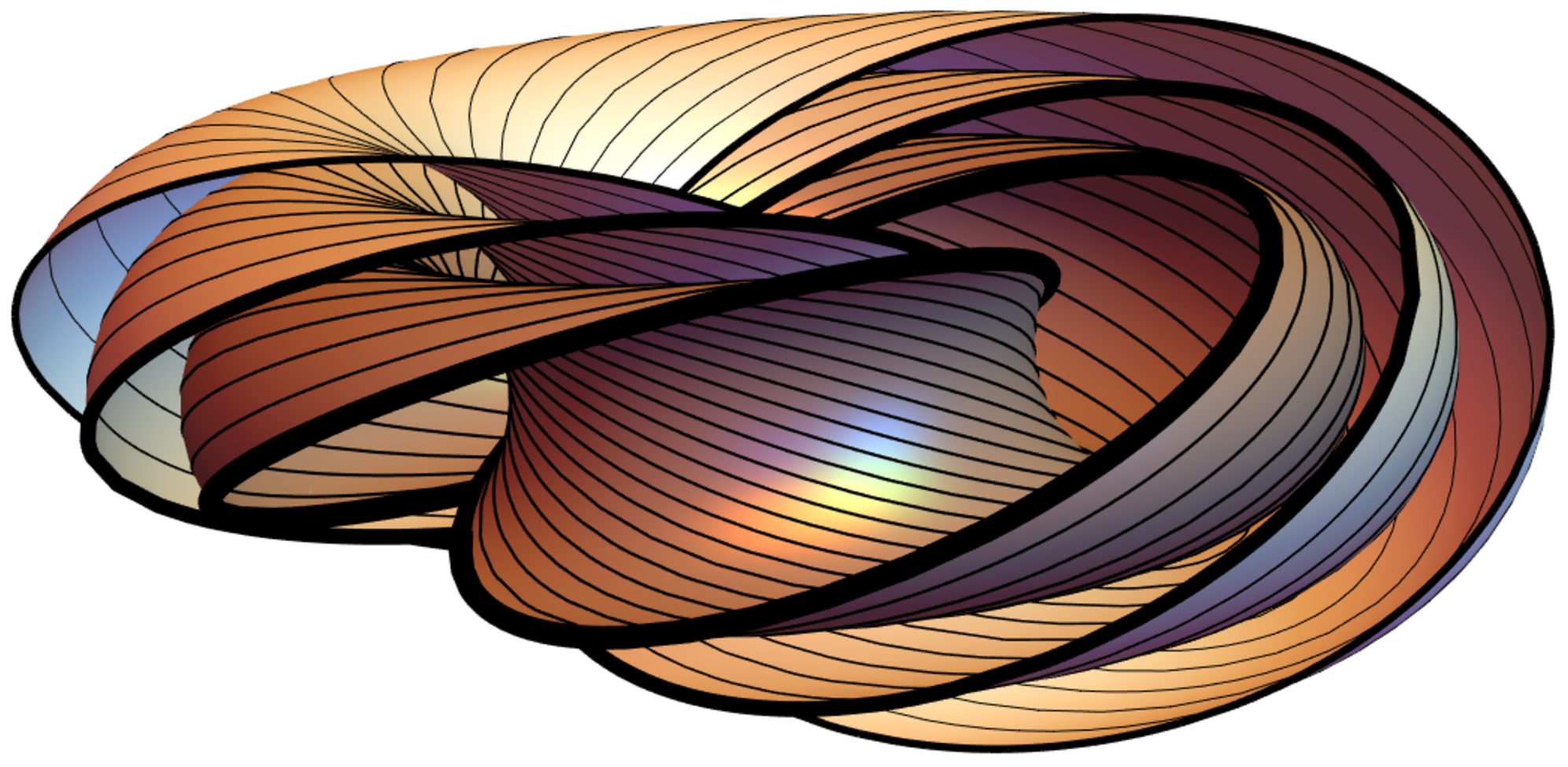}
\caption{A portion of the Hopf fibration of $S^3$ (here visualized using stereographic projection to $\mathbb{R}^3$): each fiber is a circle $S^1$ labelled by a point in an $S^2$ base space \cite{Urbantke}. The trajectories of edge modes in an isotropic 4D quantum Hall droplet coincide with such $S^1$ fibers in an $S^3$ boundary; they are localized on nested tori, each of which is itself partitioned into (chiral) Villarceau circles. By contrast, in generic {\it anisotropic} droplets, each torus is filled {\it densely} by a single edge mode trajectory.}
\label{Hopf}
\end{figure}

For definiteness we will focus on harmonic traps, where electron dynamics is integrable and the boundary is a (squashed) 3D sphere. Electronic guiding center orbits are then localized on invariant 2D tori and can be either periodic or quasi-periodic. In particular, edge modes of isotropic droplets realize the Hopf fibration of $S^3$ (see fig.\ \ref{Hopf}), while edge modes in anisotropic traps are generally ergodic in the sense that each of their orbits covers its torus densely. These results rely on the following series of arguments:
\begin{itemize}[leftmargin=*]\setlength{\itemsep}{0em}
\item[(i)] First, we shall see how some of the most striking properties of edge modes can already be inferred from the classical picture of skipping electrons whose guiding centers follow equipotentials. This will show that, at strong magnetic fields, any trapping potential produces 1D edge modes in an otherwise higher-dimensional boundary (see figs.\ \ref{fimotion}--\ref{fitorus}). This fact --- essentially restating the classical Hall conductance formula --- readily implies that the global structure of edge trajectories hinges on the degree of anisotropy of the trap, suggesting the existence of ergodic and chaotic regimes.
\item[(ii)] We will then study quantum corrections of this classical approximation by evaluating the many-body ground state correlation function of a non-interacting 4D droplet. Trapping anisotropies with rational frequency ratios will allow us to evaluate the asymptotics of correlations near the boundary (see eq.\ \eqref{main} below). As we shall see, edge correlations indeed localize on classical guiding center trajectories, along which they satisfy a power-law characteristic of 1D CFTs, while they decay in a Gaussian manner in all remaining directions. This is depicted in fig.\ \ref{fiMainPlot}; note in particular the delicate dependence of edge correlations on the trap's anisotropy.
\item[(iii)] The second step of our analysis of correlations will be concerned with `irrational' anisotropic traps, where the asymptotic methods appropriate for periodic edge modes fail to apply. Nevertheless, semiclassical intuition, supported by ample numerical evidence (see figs.\ \ref{fig:irr1}--\ref{fig:irr2}), will allow us to analyse edge correlations in the thermodynamic limit. This is expressed in eq.\ \eqref{conjec} and plotted in fig.\ \ref{fig:fractal}; it involves a function that vanishes almost everywhere but takes non-zero values on a dense subset of the torus, displaying fractal-like properties and a certain degree of self-similarity.
\item[(iv)] Finally, we shall estimate the ground state entanglement of various spatial regions crossing the edge of a 4D droplet, thus generalizing the 2D considerations of \cite{Estienne2019hmd}. The derivation will rely on well established techniques relating the entanglement spectrum of free fermions to their correlations \cite{Chung_2001,Peschel_2003,Cheong_2004,islam2015measuring,Peschel_2009}. For regions whose boundary is parallel to edge fibers, subleading contributions to entanglement entropy will turn out to vanish, confirming that edge modes on distinct fibers are genuinely independent. By contrast, the entanglement entropy of regions that cut through certain fibers will display the logarithmic behavior expected of a bundle of 1D CFTs, with a total central charge related to the bulk Fermi energy.
\end{itemize}
The plan of the paper follows this sequence: section \ref{seclas} is classical, quantum correlations of periodic and ergodic edge modes are respectively studied in sections \ref{secor}--\ref{sergo}, and entanglement is treated in section \ref{senta}. The back matter contains various technical details, such as the exact spectrum of a trapped 4D Landau Hamiltonian (appendix \ref{app_quadratic}) or the derivation of asymptotic relations needed in sections \ref{secor} (appendices \ref{appa}--\ref{appb}) and \ref{senta} (appendices \ref{appc}--\ref{appex}).

\newpage
\section{Classical edge dynamics}
\label{seclas}

Guided by classical intuition, this section serves to develop basic expectations about the edge modes of Hall droplets in any dimension, with arbitrary trapping potentials. This will rely on the well-known projected form of the Landau Hamiltonian at strong magnetic fields \cite{GirvinJach}. As an example, we introduce the harmonic 4D Landau Hamiltonian that will be used throughout this work. Technical details regarding the spectrum of this Hamiltonian are relegated to appendix \ref{app_quadratic}.

\phantomsection
\addcontentsline{toc}{subsection}{Guiding center approximation}%
\paragraph*{Guiding center approximation.}
Consider a massive charged particle in $\RR^d$ ($d$ even); we work in natural units, so mass $=$ charge $=1$. Points in $\RR^d$ are written as vectors $\bx$ with components $x^i$, $i=1,...,d$. We assume (i) that some confining potential $V(\bx)$ is present, (ii) that the system supports a magnetic field $\bB=\dd\bA$ for some vector potential $\bA$, such that the matrix $B_{ij}(\bx)=\der_iA_j-\der_jA_i$ is invertible everywhere. At very strong magnetic fields, kinetic energy may be neglected and the Lagrangian of the particle reduces to $L=\bA(\bx)\cdot\dot\bx-V(\bx)$. This is linear in velocities, so canonical momenta are not independent coordinates on phase space; they are, instead, related to positions:
\be
\label{pix}
\bp
\equiv
\frac{\der L}{\der\dot\bx}
=
\bA(\bx).
\ee
This equality may be seen as a set of $d$ constraints\footnote{The (standard) notation $\approx$ denotes equalities that hold on the constraint surface specified by \eqref{pix} \cite{HenneauxTeitelboim}.} $\bp-\bA(\bx)\approx0$ determining the components $p_i$ from the knowledge of spatial coordinates $x^i$. Because $d$ is even and $\bB$ is non-degenerate, these constraints are second class and yield a Dirac bracket \cite[sec.\ 1.3]{HenneauxTeitelboim}
\be
\label{dib}
\{x^i,x^j\}
= B^{ij}(\bx),
\ee
where $B^{ij}$ is the inverse matrix of the magnetic field $B_{ij}$. As a result, one may think of $\RR^d$ itself as a phase space with (classical) coordinates $x^i$ that fail to Poisson-commute, the magnetic field playing the role of a symplectic form. The Hamiltonian projected on this effective phase space coincides with the trapping potential:
\be
\label{hav}
H(\bx,\bp)
=
\frac{1}{2}(\bp-\bA)^2+V(\bx)
\approx
V(\bx),
\ee
so that the pure Landau term $(\bp-\bA)^2/2$ is minimized. This approximation is a classical analogue of the quantum projection to the lowest Landau level \cite{GirvinJach}. In particular, the projected Hamiltonian describes the slow motion of guiding centers of cyclotron orbits: using \eqref{hav} and the bracket \eqref{dib}, one finds
\be
\label{eom}
\dot x^i
\approx
\{x^i,V(\bx)\}
=
B^{ij}\der_jV.
\ee
Our presentation of this result used the formalism of constrained systems for brevity, but a more intuitive argument can also be provided. Indeed, the guiding center approximation stems from a separation of time scales (fast cyclotron rotations versus slow guiding center drift) in strong magnetic fields. This splitting becomes sharper as the magnetic field increases and cyclotron orbits with bounded energy shrink, eventually leading to an elimination of fast degrees of freedom and a dimensional reduction to the so-called {\it slow manifold} \cite{Burby}. Here, the latter is just the space $\RR^d$ of guiding center positions endowed with the Poisson bracket \eqref{dib} and the effective Hamiltonian $V(\bx)$, yielding the slow dynamics \eqref{eom}.

\begin{figure}[t]
\centering
\includegraphics[width=0.30\textwidth]{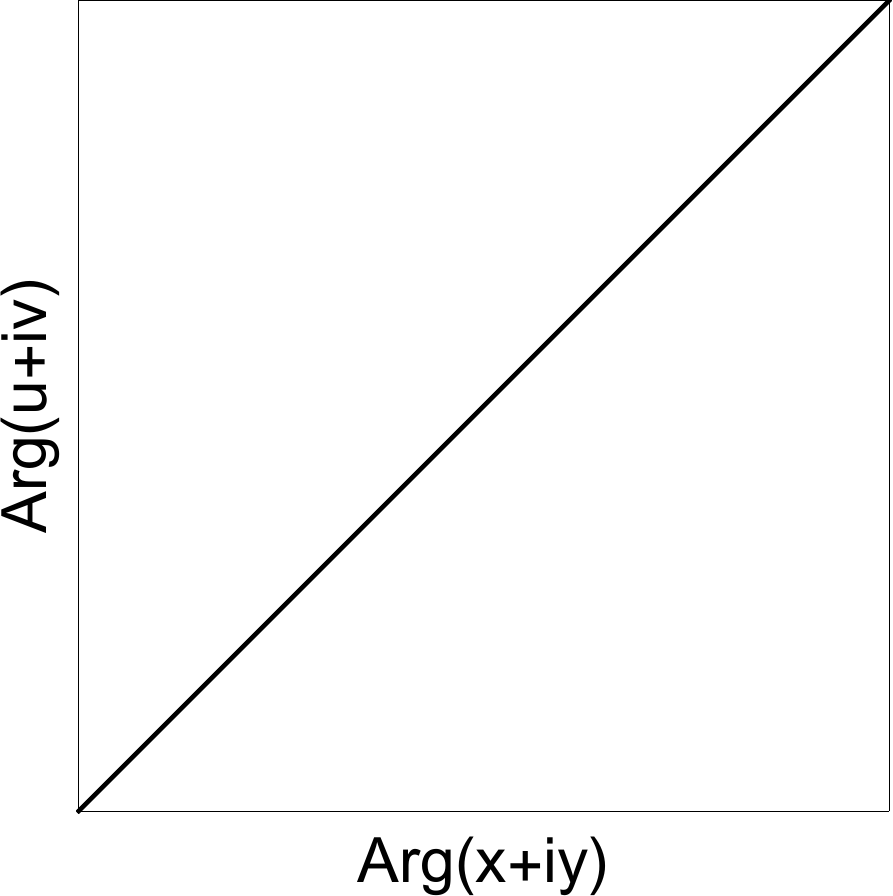}~~~~~~%
\includegraphics[width=0.30\textwidth]{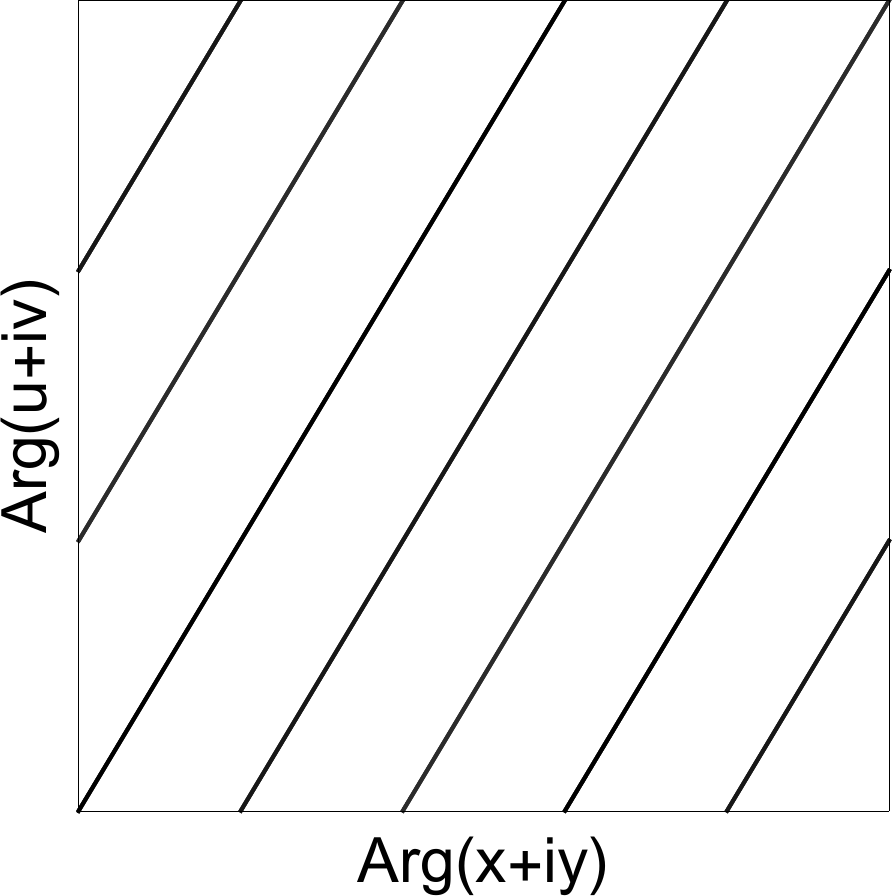}~~~~~~%
\includegraphics[width=0.30\textwidth]{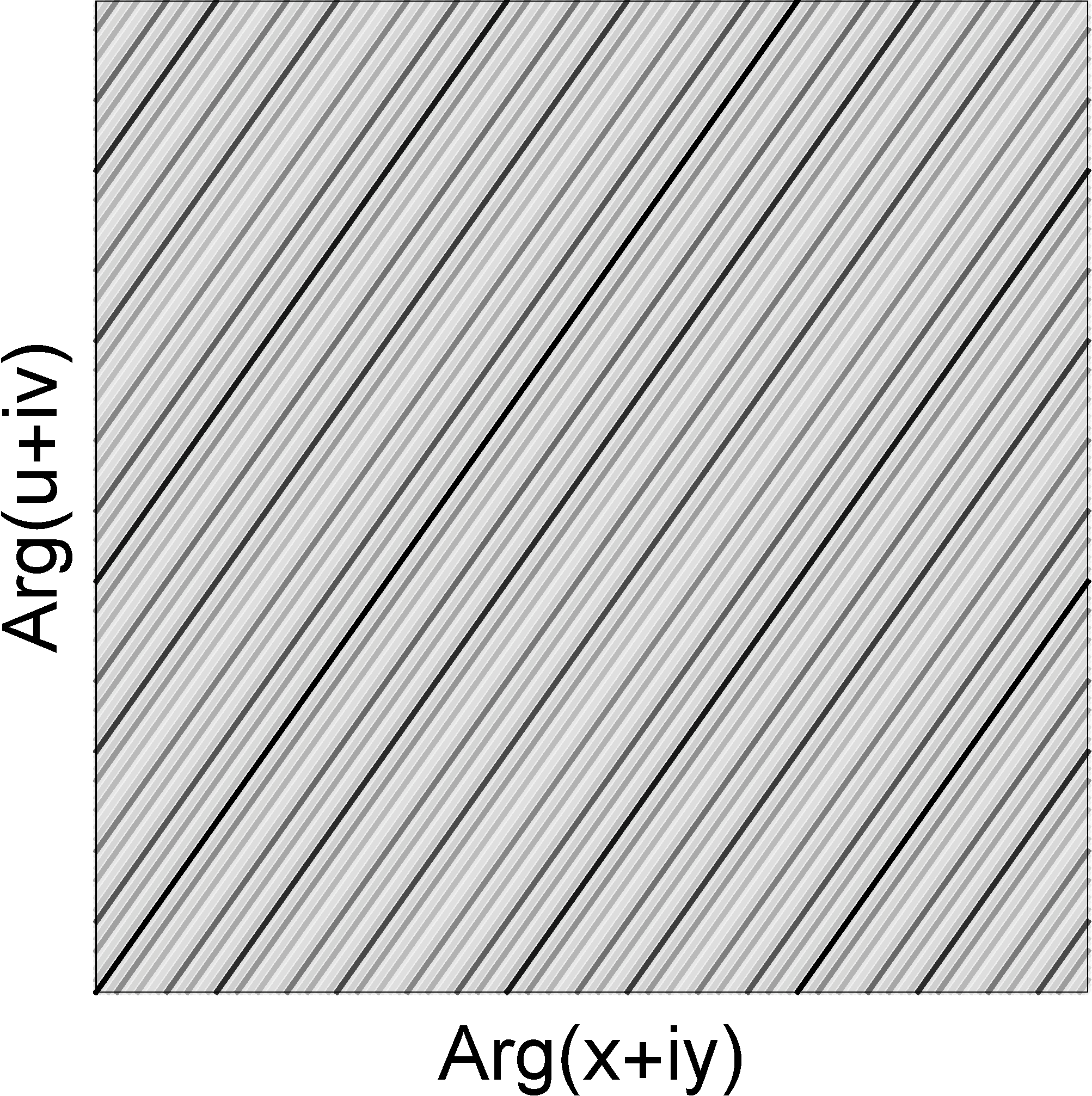}
\caption{The motion \eqref{gc} of a guiding center on a torus (at fixed $x^2+y^2$ and $u^2+v^2$), initially located in the square's lower left corner. From left to right, the frequency ratio \eqref{rr} takes values $\cR=1$, $5/3$ and $\sqrt{2}$. The first and second cases ($\cR=1$ or $5/3$) are rational, producing periodic guiding center motion; $\cR=1$ is even isotropic. By contrast, the third case ($\cR=\sqrt{2}$) is not only anisotropic, but irrational. As a result, even a single guiding center trajectory covers the torus densely; the color-coding is such that the trajectory is initially black, then progressively fades to white. These pictures should be compared with their quantum counterparts, namely the edge correlations of figs.\ \ref{fiMainPlot}--\ref{fig:irr1}.}
\label{fimotion}
\end{figure}

\begin{figure}[t]
\centering
\includegraphics[width=0.30\textwidth]{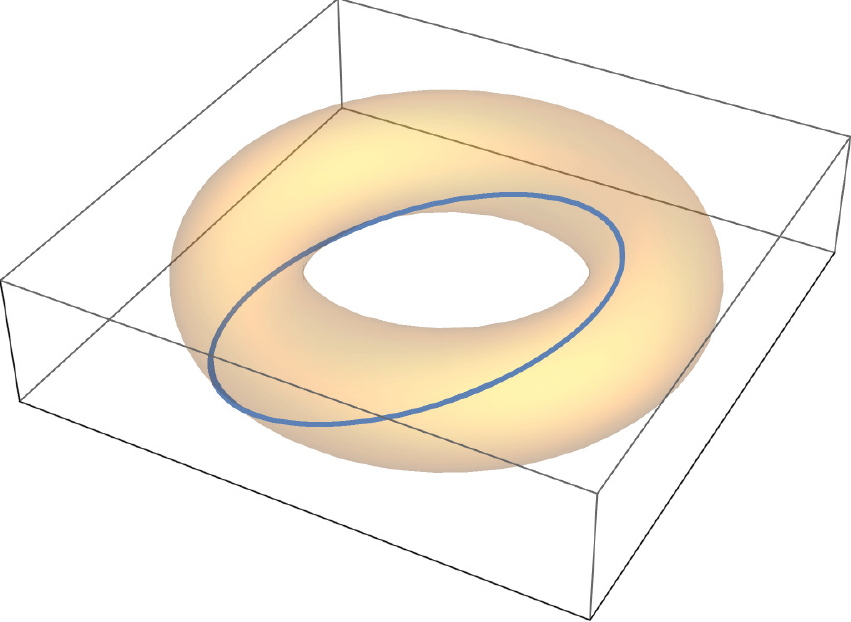}~~~~~~%
\includegraphics[width=0.30\textwidth]{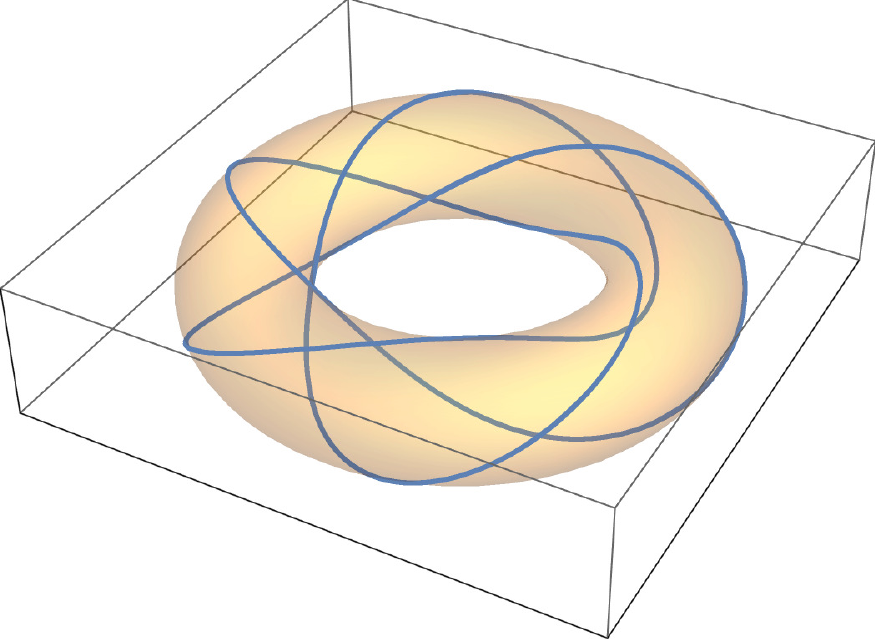}~~~~~~%
\includegraphics[width=0.30\textwidth]{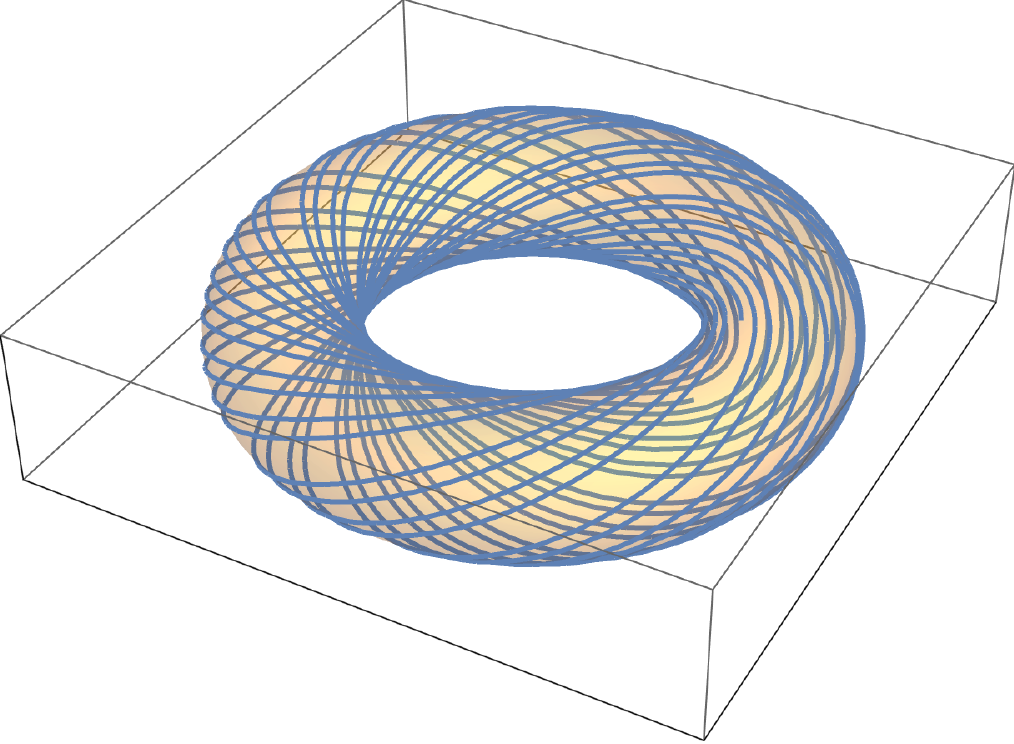}
\caption{The guiding center paths of fig.\ \ref{fimotion} represented on tori embedded in a three-sphere $S^3$ (whose points satisfy $x^2+y^2+u^2+v^2=\text{cst}$). Through stereographic projection, one can think of this $S^3$ as $\RR^3$ plus a point at infinity, so each of the three plots in this figure depicts one edge mode in what will, eventually, be the boundary of a Hall droplet. As in fig.\ \ref{fimotion}, the anisotropy \eqref{rr} is $\cR=1$, $5/3$ and $\sqrt{2}$ from left to right.}
\label{fitorus}
\end{figure}

In 2D, the equation of motion \eqref{eom} is necessarily integrable, so its implications are somewhat trivial: guiding centers follow 1D equipotentials of $V(\bx)$, merely restating the classical Hall law (namely that the current $B^{ij}\der_jV$ is perpendicular to the electric field $\der_jV$). The situation is much richer in higher dimensions, where periodic trajectories only occur in highly symmetric potentials, while generic traps (with compact equipotentials) produce ergodic, or even {\it chaotic}, guiding center dynamics. We do not investigate the fully chaotic situation in this work. Instead, we focus on a family of integrable setups whose edge modes lie on Liouville-Arnold tori, in which case the possibility of resonances entails higher-dimensional subtleties that never affect 2D droplets. For instance, in 4D, integrable guiding center motion is specified by two frequencies; rational frequency ratios then produce periodic trajectories (fig.\ \ref{Hopf} $+$ left and center panels of figs.\ \ref{fimotion}--\ref{fitorus}), but irrational ratios entail quasi-periodic, {\it ergodic} trajectories, each of which is everywhere dense on its torus (rightmost panels of figs.\ \ref{fimotion}--\ref{fitorus}). Sections \ref{secor} and \ref{sergo} will respectively describe the quantum correlations of edge modes corresponding to those two cases. For now, however, we remain in the classical realm.

\phantomsection
\addcontentsline{toc}{subsection}{Harmonic potential in 4D}%
\paragraph*{Harmonic potential in 4D.}
\label{sehar}
Let us illustrate the general arguments above with a simple example to which we shall refer repeatedly. Assume $d=4$ and write $\bx=(x,y,u,v)$, with a uniform magnetic field $\bB=B(\dd x\wedge\dd y+\dd u\wedge\dd v)$. Let the trapping potential be harmonic and partially isotropic, with stiffnesses $k,k'>0$:
\be
\label{trap}
V(x)=\frac{k}{2}(x^2+y^2)+\frac{k'}{2}(u^2+v^2).
\ee
In the slow guiding center phase space $\RR^4$ with Poisson brackets \eqref{dib}, this trap may be viewed as the Hamiltonian of a 2D harmonic oscillator. This is the canonical example of a classical integrable system: one trivially has two conserved quantities ($x^2+y^2$ and $u^2+v^2$), so guiding center trajectories lie on tori at constant $x^2+y^2$ and $u^2+v^2$. Furthermore, the `Hamiltonian' \eqref{trap} is quadratic, so the slow equation of motion \eqref{eom} is linear:
\begin{align}
\label{gc}
\dot{x}=-\omega y, \qquad \dot{y}=\omega x, \qquad \dot{u}=-\omega' v, \qquad \dot{v}=\omega' u,
\end{align}
with $\omega\equiv k/B$, $\omega'\equiv k'/B$, both assumed to be much smaller than the cyclotron frequency $\omega_c=B$. Guiding centers thus rotate in the $(x,y)$ and $(u,v)$ planes with respective frequencies $\omega$ and $\omega'$; their winding around the torus depends on the ratio
\be
\label{rr}
\cR\equiv\omega'/\omega.
\ee
Note that the quadratic potential \eqref{trap} is so simple that the dynamics of the full Hamiltonian $H=(\bp-\bA)^2/2+V(\bx)$ is actually integrable, even without projecting to the slow manifold as was done here. This derivation is exposed in appendix \ref{app_quadratic}. Its results only differ from those just displayed by small corrections, proportional to the (dimensionless) parameter $k/B^2\ll1$. For instance, the actual ratio of guiding center frequencies on the torus is $(\sqrt{1+4k'/B^2}-1)/(\sqrt{1+4k/B^2}-1)$, which indeed reduces to \eqref{rr} at large $B^2/k$.

It is worth pausing to appreciate the different regimes that occur depending on the value of the ratio \eqref{rr}. The simplest, fully isotropic setup has $\cR=1$; guiding center trajectories then partition their $S^3$ equipotential into linked circles $S^1$, each labelled by a point in a Bloch sphere $S^2$, producing the Hopf fibration $S^1\to S^3\to S^2$ (fig.\ \ref{Hopf}).\footnote{The $S^2$ labelling of trajectories arises as follows (see \eg \cite{Urbantke}). In complex coordinates $z\propto x+iy$, $w\propto u+iv$, the solution of \eqref{gc} with $k=k'$ reads $z(t)=z_0e^{i\omega t}$,  $w(t)=w_0e^{i\omega t}$, so the vector $\pi(z,w)\equiv(2\bar zw,|z|^2-|w|^2)$ is conserved. The resulting projection $\pi:S^3\to S^2$ has preimages which coincide with guiding center trajectories and is locally trivial, confirming that $S^3$ is an $S^1$ bundle over $S^2$.} More generally, any rational ratio \eqref{rr} leads to periodic trajectories in a (squashed) $S^3$, generalizing the isotropic case so that each equipotential is a (covering of a) lens space \cite{Bartsch}, with distinct fibers again labelled smoothly by points in $S^2$. By contrast, when $\cR$ is irrational, the motion of $\bx(t)$ is ergodic in the sense that it fills its torus densely, regardless of the initial condition $\bx(0)$. This range of $\cR$-dependent behaviors is displayed in figs.\ \ref{fimotion}--\ref{fitorus}.

Classical considerations of this kind readily extend to many-body quantum Hall droplets, whose low-energy excitations should indeed be localized on guiding center trajectories. In a harmonic trap \eqref{trap}, a droplet's boundary is a (squashed) three-sphere that may be seen as a collection of nested tori (as in fig.\ \ref{Hopf}), each supporting many independent edge modes whose winding is determined by the ratio of stiffnesses \eqref{rr}. The remainder of this paper investigates this expectation in a fully fledged microscopic quantum theory.

\section{Quantum correlations}
\label{secor}

This section is the first step in our study of quantum aspects of 4D edge modes. After a brief preliminary on one-body quantum mechanics, we consider an isotropic droplet, whose bulk correlation function is an incomplete gamma function (similarly to the isotropic 2D QHE). In the thermodynamic limit, the asymptotics of this expression near the boundary confirm that edge modes are indeed gapless and localized, to within a magnetic length $\ell_B= \sqrt{\hbar/B}$, on guiding center trajectories. A similar result is then shown to hold for anisotropic droplets whose ratio \eqref{rr} is rational, despite the more complicated form of their correlations and the impossibility to rely on known special functions. In order to streamline the presentation, detailed asymptotic computations are relegated to appendices \ref{appa}--\ref{appb}. Correlations of ergodic edge modes are treated separately in section \ref{sergo}.

\phantomsection
\addcontentsline{toc}{subsection}{One-body spectrum}%
\paragraph*{One-body spectrum.} 
\label{sebody}
We begin by considering the one-body Hamiltonian $H$ of eq.\ \eqref{hav} with a harmonic trap \eqref{trap} and magnetic field $\bB=\dd\bA=B(\dd x\wedge\dd y+\dd u\wedge\dd v)$. Owing to the quadratic form of the complete Hamiltonian, including both the Landau term $(\bp-\bA)^2/2$ and the potential $V(\bx)$, it is in fact possible to diagonalize it exactly: this is shown in appendix \ref{app_quadratic}. However, in keeping with section \ref{seclas}, our approach here will rely on projected operators in the lowest Landau level (LLL). The ensuing formulas only differ from the exact results of appendix \ref{app_quadratic} by corrections that are tiny in the limit of strong magnetic fields, so they are eventually harmless for our main points below regarding correlations (this section and section \ref{sergo}) and entanglement (section \ref{senta}).

Choosing symmetric gauge, let $\bA=B(x\,\dd y-y\,\dd x+u\,\dd v-v\,\dd u)/2$. To diagonalize the quantum Hamiltonian $H$, it is then convenient to define dimensionless complex coordinates $z\equiv(x+iy)/\sqrt{2}\ell_B$, $w\equiv(u+iv)/\sqrt{2}\ell_B$ and annihilation operators
\be
\label{annihop}
a\equiv\frac{z}{2}+\der_{\bar z},
\qquad
b\equiv\frac{\bar z}{2}+\der_z,
\qquad
c\equiv\frac{w}{2}+\der_{\bar w},
\qquad
d\equiv\frac{\bar w}{2}+\der_w.
\ee
In these terms, the second class constraint \eqref{pix} at strong magnetic fields reads $a\approx c\approx0$ (in addition to their Hermitian conjugates). This is enforced in the quantum theory by restricting attention to those states for which $a|\phi\rangle=c|\phi\rangle=0$ \cite[chap.\ 13]{HenneauxTeitelboim}. Equivalently, such states minimize the pure Landau Hamiltonian $(\bp-\bA)^2/2=\hbar\cyclotron(a^{\dag}a+c^{\dag}c+1)$ and span the LLL. The trapping potential \eqref{trap} is then treated in the framework of degenerate perturbation theory, where its LLL projection reads $V\approx\hbar\omega\left(b^{\dag}b+1\right)+\hbar\omega' \left(d^{\dag}d+1\right)$. The ensuing approximate one-body energy spectrum is
\be
E_{m,n}
=
\hbar\omega\,m+\hbar \omega'\,n
\label{lolspec}
\ee
(up to an irrelevant additive constant) and the corresponding normalized eigenfunctions
\be
\phi_{mn}(z,\bar z,w,\bar w)
=
\frac{1}{\pi}\frac{z^mw^n}{\sqrt{m!\,n!}}\,
e^{-(|z|^2+|w|^2)/2}
\label{lola}
\ee
generalize standard LLL wavefunctions in 2D symmetric gauge. (This also holds without LLL projection, up to slightly different definitions of $(z,w)$: see the end of appendix \ref{app_quadratic}.)

Note that the probability density $|\phi_{mn}|^2$ is maximal on the torus where $(|z|,|w|)=(\sqrt{m},\sqrt{n})$. This is consistent with the fact that all guiding center trajectories lie in such tori (recall section \ref{sehar}). To some extent, it is even possible to build more localized eigenstates that exhibit {\it individual} trajectories, at least provided the ratio \eqref{rr} is rational: the energy \eqref{lolspec} only depends on the sum $\omega m+\omega'n$, so Fourier-transforming the wavefunctions \eqref{lola} along a fixed high-energy shell produces states localized on curves of the form $\beta=\cR\,\alpha+\text{cst}$. The remainder of this section is devoted to the proof of a similar localization affecting the correlation function of non-interacting Hall droplets.

\begin{figure}[t]
\centering
\begin{tikzpicture}[rotate=45,scale=.6]
\draw[->] (0,0) -- (8,0);
\draw (8,0) node[above] {$m$};
\draw[->] (0,0) -- (0,8);
\draw (0,8) node[above] {$n$};
\fill[MyRed] (0,0) circle (.08);
\fill[MyRed] (0.2,0) circle (.08);
\fill[MyRed] (0,0.2) circle (.08);
\fill[MyRed] (0.4,0) circle (.08);
\fill[MyRed] (0.2,0.2) circle (.08);
\fill[MyRed] (0,0.4) circle (.08);
\foreach \i in {0,...,3}{\fill[MyRed] (0.6-0.2*\i,0.2*\i) circle (.08);}
\foreach \i in {0,...,4}{\fill[MyRed] (0.8-0.2*\i,0.2*\i) circle (.08);}
\foreach \i in {0,...,5}{\fill[MyRed] (1-0.2*\i,0.2*\i) circle (.08);}
\foreach \i in {0,...,6}{\fill[MyRed] (1.2-0.2*\i,0.2*\i) circle (.08);}
\foreach \i in {0,...,7}{\fill[MyRed] (1.4-0.2*\i,0.2*\i) circle (.08);}
\foreach \i in {0,...,8}{\fill[MyRed] (1.6-0.2*\i,0.2*\i) circle (.08);}
\foreach \i in {0,...,9}{\fill[MyRed] (1.8-0.2*\i,0.2*\i) circle (.08);}
\foreach \i in {0,...,10}{\fill[MyRed] (2-0.2*\i,0.2*\i) circle (.08);}
\foreach \i in {0,...,11}{\fill[MyRed] (2.2-0.2*\i,0.2*\i) circle (.08);}
\foreach \i in {0,...,12}{\fill[MyRed] (2.4-0.2*\i,0.2*\i) circle (.08);}
\foreach \i in {0,...,13}{\fill[MyRed] (2.6-0.2*\i,0.2*\i) circle (.08);}
\foreach \i in {0,...,14}{\fill[MyRed] (2.8-0.2*\i,0.2*\i) circle (.08);}
\foreach \i in {0,...,15}{\fill[MyRed] (3.0-0.2*\i,0.2*\i) circle (.08);}
\foreach \i in {0,...,16}{\fill[MyRed] (3.2-0.2*\i,0.2*\i) circle (.08);}
\foreach \i in {0,...,17}{\fill[MyRed] (3.4-0.2*\i,0.2*\i) circle (.08);}
\foreach \i in {0,...,18}{\fill[MyRed] (3.6-0.2*\i,0.2*\i) circle (.08);}
\foreach \i in {0,...,19}{\fill[MyRed] (3.8-0.2*\i,0.2*\i) circle (.08);}
\foreach \i in {0,...,20}{\fill[MyRed] (4.0-0.2*\i,0.2*\i) circle (.08);}
\foreach \i in {0,...,21}{\fill[MyRed] (4.2-0.2*\i,0.2*\i) circle (.08);}
\foreach \i in {0,...,22}{\fill[MyRed] (4.4-0.2*\i,0.2*\i) circle (.08);}
\foreach \i in {0,...,23}{\fill[MyRed] (4.6-0.2*\i,0.2*\i) circle (.08);}
\foreach \i in {0,...,24}{\fill (4.8-0.2*\i,0.2*\i) circle (.08);}
\foreach \i in {0,...,25}{\fill (5.0-0.2*\i,0.2*\i) circle (.08);}
\foreach \i in {0,...,26}{\fill (5.2-0.2*\i,0.2*\i) circle (.08);}
\foreach \i in {0,...,27}{\fill (5.4-0.2*\i,0.2*\i) circle (.08);}
\foreach \i in {0,...,28}{\fill (5.6-0.2*\i,0.2*\i) circle (.08);}
\foreach \i in {0,...,29}{\fill (5.8-0.2*\i,0.2*\i) circle (.08);}
\foreach \i in {0,...,30}{\fill (6.0-0.2*\i,0.2*\i) circle (.08);}
\foreach \i in {0,...,31}{\fill (6.2-0.2*\i,0.2*\i) circle (.08);}
\foreach \i in {0,...,32}{\fill (6.4-0.2*\i,0.2*\i) circle (.08);}
\foreach \i in {0,...,33}{\fill (6.6-0.2*\i,0.2*\i) circle (.08);}
\foreach \i in {1,...,33}{\fill (6.8-0.2*\i,0.2*\i) circle (.08);}
\foreach \i in {2,...,33}{\fill (7.0-0.2*\i,0.2*\i) circle (.08);}
\foreach \i in {3,...,33}{\fill (7.2-0.2*\i,0.2*\i) circle (.08);}
\foreach \i in {4,...,33}{\fill (7.4-0.2*\i,0.2*\i) circle (.08);}
\foreach \i in {5,...,33}{\fill (7.6-0.2*\i,0.2*\i) circle (.08);}
\foreach \i in {6,...,33}{\fill (7.8-0.2*\i,0.2*\i) circle (.08);}
\foreach \i in {7,...,33}{\fill (8.0-0.2*\i,0.2*\i) circle (.08);}
\foreach \i in {7,...,34}{\node at (8.2-0.2*\i,0.2*\i) {$\cdot$};}
\foreach \i in {7,...,35}{\node at (8.4-0.2*\i,0.2*\i) {$\cdot$};}
\foreach \i in {7,...,36}{\node at (8.6-0.2*\i,0.2*\i) {$\cdot$};}
\foreach \i in {8,...,36}{\node at (8.8-0.2*\i,0.2*\i) {$\cdot$};}
\foreach \i in {0,...,6}{\node at (6.8,0.2*\i) {$\cdot$};}
\foreach \i in {0,...,6}{\node at (7.0,0.2*\i) {$\cdot$};}
\foreach \i in {0,...,6}{\node at (7.2,0.2*\i) {$\cdot$};}
\foreach \i in {0,...,6}{\node at (0.2*\i,6.8) {$\cdot$};}
\foreach \i in {0,...,6}{\node at (0.2*\i,7.0) {$\cdot$};}
\foreach \i in {0,...,6}{\node at (0.2*\i,7.2) {$\cdot$};}
\draw (4.8,-0.1) node[right] {$\cN$};
\draw[->] (-4.5,4.5)--(-.5,8.5);
\draw (-2.5,6.5) node[right] {Energy};
\end{tikzpicture}
\caption{The spectrum \eqref{lolspec} of LLL states \eqref{lola} in an isotropic trap \eqref{trap}. In contrast to the 2D QHE, degeneracies persist despite the trap: wavefunctions $\phi_{mn}$ with the same value of $m+n$ have the same energy. The red dots are states that contribute to the many-body ground state \eqref{omegan}. This should be compared with the anisotropic spectrum in fig.\ \ref{find}.}
\label{fispectrum}
\end{figure}
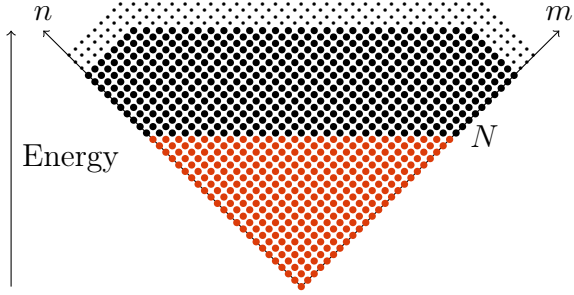

\phantomsection
\addcontentsline{toc}{subsection}{Isotropic droplets}%
\paragraph*{Isotropic droplets.}
\label{sesotrap}
Consider a droplet of non-interacting fermions subjected to the Landau Hamiltonian \eqref{hav} in an {\it isotropic} trap \eqref{trap}, so that the frequency ratio \eqref{rr} is $\cR=1$. As in section \ref{seclas}, we assume that the magnetic field is so large that low energy physics is entirely described by the LLL. The many-body ground state (see fig.\ \ref{fispectrum}) then reads
\be
|\Omega\rangle
=
\prod_{\substack{m,n\in\NN\\ m+n<\cN}}a^{\dagger}_{mn}|0\rangle
\label{omegan}
\ee
for some integer $\cN\gg1$ determined by the Fermi energy, with $|0\rangle$ denoting the empty state and $a^{\dagger}_{mn}$ a second-quantized creation operator for \eqref{lola}. The number of particles in the droplet is thus $\cN(\cN+1)/2\sim\cN^2/2$.

Now let $\bx$ and $\bx'$ be two points in $\RR^4$ with respective complex coordinates $(z,w)$ and $(z',w')$, as defined above eqs.\ \eqref{annihop}. Owing to the form \eqref{lola} of energy eigenstates, the correlation between these points in the ground state \eqref{omegan} is a sum
\be
\CC(\bx,\bx')
=
\sum_{\substack{m,n\in\NN\\ m+n<\cN}}
\phi_{mn}^*(\bx)\,\phi_{mn}(\bx')
\propto
\sum_{\substack{m,n\in\NN\\m+n<\cN}}
\frac{(\bar zz')^m\,(\bar ww')^n}{m!\,n!}.
\label{contra}
\ee
Changing summation variables into $m'\equiv m+n$ and $n'\equiv n$, one recognizes a binomial expansion in $n'$. The correlator can thus be recast as an incomplete gamma function
\be
\CC(Z,Z')
=
\frac{1}{\pi^2}
e^{-|Z-Z'|^2/2}\,
e^{(Z^{\dagger}Z'-Z'{}^{\dagger}Z)/2}\,
\frac{\Gamma(\cN,Z^{\dagger}Z')}{\Gamma(\cN)}
\label{cofaz}
\ee
in terms of complex column vectors $Z\equiv(z\;w)^t$ and $Z'\equiv(z'\;w')^t$. Similarly to 2D droplets, bulk correlations decay in a Gaussian manner and the density $\CC(\bx,\bx)$ is nearly constant in a 4D ball whose boundary is the sphere $S^3\subset\mathbb{C}^2$ where $|Z|=\sqrt{\cN}$. 

As in the 2D QHE, the correlator \eqref{cofaz} is most interesting near the boundary of the droplet, where it is sensitive to edge modes. Our goal is thus to find an asymptotic formula for \eqref{cofaz} in the thermodynamic limit, in the regime where both $\bx$ and $\bx'$ are close to the sphere at $|Z|=\sqrt{\cN}$. One can readily anticipate that the 4D result will differ from its 2D cousin: in 4D, the incomplete gamma function \eqref{cofaz} implies that non-zero edge correlations occur when $|Z|\sim\sqrt{\cN}$, $|Z'|\sim\sqrt{\cN}$ and $|Z^{\dagger}Z'|\sim\cN$ all at once. The last condition would follow from the two first ones in 2D, but this is not so in higher dimensions, where the Cauchy-Schwarz inequality $|Z^{\dagger}Z'|\leq|Z||Z'|$ is generally not saturated. Non-zero edge correlations thus occur when $Z\sim e^{i\phii}\,Z'$ for some $\phii\in\RR$, which is to say that $Z$ and $Z'$ lie on the same guiding center trajectory \eqref{gc}. This already confirms the classical intuition of section \ref{seclas}; we now dig deeper by zooming in on correlations near the edge.

To begin, recall a general result on the asymptotics of the incomplete gamma function. Using steepest-descent methods based on the definition of the gamma function as an integral in the complex plane, a straightforward but lengthy computation shows that the large $\cN$ limit of $\Gamma(\cN,\cN\lambda e^{i\phii})$, with fixed $\lambda>0$ and $\phii\neq0$, is given by\footnote{Eq.\ \eqref{mare} actually holds only if $\lambda\leq1$, or if $\lambda>1$ and $|\phii|>\phii_c$ for some critical value $\phii_c$ (see appendix \ref{appa}). When, instead, $\lambda>1$ and $|\phii|\leq\phii_c$, the $\Gamma(\cN)$ on the right-hand side of \eqref{mare} disappears, but this does not affect the asymptotics of correlations \eqref{cofaz}, where $\Gamma(\cN)$ turns out to be negligible anyway.\label{funnyfoo}}
\be
\Gamma(\cN,\cN\lambda e^{i\phii})
\stackrel{\cN\to\infty}{\sim}
\Gamma(\cN)
-\frac{1}{\cN}
\left[\cN\lambda e^{i\phii-\lambda e^{i\phii}}\right]^{\cN}
\frac{1-\lambda e^{-i\phii}}{\lambda^2+1-2\lambda\cos\phii}.
\label{mare}
\ee
The details of this argument are relegated to appendix \ref{appa}. Our goal is now to apply \eqref{mare} to the correlation function \eqref{cofaz} in the thermodynamic limit $\cN\to\infty$, with $\lambda=|Z^{\dagger}Z'|/\cN\sim1$. In order to relate the result to the classical trajectories of section \ref{sehar}, we write the complex coordinates of $\RR^4\cong\mathbb{C}^2$ as
\be
Z
=
\begin{pmatrix}
z\\ w
\end{pmatrix}
\equiv
r\,
\begin{pmatrix}
e^{i\alpha}\cos(\theta/2)\\
e^{i\beta}\sin(\theta/2)
\end{pmatrix},
\label{hopfco}
\ee
where $r\geq0$ and $\theta\in[0,\pi]$. The same notation applies to $Z'$ (with extra primes), but we assume without loss of generality that $\alpha'=\beta'=0$ since the correlation \eqref{contra} only depends on differences of phases between $(z,w)$ and $(z',w')$. One can then express the various ingredients of eq.\ \eqref{cofaz} in these terms. In order to approach the edge, let $r\equiv\sqrt{\cN}+a$ and $r'\equiv\sqrt{\cN}+b$, being understood that $a,b$ are finite and fixed in the limit $\cN\to\infty$. As for angular coordinates, section \ref{seclas} suggests that edge modes propagate along $(\alpha+\beta)/2$, while the remaining angles $(\theta,\alpha-\beta)$ parametrize a two-sphere $S^2$. Accordingly, we assume that $\alpha-\beta$ and $\delta\theta\equiv\theta'-\theta$ are both of order $\cO(1/\sqrt{\cN})$, whereupon \eqref{mare} yields
\be
\CC(Z,Z')
\sim
-
\frac{1}{\pi^2}\,
\frac{e^{-i(\cN-1/2)\alpha}}{i\sqrt{8\pi\cN}\sin(\alpha/2)}\,
e^{-a^2-b^2-\frac{\cN}{8}[(\alpha-\beta)^2\sin^2\theta+\delta\theta^2]}.
\label{asycofo}
\ee
This expression is a key formula for edge correlations in an isotropic droplet. It holds in the thermodynamic limit $\cN\gg1$ and for $\alpha\sim\beta\neq0$, excluding in particular the density regime $\alpha=\beta=0$ (whose asymptotics are radically different).

Eq.\ \eqref{asycofo} is reminiscent of its 2D cousin (see \eg \cite[eq.~(27)]{Estienne2019hmd}) in two respects: first, it decays in a Gaussian manner $\propto e^{-a^2-b^2}$ in the radial direction, confirming that edge correlations are localized, to within a magnetic length, on the $S^3$ boundary. Second, it exhibits a power-law correlation $\propto[\sin(\alpha/2)]^{-1}$, characteristic of fermionic 1D CFTs. What distinguishes \eqref{asycofo} from its 2D analogue is the additional localization $\propto e^{-\frac{\cN}{8}[(\alpha-\beta)^2\sin^2\theta+\delta\theta^2]}$ on the guiding center trajectory \eqref{gc} where $\alpha=\beta$, which becomes sharper as $\cN$ increases. This confirms that 4D edge modes are gapless fermions that effectively propagate along 1D circles embedded in a 3D sphere. Note that the generalization to $d$-dimensional isotropic droplets is straightforward: the correlator is still given by eq. \eqref{cofaz} with $Z=(z_1\; \cdots \; z_{d/2})^t$, and edge modes are still supported on circles. We shall now extend this result to less symmetric traps, where the embedding of edge modes in the $S^3$ boundary is more tangled than in the Hopf fibration. Along the way we will derive the edge correlation \eqref{asycofo} without relying on the incomplete gamma function.

\begin{figure}[t]
\centering
\begin{tikzpicture}[rotate=31,scale=.6]
\draw[->] (0,0) -- (11,0);
\draw (11,0) node[above] {$m$};
\draw[->] (0,0) -- (0,7);
\draw (0,7) node[above] {$n$};
\foreach \i in {0,...,52}{\node at (0.2*\i,0) {$\cdot$};}
\foreach \i in {0,...,50}{\node at (0.2*\i,0.2) {$\cdot$};}
\foreach \i in {0,...,48}{\node at (0.2*\i,0.4) {$\cdot$};}
\foreach \i in {0,...,47}{\node at (0.2*\i,0.6) {$\cdot$};}
\foreach \i in {0,...,45}{\node at (0.2*\i,0.8) {$\cdot$};}
\foreach \i in {0,...,43}{\node at (0.2*\i,1.0) {$\cdot$};}
\foreach \i in {0,...,42}{\node at (0.2*\i,1.2) {$\cdot$};}
\foreach \i in {0,...,40}{\node at (0.2*\i,1.4) {$\cdot$};}
\foreach \i in {0,...,38}{\node at (0.2*\i,1.6) {$\cdot$};}
\foreach \i in {0,...,37}{\node at (0.2*\i,1.8) {$\cdot$};}
\foreach \i in {0,...,35}{\node at (0.2*\i,2.0) {$\cdot$};}
\foreach \i in {0,...,33}{\node at (0.2*\i,2.2) {$\cdot$};}
\foreach \i in {0,...,32}{\node at (0.2*\i,2.4) {$\cdot$};}
\foreach \i in {0,...,30}{\node at (0.2*\i,2.6) {$\cdot$};}
\foreach \i in {0,...,28}{\node at (0.2*\i,2.8) {$\cdot$};}
\foreach \i in {0,...,27}{\node at (0.2*\i,3.0) {$\cdot$};}
\foreach \i in {0,...,25}{\node at (0.2*\i,3.2) {$\cdot$};}
\foreach \i in {0,...,23}{\node at (0.2*\i,3.4) {$\cdot$};}
\foreach \i in {0,...,22}{\node at (0.2*\i,3.6) {$\cdot$};}
\foreach \i in {0,...,20}{\node at (0.2*\i,3.8) {$\cdot$};}
\foreach \i in {0,...,18}{\node at (0.2*\i,4.0) {$\cdot$};}
\foreach \i in {0,...,17}{\node at (0.2*\i,4.2) {$\cdot$};}
\foreach \i in {0,...,15}{\node at (0.2*\i,4.4) {$\cdot$};}
\foreach \i in {0,...,13}{\node at (0.2*\i,4.6) {$\cdot$};}
\foreach \i in {0,...,12}{\node at (0.2*\i,4.8) {$\cdot$};}
\foreach \i in {0,...,10}{\node at (0.2*\i,5.0) {$\cdot$};}
\foreach \i in {0,...,8}{\node at (0.2*\i,5.2) {$\cdot$};}
\foreach \i in {0,...,7}{\node at (0.2*\i,5.4) {$\cdot$};}
\foreach \i in {0,...,5}{\node at (0.2*\i,5.6) {$\cdot$};}
\foreach \i in {0,...,3}{\node at (0.2*\i,5.8) {$\cdot$};}
\foreach \i in {0,...,2}{\node at (0.2*\i,6.0) {$\cdot$};}
\node at (0,6.2) {$\cdot$};
\foreach \i in {0,...,49}{\fill (0.2*\i,0) circle (.08);}
\foreach \i in {0,...,47}{\fill (0.2*\i,0.2) circle (.08);}
\foreach \i in {0,...,45}{\fill (0.2*\i,0.4) circle (.08);}
\foreach \i in {0,...,44}{\fill (0.2*\i,0.6) circle (.08);}
\foreach \i in {0,...,42}{\fill (0.2*\i,0.8) circle (.08);}
\foreach \i in {0,...,40}{\fill (0.2*\i,1.0) circle (.08);}
\foreach \i in {0,...,39}{\fill (0.2*\i,1.2) circle (.08);}
\foreach \i in {0,...,37}{\fill (0.2*\i,1.4) circle (.08);}
\foreach \i in {0,...,35}{\fill (0.2*\i,1.6) circle (.08);}
\foreach \i in {0,...,34}{\fill (0.2*\i,1.8) circle (.08);}
\foreach \i in {0,...,32}{\fill (0.2*\i,2.0) circle (.08);}
\foreach \i in {0,...,30}{\fill (0.2*\i,2.2) circle (.08);}
\foreach \i in {0,...,29}{\fill (0.2*\i,2.4) circle (.08);}
\foreach \i in {0,...,27}{\fill (0.2*\i,2.6) circle (.08);}
\foreach \i in {0,...,25}{\fill (0.2*\i,2.8) circle (.08);}
\foreach \i in {0,...,24}{\fill (0.2*\i,3.0) circle (.08);}
\foreach \i in {0,...,22}{\fill (0.2*\i,3.2) circle (.08);}
\foreach \i in {0,...,20}{\fill (0.2*\i,3.4) circle (.08);}
\foreach \i in {0,...,19}{\fill (0.2*\i,3.6) circle (.08);}
\foreach \i in {0,...,17}{\fill (0.2*\i,3.8) circle (.08);}
\foreach \i in {0,...,15}{\fill (0.2*\i,4.0) circle (.08);}
\foreach \i in {0,...,14}{\fill (0.2*\i,4.2) circle (.08);}
\foreach \i in {0,...,12}{\fill (0.2*\i,4.4) circle (.08);}
\foreach \i in {0,...,10}{\fill (0.2*\i,4.6) circle (.08);}
\foreach \i in {0,...,9}{\fill (0.2*\i,4.8) circle (.08);}
\foreach \i in {0,...,7}{\fill (0.2*\i,5.0) circle (.08);}
\foreach \i in {0,...,5}{\fill (0.2*\i,5.2) circle (.08);}
\foreach \i in {0,...,4}{\fill (0.2*\i,5.4) circle (.08);}
\foreach \i in {0,...,2}{\fill (0.2*\i,5.6) circle (.08);}
\fill (0,5.8) circle (.08);
\foreach \i in {0,...,30}{\fill[MyRed] (0.2*\i,0) circle (.08);}
\foreach \i in {0,...,28}{\fill[MyRed] (0.2*\i,0.2) circle (.08);}
\foreach \i in {0,...,26}{\fill[MyRed] (0.2*\i,0.4) circle (.08);}
\foreach \i in {0,...,25}{\fill[MyRed] (0.2*\i,0.6) circle (.08);}
\foreach \i in {0,...,23}{\fill[MyRed] (0.2*\i,0.8) circle (.08);}
\foreach \i in {0,...,21}{\fill[MyRed] (0.2*\i,1.0) circle (.08);}
\foreach \i in {0,...,20}{\fill[MyRed] (0.2*\i,1.2) circle (.08);}
\foreach \i in {0,...,18}{\fill[MyRed] (0.2*\i,1.4) circle (.08);}
\foreach \i in {0,...,16}{\fill[MyRed] (0.2*\i,1.6) circle (.08);}
\foreach \i in {0,...,15}{\fill[MyRed] (0.2*\i,1.8) circle (.08);}
\foreach \i in {0,...,13}{\fill[MyRed] (0.2*\i,2.0) circle (.08);}
\foreach \i in {0,...,11}{\fill[MyRed] (0.2*\i,2.2) circle (.08);}
\foreach \i in {0,...,10}{\fill[MyRed] (0.2*\i,2.4) circle (.08);}
\foreach \i in {0,...,8}{\fill[MyRed] (0.2*\i,2.6) circle (.08);}
\foreach \i in {0,...,6}{\fill[MyRed] (0.2*\i,2.8) circle (.08);}
\foreach \i in {0,...,5}{\fill[MyRed] (0.2*\i,3.0) circle (.08);}
\foreach \i in {0,...,3}{\fill[MyRed] (0.2*\i,3.2) circle (.08);}
\foreach \i in {0,...,1}{\fill[MyRed] (0.2*\i,3.4) circle (.08);}
\fill[MyRed] (0,3.6) circle (.08);
\draw[->] (-6,4)--(-3,9);
\draw (-4.2,7) node[right] {Energy};
\end{tikzpicture}
\caption{The spectrum \eqref{lolspec} for an anisotropic trap with stiffness ratio $\cR=5/3$. As in fig.\ \ref{fispectrum}, red dots highlight the occupied states of the many-body ground state \eqref{moc}.}
\label{find}
\end{figure}
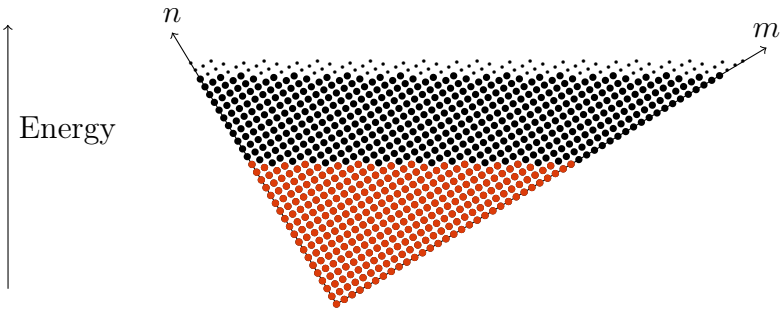

\phantomsection
\addcontentsline{toc}{subsection}{Rational anisotropic droplets}%
\paragraph*{Rational anisotropic droplets.}
\label{senisotro}
Let us generalize the asymptotics \eqref{asycofo} to arbitrary anisotropic traps with \emph{periodic} guiding center trajectories. In the language of section \ref{sehar}, this corresponds to a rational ratio \eqref{rr}, which we write as $\cR=p/q$ in terms of coprime integers $p,q$. We assume, as before, that the magnetic field is so strong that the many-body ground state only contains one-body states in the LLL, with a spectrum \eqref{lolspec}. However, because of the anisotropy, the ground state is no longer given by eq.\ \eqref{omegan}; it consists instead of wavefunctions \eqref{lola} whose indices $m,n$ satisfy (see fig.\ \ref{find})
\be
0
\leq
m+\cR \,n
<
\cN
\label{moc}
\ee
for some integer $\cN\gg1$ proportional to the Fermi energy. The number of electrons is thus $\cN^2/(2\cR)$ in the thermodynamic limit. Owing to \eqref{lola}, the correlation function is
\be
\CC(\bx,\bx')
=
\!\!\sum_{\substack{m,n\in\NN \\ m+\cR \,n<\cN}}\!\!
\phi_{mn}^*(\bx)\phi_{mn}(\bx')
\propto
\!\!\sum_{\substack{m,n\in\NN \\ m+\cR \,n<\cN}}\!\!
\frac{(\bar zz')^m\,(\bar ww')^n}{m!\,n!},
\label{t3}
\ee
which reduces to \eqref{contra} in the isotropic case $\cR=1$. The density $\CC(\bx,\bx)$ is almost constant in the squashed 4D ball where $|z|^2+\cR |w|^2<\cN$, outside of which it essentially vanishes; the edge of the droplet is the squashed three-sphere where $|z|^2+ \cR |w|^2=\cN$.

\begin{SCfigure}[2][t]
\begin{tikzpicture}[scale=.7]
\draw[->] (-.2,0) -- (7,0);
\draw (7,0) node[above] {$\alpha$};
\draw[->] (0,-.2) -- (0,11);
\draw (0,11) node[right] {$\beta$};
\draw (2,-.2)--(2,10);
\draw (4,-.2)--(4,10);
\draw (6,-.2)--(6,10);
\draw (-.2,2)--(6,2);
\draw (-.2,4)--(6,4);
\draw (-.2,6)--(6,6);
\draw (-.2,8)--(6,8);
\draw (-.2,10)--(6,10);
\draw (2,0) node[below] {$2\pi$};
\draw (4,0) node[below] {$4\pi$};
\draw (6,0) node[below] {$6\pi$};
\draw (0,2) node[left] {$2\pi$};
\draw (0,4) node[left] {$4\pi$};
\draw (0,6) node[left] {$6\pi$};
\draw (0,8) node[left] {$8\pi$};
\draw (0,10) node[left] {$10\pi$};
\begin{scope}
\clip (0,0)--(2,0)--(2,2)--(0,2)--cycle;
\draw[MyRed,ultra thick] (0,0)--(6,10);
\draw[MyOrange,ultra thick] (0,-2)--(6,8);
\draw[MyYellow,ultra thick] (-2,-2)--(4,8);
\draw[MyGreen,ultra thick] (-2,-4)--(4,6);
\draw[MyTurk,ultra thick] (-2,-6)--(4,4);
\draw[MyBlue,ultra thick] (-4,-6)--(2,4);
\draw[MyPurple,ultra thick] (-4,-8)--(2,2);
\end{scope}
\begin{scope}
\clip (0,2)--(2,2)--(2,4)--(0,4)--cycle;
\draw[MyOrange,ultra thick] (0,0)--(6,10);
\end{scope}
\begin{scope}
\clip (2,2)--(4,2)--(4,4)--(2,4)--cycle;
\draw[MyYellow,ultra thick] (0,0)--(6,10);
\end{scope}
\begin{scope}
\clip (2,4)--(4,4)--(4,6)--(2,6)--cycle;
\draw[MyGreen,ultra thick] (0,0)--(6,10);
\end{scope}
\begin{scope}
\clip (2,6)--(4,6)--(4,8)--(2,8)--cycle;
\draw[MyTurk,ultra thick] (0,0)--(6,10);
\end{scope}
\begin{scope}
\clip (4,6)--(6,6)--(6,8)--(4,8)--cycle;
\draw[MyBlue,ultra thick] (0,0)--(6,10);
\end{scope}
\begin{scope}
\clip (4,8)--(6,8)--(6,10)--(4,10)--cycle;
\draw[MyPurple,ultra thick] (0,0)--(6,10);
\end{scope}
\draw[ultra thick,rounded corners=1] (0,0)--(2,0)--(2,2)--(0,2)--cycle;
\draw[ultra thick,rounded corners=1] (0,0)--(6,0)--(6,10)--(0,10)--cycle;
\end{tikzpicture}
\caption{Periodic motion on a torus, with rational winding $p/q$, may be seen as the projection of a trajectory with unit winding on a $pq$-fold larger torus. Here we take $p/q=5/3$, as in the central panels of figs.\ \ref{fimotion}--\ref{fitorus}. As a result, the small torus where $(\alpha,\beta)\in[0,2\pi]\times[0,2\pi]$ may be seen as a quotient of the larger torus where $(\alpha,\beta)\in[0,6\pi]\times[0,10\pi]$ under the action of the group $\ZZ_3\times\ZZ_5$ that maps $(\alpha,\beta)$ on $(\alpha+2\pi\mu,\beta+2\pi\nu)$ with $\mu=0,1,2$ and $\nu=0,1,...,4$. The method of images mentioned in the main text, and exposed in greater detail in appendix \ref{appb}, exploits this picture to recast edge correlations in an {\it anisotropic} trap as a sum of shifted correlations in an {\it isotropic} trap.}
\label{fimeth}
\end{SCfigure}

As before, we wish to find the large $\cN$ asymptotics of the correlator \eqref{t3} near the edge. This can be done in three steps: (i) use a `method of images' to simplify the sum \eqref{t3}, (ii) establish the asymptotics of the summand in \eqref{t3}, (iii) approximate the sum by a series and evaluate it. This approach is explained in detail in appendix \ref{appb}, and the method of images is depicted in fig.\ \ref{fimeth}; here, we merely state the result. Namely, let the points $\bx,\bx'$ lie on the boundary and write their complex coordinates as
\be
\label{zz}
\begin{split}
z&=\sqrt{\cN}\,e^{i\alpha}\,\cos(\theta/2),\\
w&=\sqrt{\cN/\cR}\,e^{i\beta}\,\sin(\theta/2),
\end{split}
\qquad
\begin{split}
z'&=\sqrt{\cN}\,\cos(\theta/2),\\
w'&=\sqrt{\cN/\cR}\,\sin(\theta/2).
\end{split}
\ee
Note that we choose the same radii and the same azimuth $\theta$ for both $\bx$ and $\bx'$; this ensures that both points belong to the same torus in the edge, although radial and azimuthal corrections can be included similarly to what we did in \eqref{asycofo} for isotropic droplets. We also assume without loss of generality that the complex coordinates of $\bx'$ are both real (this is again because \eqref{t3} is invariant under independent rotations of $z$ and $w$), and we avoid singular cases by restricting attention to $\theta\in(0,\pi)$. Under these assumptions, edge correlations are found to be
\begin{empheq}[box=\othermathbox]{equation}
\label{main}
\CC(\bx,\bx')\;
\sim \;
\sum_{\mu=0}^{q-1}
\sum_{\nu=0}^{p-1}
\frac{e^{-i\frac{\alpha+2\pi\mu}{2q}}}%
{\sqrt{\cN}\, q\sin\big(\frac{\alpha+2\pi\mu}{2q}\big)}\,
e^{-\frac{\cN}{8}\left(\alpha+2\pi\mu
-\frac{\beta+2\pi\nu}{\cR}\right)^2
\sin^2\theta\,F(\theta)}
\end{empheq}
where we omit an overall prefactor (see the exact formula (\ref{mainbis})) and $F(\theta)\equiv[\sin^2(\theta/2)+\cos^2(\theta/2)/\cR]^{-1}$. The double sum runs over the aforementioned images: periodic motion with winding $p/q$ on the unit torus lifts into a trajectory with unit winding on a larger covering torus consisting of $pq$ copies of the initial one (see fig.\ \ref{fimeth}). The coordinates $(\alpha,\beta)$ then have periods $(2\pi q,2\pi p)$ on the larger torus; the unit torus follows by identifying points that differ by an action $(\alpha,\beta)\mapsto(\alpha+2\pi\mu,\beta+2\pi\nu)$ of the group $\ZZ_q\times\ZZ_p$.

\begin{figure}[t]
\includegraphics[width=0.32\textwidth]{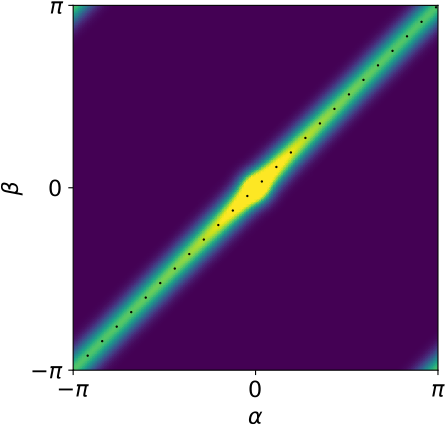}\hfill
\includegraphics[width=0.32\textwidth]{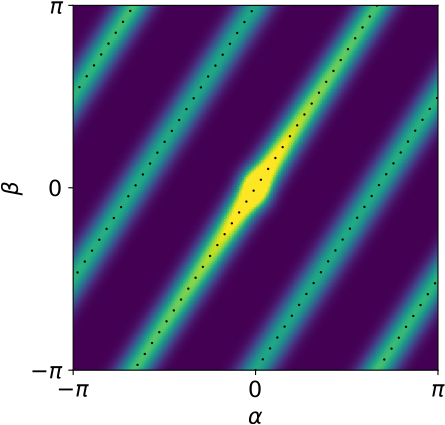}\hfill
\includegraphics[width=0.32\textwidth]{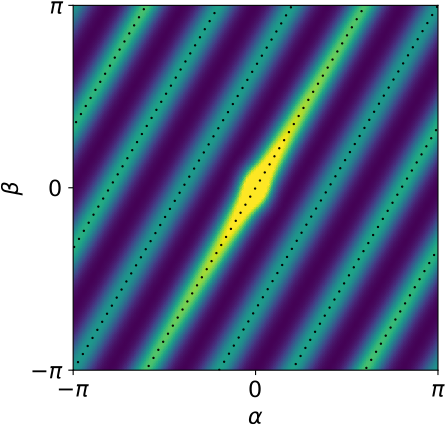}
\caption{The norm $|\CC(\bx,\bx')|$, computed numerically thanks to eq.\ \eqref{t3}, for boundary points whose complex coordinates \eqref{zz} have $\theta=\pi/2$. All three plots represent a torus whose horizontal and vertical axes are respectively $\alpha,\beta$. Yellow denotes higher norms, navy blue lower ones; intermediate colors interpolate. From left to right, anisotropy ratios are $\cR=1$, $3/2$, $5/3$ (respectively with $45150$, $30200$, $27210$ particles). The localization on the guiding center trajectories of fig.\ \ref{fimotion} (here shown as black dots) is manifest.}
\label{fiMainPlot}
\end{figure}

Eq.\ \eqref{main} is one of our main results; it is an explicit formula for edge correlations (at fixed $\theta$) in the case of rational ratios \eqref{rr}. Such correlations are plotted in fig.\ \ref{fiMainPlot}. When $p=q=1$, eq.\ \eqref{main} reproduces the isotropic expression \eqref{asycofo} with $a=b=\delta\theta=0$. More generally, \eqref{main} implies that edge modes in rational droplets have power-law correlations localized on classical guiding center trajectories (where $\beta=\cR\,\alpha$), away from which they decay in a Gaussian manner. More precisely, if $\bx$ and $\bx'$ do not lie on the same classical trajectory, the correlator decays exponentially with $N$. Meanwhile, on a classical trajectory, the correlator reduces to that of a (1+1)D chiral massless fermion on a circle:
\begin{equation}
\label{CFT_correlator_circle}
     \left.\CC(\bx,\bx')\right|_{\beta\,=\,\Delta\, \alpha }\; \sim  \langle \Psi^{\dag}(\bx) \Psi(\bx') \rangle_{\textrm{CFT}} = \frac{1}{\frac{L}{\pi}\sin\left(\pi\frac{\ell}{L}\right)} 
\end{equation}
where $\ell = \ell(\bx,\bx')$ is the distance separating $\bx$ and $\bx'$ \emph{along the classical trajectory}; the latter is diffeomorphic to a circle with total length $L= 2\pi q\sqrt{N} \sqrt{\cos^2(\theta/2)+\cR\sin^{2}(\theta/2)}$ (in units of the magnetic length). Note that $\ell$ is indeed a correct conformal parametrization of the classical trajectory \eqref{gc}, since the guiding center velocity has constant norm.

\section{Ergodic edge modes}
\label{sergo}

Edge correlations in irrational anisotropic traps are challenging: guiding center trajectories are no longer periodic and the sum over images \eqref{main} turns into a series. To resolve this puzzle, the present section reports a numerical exploration of edge correlations in irrational traps. A natural proposal will thus emerge for their thermodynamic limiting form. It involves a function with a fractal-like graph, not unlike the Thoma\'e function, that may be seen as an irrational limit of the strict $\cN=\infty$ version of the rational formula \eqref{main}.  In practice, the correlation function of ergodic edge modes then decays as a power law along classical trajectories, while it falls off sharply along transverse directions.

\phantomsection
\addcontentsline{toc}{subsection}{Continued fractions and numerics}%
\paragraph*{Continued fractions and numerics.} In order to build intuition on irrational droplets, it will be helpful to approximate irrational ratios \eqref{rr} by sequences of rational numbers; this can be done with continued fractions. An example that we will use throughout this section is the continued fraction representation of $\cR=\sqrt{2}$:
\begin{align}
\sqrt{2}&
=
1+\frac{1}{2+\frac{1}{2+\frac{1}{2+\cdots}}}
\equiv
[1,2,2,2,2,\ldots].
\end{align}
Successive approximations of $\sqrt{2}$ are obtained by truncating this sequence, giving $3/2$, $7/5$, $17/12$, $41/29$, etc. A similar rewriting applies to any irrational number; it is optimal in the sense that (i) truncated approximations given by continued fractions converge exponentially fast with the order of truncation, (ii) if a rational approximation $p'/q'$ is closer to $\cR\notin\mathbb{Q}$ than some truncated continued fraction $p/q$, then $q'>q$ \cite{Khinchin}.

\begin{figure}[t]
\includegraphics[width=0.32\textwidth]{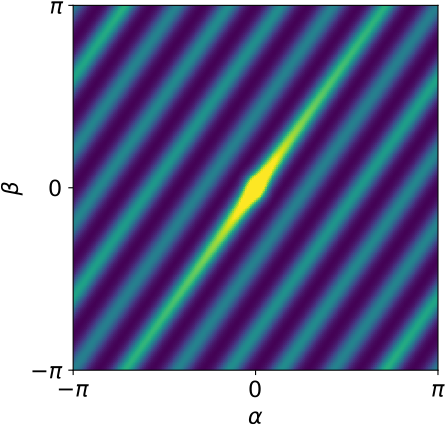}\hfill
\includegraphics[width=0.32\textwidth]{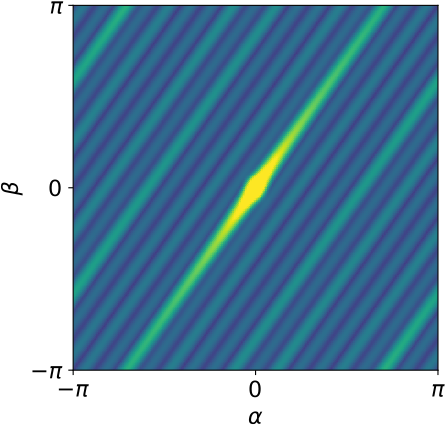}\hfill
\includegraphics[width=0.32\textwidth]{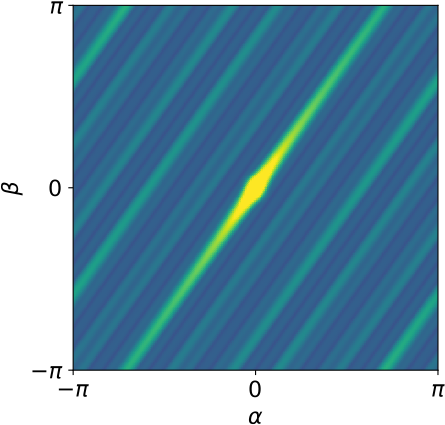}
\caption{Numerical computations of the norm $|\CC(\bx,\bx')|$ given by \eqref{t3}, for points \eqref{zz} on an edge torus at $\theta=\pi/2$. The color coding is the same as in fig.\ \ref{fiMainPlot}. From left to right, the ratio \eqref{rr} provides successive continued fraction approximations of $\sqrt{2}$, namely $\cR=7/5$, $\cR=17/12$, and $\cR=41/29$ (respectively with 129043, 127553 and 127822 particles). Classical trajectories become more and more difficult to distinguish as $\cR$ converges to an irrational value, motivating the more accurate `slice' plots in fig.\ \ref{fig:irr2} below.}
\label{fig:irr1}
\end{figure}

\begin{figure}[t]
\includegraphics[width=0.48\textwidth]{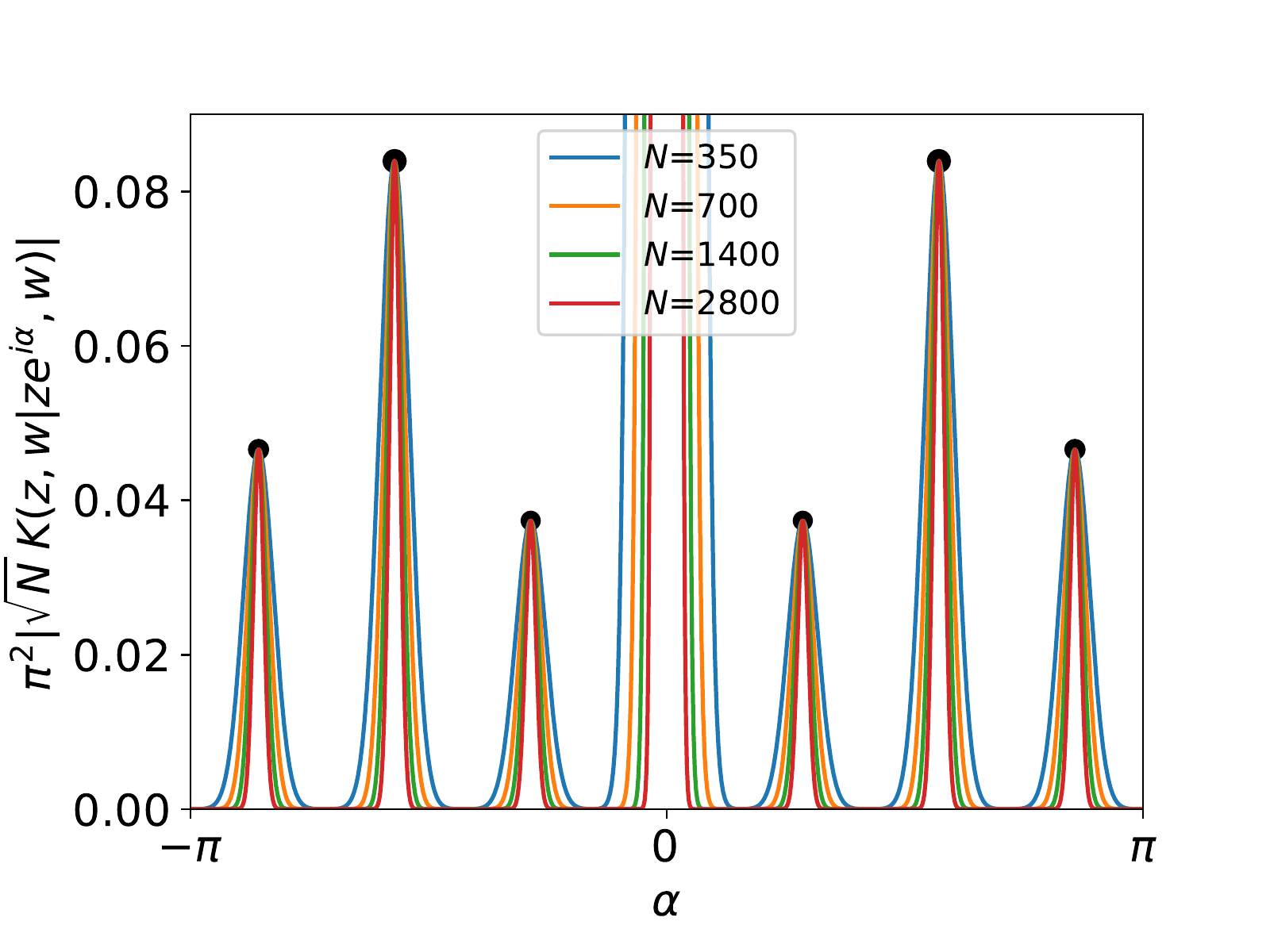}\hfill
\includegraphics[width=0.48\textwidth]{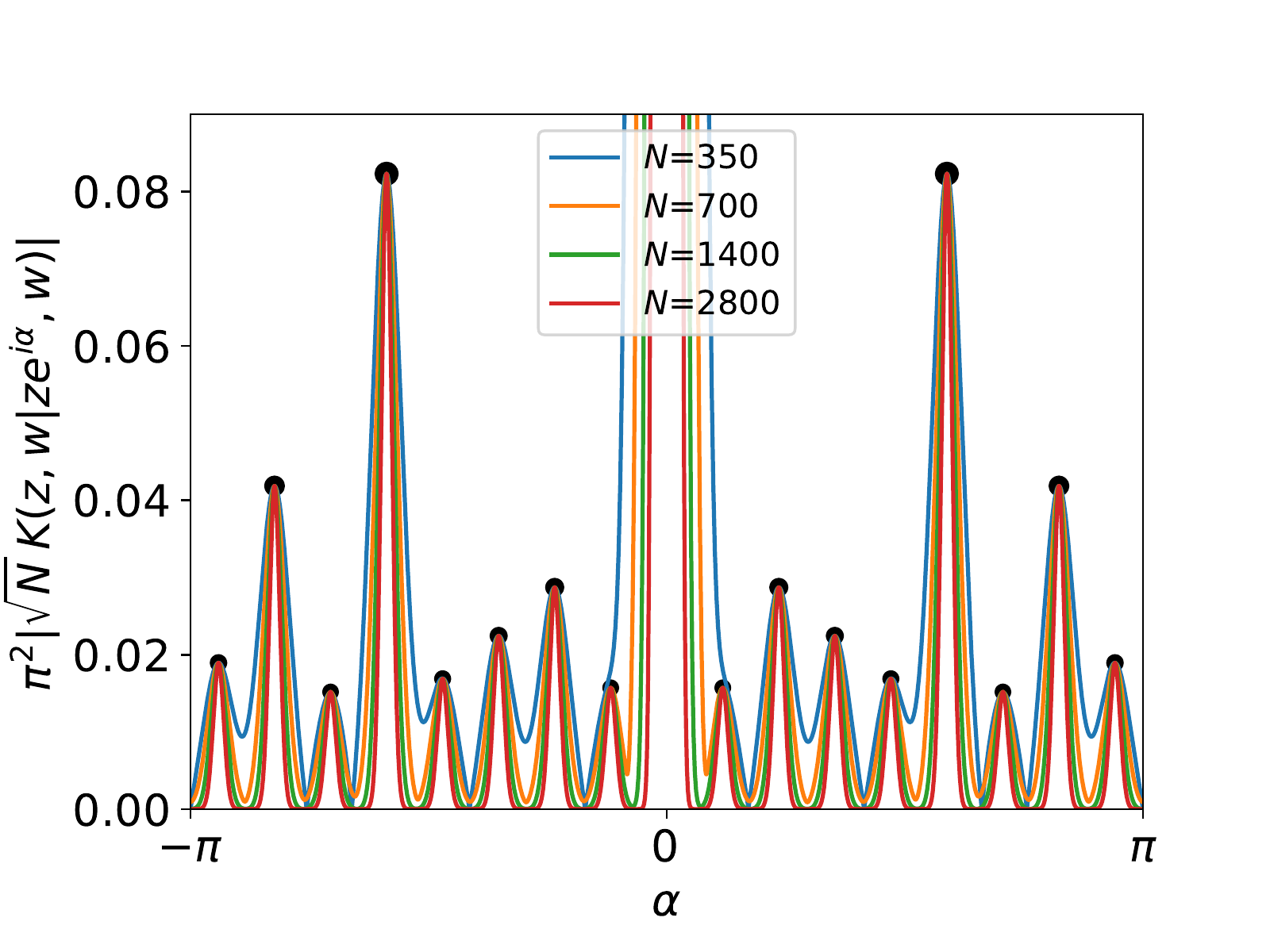}
\includegraphics[width=0.48\textwidth]{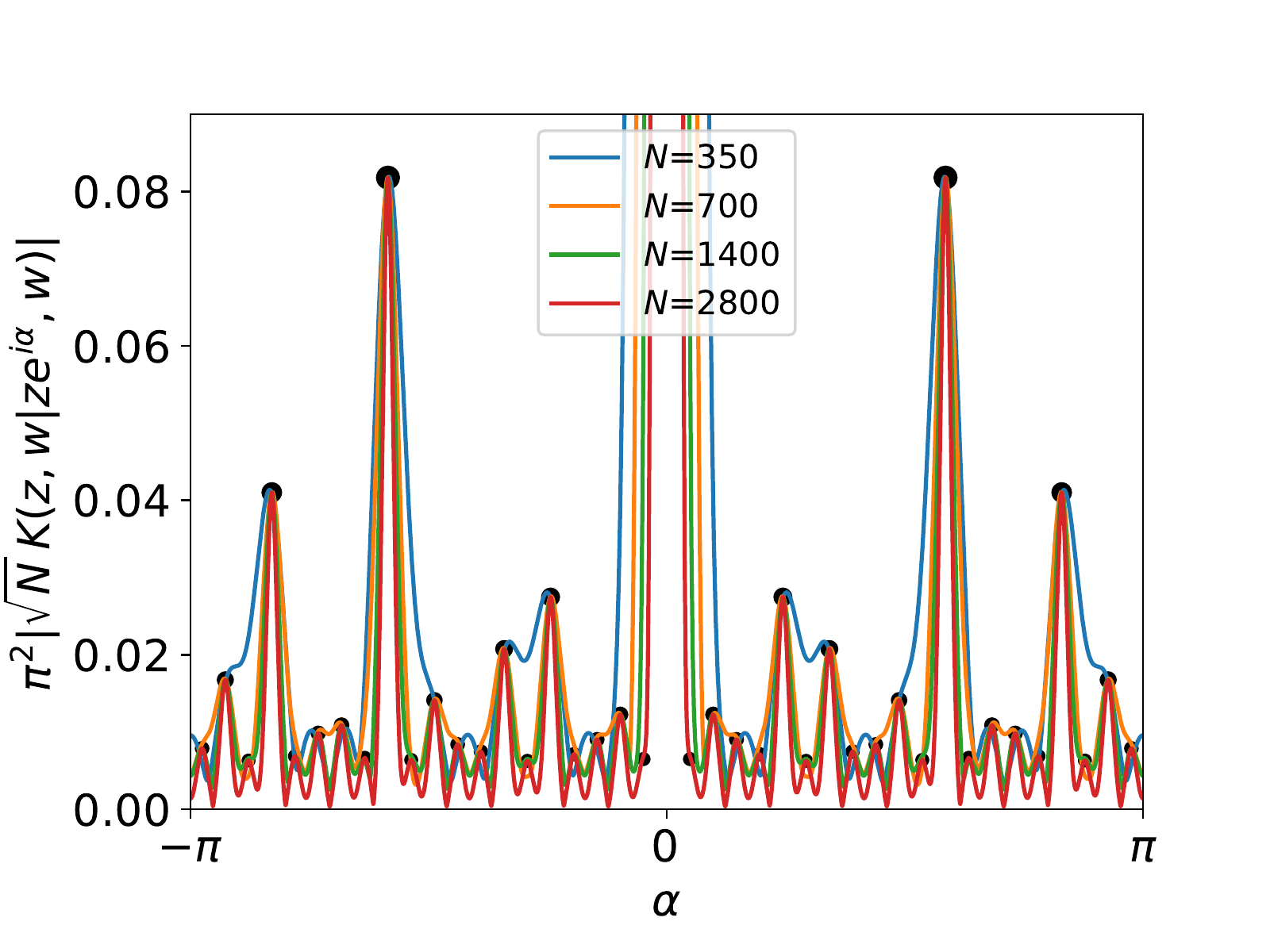}\hfill
\includegraphics[width=0.48\textwidth]{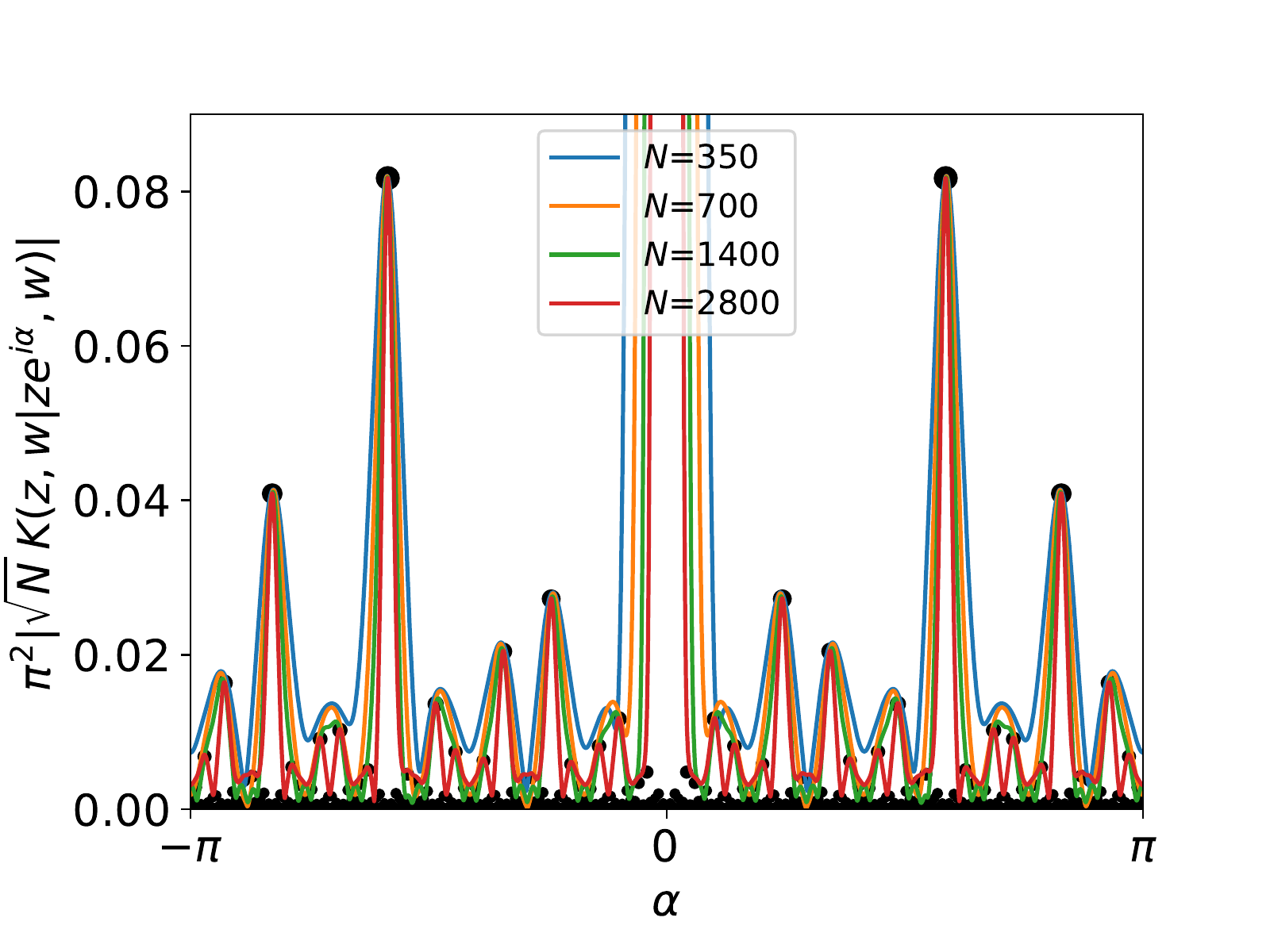}
\caption{The rescaled correlator $\sqrt{\cN}|\CC(\bx,\bx')|$ given by \eqref{t3}, for points \eqref{zz} on the slice $\beta=0$ of an edge torus at $\theta=\pi/2$. Black dots denote the maxima of correlations predicted by the limiting function \eqref{limain}. The four panels have anisotropy ratios providing different rational approximations of $\sqrt{2}$: from left to right, $\cR=7/5$ and $\cR=17/12$ in the top panels, while $\cR=41/29$ and $\cR=\sqrt{2}$ in the bottom panels. Each plot displays several values of $\cN$; the largest corresponds to over 2 million particles. The correlator does not seem to converge to a smooth function in the thermodynamic limit $N\to\infty$; it becomes instead more ragged at shorter distance scales.}
\label{fig:irr2}
\end{figure}

This approximation algorithm applies to edge correlations in anisotropic droplets: it is illustrated in figure \ref{fig:irr1} for three truncations of the continued fraction representation of $\cR=\sqrt{2}$. Notice that correlations involving successive approximations of $\sqrt{2}$ become more and more ergodic, with the true limit (if it exists at all) difficult to guess. In fact, much sharper statements about edge correlations are obtained by analysing density plots through one-dimensional `slices' --- \eg by setting $\beta$ to some constant value --- which one can roughly think of as quantum analogues of Poincar\'e sections. This is done in fig.\ \ref{fig:irr2} at $\beta=0$. In contrast to fig.\ \ref{fig:irr1}, it is then apparent that correlations in the thermodynamic limit are far from uniform, even in irrational droplets. The problem thus becomes to find the limiting form of these correlations.

\phantomsection
\addcontentsline{toc}{subsection}{Fractal correlations}%
\paragraph*{Fractal correlations.}
The correlations of ergodic edge modes with anisotropy $\cR\notin\mathbb{Q}$ can be inferred as follows. Starting from the rational formula \eqref{main}, normalize it in such a way that its strict thermodynamic limit ($\cN=\infty$) be finite and non-trivial. Then consider a sequence of rational anisotropies that converges to $\cR$; the limit of their correlation functions at $\cN=\infty$ is the sought-for ergodic correlation function. The remainder of this section explains this procedure in greater detail.

Recall that eq.\ \eqref{main} exhibits a Gaussian localization of edge modes on classical trajectories, with width $1/\sqrt{\cN}$. Accordingly, define a rescaled (norm of the) correlator 
\be
\cK(\alpha,\beta)
\equiv
\lim_{\cN\to\infty}\sqrt{\cN}\,\big|\CC(\mathbf{x},\mathbf{x}')\big|,
\label{limico}
\ee
where it is understood that $\bx,\bx'$ are two points on the same edge torus, with complex coordinates \eqref{zz}. This scaling may be seen as a 4D analogue of the well known (chord length) conformal scaling of edge correlations in circular 2D droplets. Our goal is thus to evaluate the limiting correlation \eqref{limico}, seen as a function of $\alpha$ and labelled parametrically by $(\Delta,\theta,\beta)$. The hope is that this limit works in both rational and irrational droplets, eventually reproducing the `slice' plots of fig.\ \ref{fig:irr2}.

In rational droplets, the limit \eqref{limico} is readily found thanks to eq.\ \eqref{main} for edge correlations. Indeed, at large $\cN$, the various Gaussians appearing in the sum over images \eqref{main} have separate supports, so the limit 
$\cN\to\infty$ turns each such Gaussian into a pointwise indicator function. The rescaled correlator \eqref{limico} thus takes non-zero values only when $\alpha=(\beta+2\pi\nu)/\Delta\mod{2\pi}$ for some integer $\nu$, that is, exactly on the classical trajectory. The actual value is determined by the power-law decay of \eqref{main}, yielding
\be
\cK(\alpha,\beta)
\propto
\begin{cases}
\frac{\ds1}{\ds q|\sin(\tfrac{\beta+2\pi\nu}{2p})|} & \text{if }\alpha=\frac{\beta+2\pi\nu}{\Delta}\!\!\!\!\mod{2\pi}\text{ for some $\nu\in\{0,\ldots,p-1\}$},\\[1em]
0 & \text{otherwise}
\end{cases}
\label{limain}
\ee
up to a $\theta$-dependent normalization (the exact result, shown in eq.\ \eqref{limicobis}, follows from the detailed rational formula \eqref{mainbis}). This expression can trivially be extended to a $2\pi$-periodic function on $\mathbb{R}$. On $[0,2\pi)$, it vanishes for almost all $\alpha$, except at $\max(p,q)$ points where it is finite. A mild exception occurs for $\beta=0\!\!\mod{2\pi}$, since the points $\bx,\bx'$ then coincide at $\alpha=0\!\!\mod{2\pi}$, where $\cK(\alpha,\beta)=\infty$. 

\begin{figure}[htbp]
\centering\includegraphics[width=0.64\textwidth]{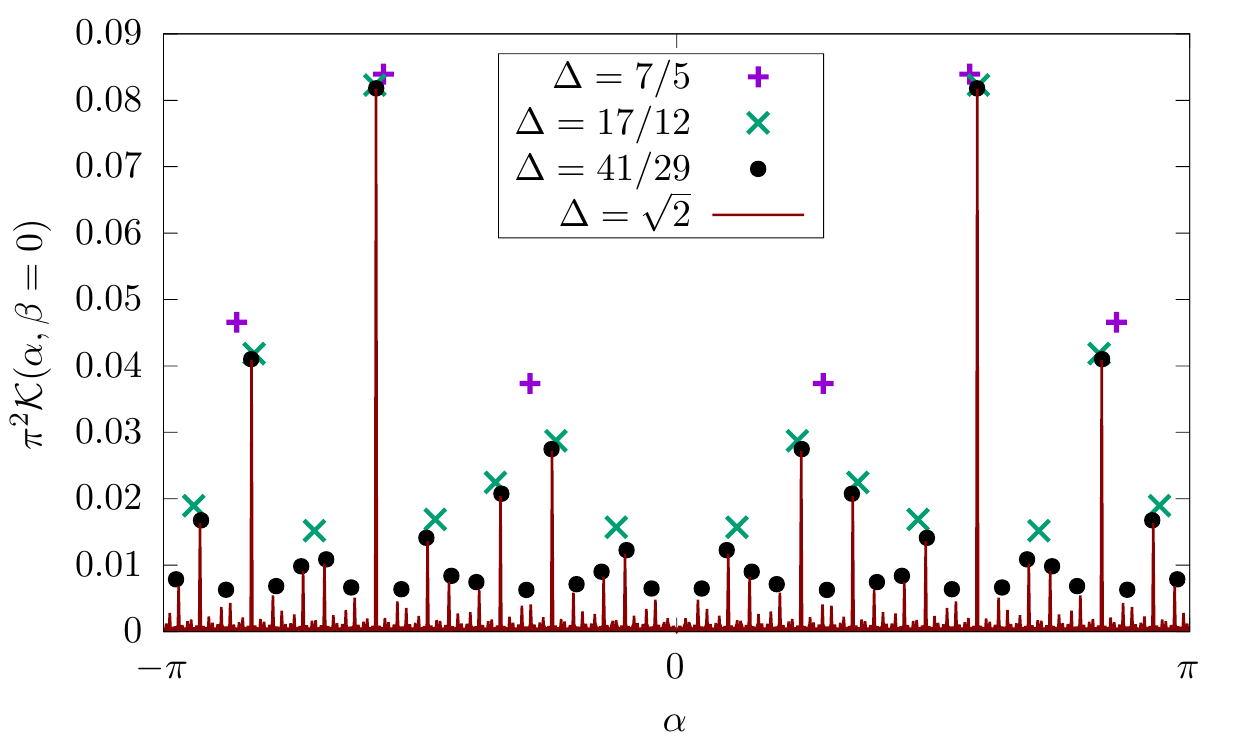}
\caption{Conjecture \eqref{conjec} for the rescaled correlator $\cK$ (defined in \eqref{limico}) in the irrational case, here for $\cR=\sqrt{2}$ and $\theta=\pi/2$, and comparison with continued fraction approximations. Notice the fractal-like structure of the graph: sequences of three successive peaks repeat themselves at different scales throughout the plot.}
\label{fig:fractal}
\end{figure}

The function \eqref{limain} at $\beta=0$ is displayed with black bullets in fig.\ \ref{fig:irr2}, for various $\Delta$'s and $\alpha\in[-\pi,\pi]$. As can be seen, the agreement with numerical data for large but finite $\cN$ is perfect. The agreement even seems to extend to irrational droplets, so it is tempting to still rely on \eqref{limain} in that case. This can be done in two ways: either take a sequence of rational ratios $\cR_k$ converging to $\cR$ when $k\to\infty$ (\eg using continued fractions) and conjecture that $\cK_{\cR}=\lim_{k\to\infty}\cK_{\cR_k}$; or take the actual irrational limit of \eqref{limain} to obtain
\begin{empheq}[box=\othermathbox]{equation}
\label{conjec}
 \cK(\alpha,\beta)
\propto
\begin{cases}
\frac{\ds1}{\ds|\beta+2\pi\nu|} & \text{if }\alpha=\frac{\beta+2\pi\nu}{\Delta}\!\!\!\!\mod{2\pi}\text{ for some $\nu\in\ZZ$},\\[1em]
0 & \text{otherwise}.
\end{cases}
\end{empheq}
(Normalization is omitted as in \eqref{limain}; the exact result is given in eq.\ \eqref{cirrbis}.) 
The function \eqref{conjec} is thus well-defined but highly irregular: it vanishes almost everywhere, except on the countable intersection between the $\beta$ slice and a guiding center trajectory, where it is discontinuous. This is reminiscent of the Thoma\'e function and other fractal-like graphs. A plot of $\cK$ for $\cR=\sqrt{2}$ is shown in fig.\ \ref{fig:irr2} (bottom right) and fig.\ \ref{fig:fractal}, confirming the perfect agreement between numerical correlations at finite $\cN$ and the prediction \eqref{conjec}.

\begin{figure}[htbp]
\includegraphics[width=0.5\textwidth]{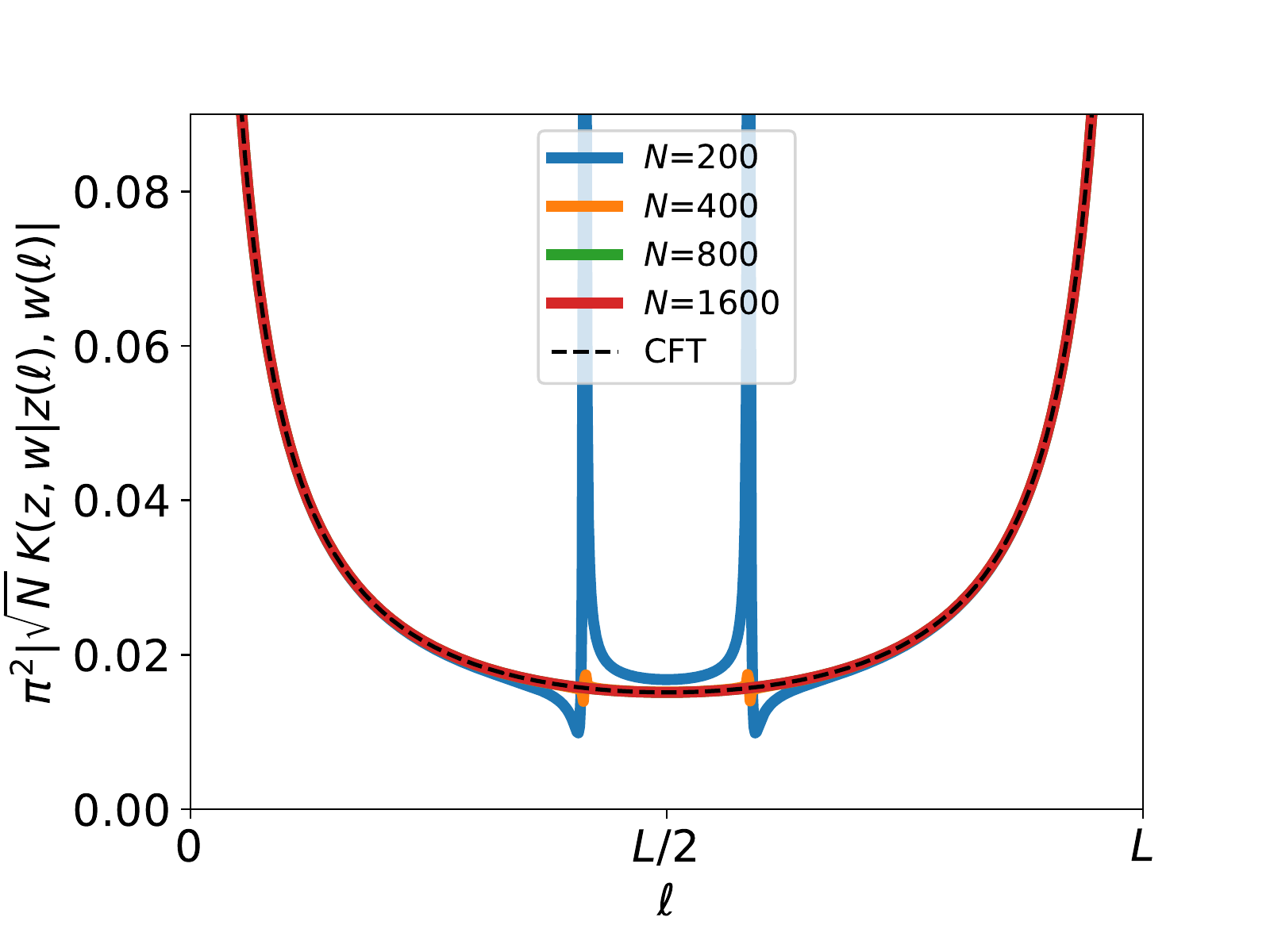}\hfill
\includegraphics[width=0.5\textwidth]{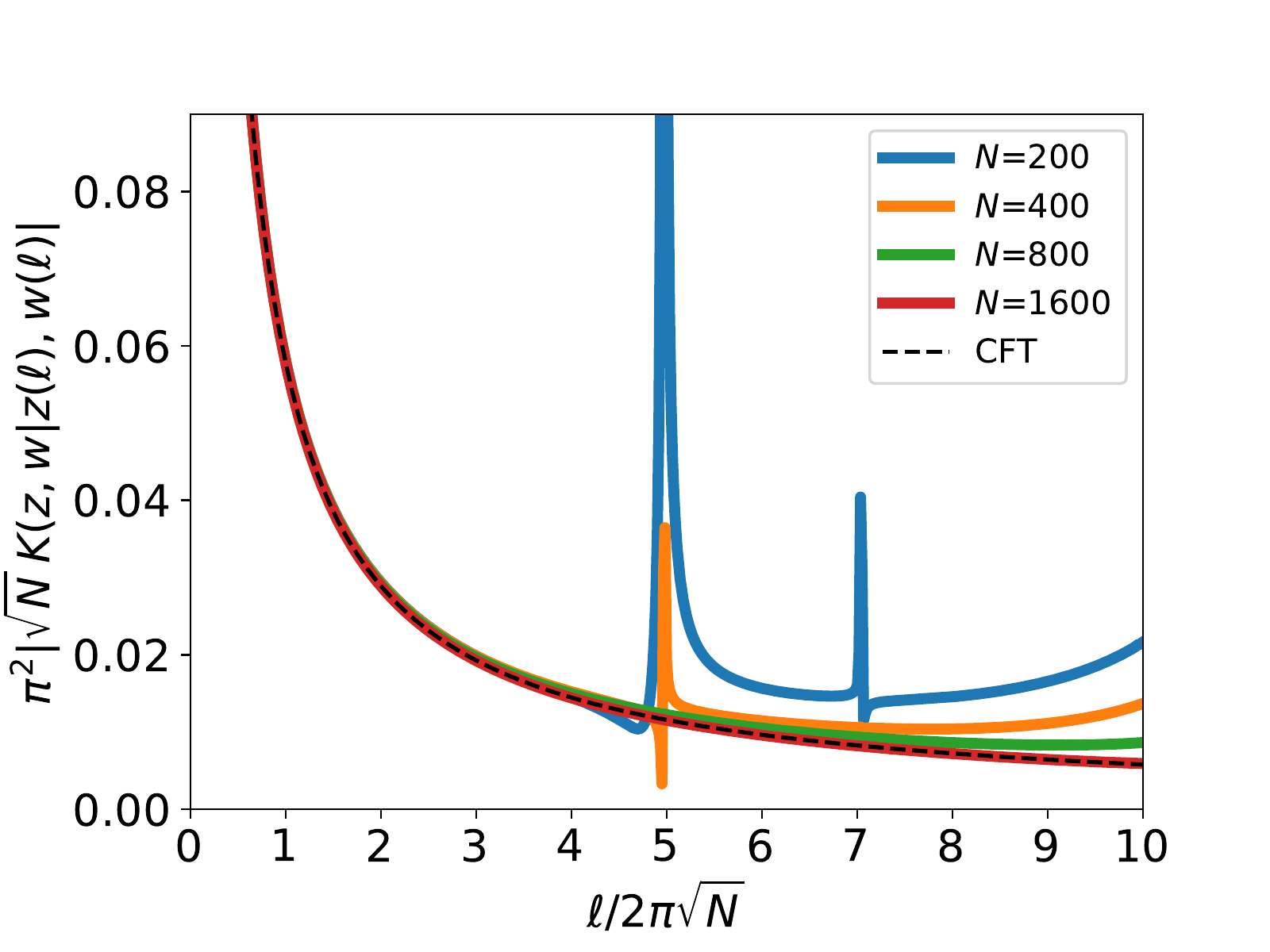}\\
\centering \includegraphics[width=0.5\textwidth]{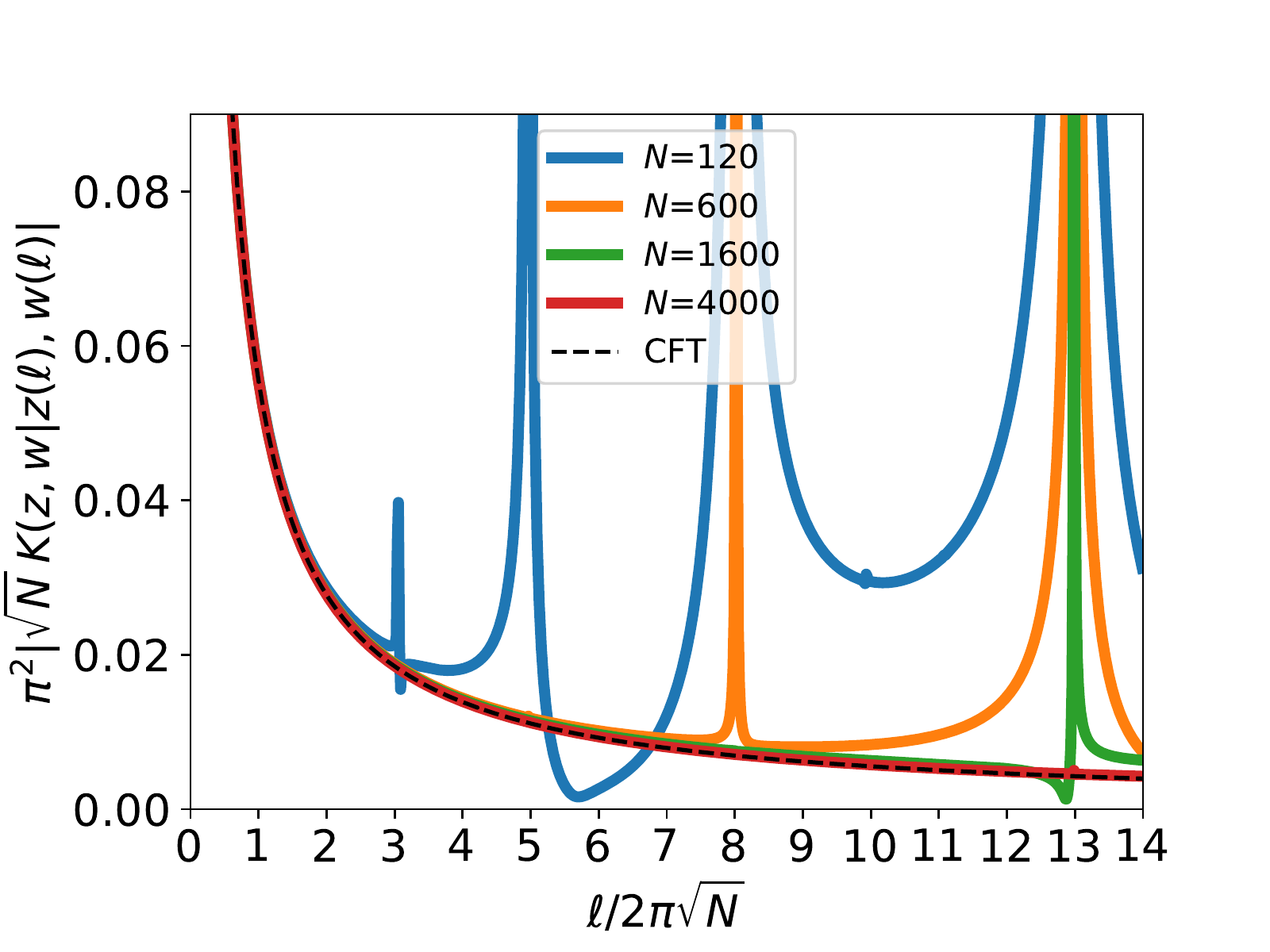}
\caption{Numerical computation of the rescaled correlator along the classical trajectory, for several values of $\cN$. Left: rational case $\Delta=17/12$, perfectly reproducing the CFT correlator \eqref{CFT_correlator_circle} on a circle. Right: irrational case $\cR=\sqrt{2}$, with integer abscissae indicating the number of times the trajectory wraps around the torus. Bottom: golden ratio $\cR=\frac{1+\sqrt{5}}{2}$. The agreement with the CFT correlator \eqref{CFT_correlator_line} on an infinite line is excellent. Notice the bumps in the curves with relatively small $N$: these are finite-size effects, due to the fact that the Gaussian spread is too large for small $\cN$ and that the classical trajectory is not yet sharply resolved. Intriguingly, the locations of these bumps coincide with the denominators obtained by truncating the continued fraction. This is manifest in the Fibonacci sequence $1,1,2,3,5,8,13,...$ of the bottom plot. (The first few denominators are not visible because particle number is already too large).
}
\label{fig:follow}
\end{figure}

Similar to before, the correlator (\ref{conjec}) boils down to the one-dimensional CFT result for free fermions on the infinite line, as long as $\bx$ and $\bx'$ lie on the same classical trajectory:
\begin{equation}
\label{CFT_correlator_line}
     \CC(\bx,\bx')\; \sim  \langle \Psi^{\dag}(\bx) \Psi(\bx') \rangle_{\textrm{CFT}} = \frac{1}{\ell}
\end{equation}
where $\ell$ is the distance separating $\bx$ and $\bx'$ \emph{along the trajectory}. This is nothing but the $L \to \infty$ limit of eq.\ \eqref{CFT_correlator_circle}. The interpretation is straightforward: an edge excitation born at $(0,0)$ propagates along its torus at constant velocity, and the correlation \eqref{conjec} is proportional to the inverse of the time it takes for it to reach the point $(\alpha,\beta)$; this time is infinite if $(\alpha,\beta)$ does not belong to the guiding center trajectory passing through the origin. This is illustrated in fig.\ \ref{fig:follow}, which shows side by side a numerical evaluation of the edge correlator along the classical trajectory, for $\cR=17/12$, $\cR=\sqrt{2}$ and the golden ratio $\cR=\frac{1+\sqrt{5}}{2}$. As can be seen, the agreement is excellent.

We conclude this section with two technical remarks. First, we stress that the ragged, fractal-like function \eqref{conjec} only emerges in the strict thermodynamic limit. This implies in practice that any finite droplet, no matter how large, has edge correlations that will at best be {\it approximately} given by \eqref{conjec}; for any finite $\cN$, one may in fact assume that the anisotropy ratio \eqref{rr} is rational (albeit with a numerator and denominator that grow with $\cN$). Second, notice that we refer to \eqref{conjec} as a {\it conjecture} rather than a proven statement. This is because our derivation of \eqref{conjec} relied on the rational correlations \eqref{main}, followed by a strict large $\cN$ limit, followed in turn by an irrational limit. By contrast, a proper proof of \eqref{conjec} would assume an irrational ratio \eqref{rr} at the outset, then find asymptotics of correlations in the thermodynamic limit. We will not attempt to perform such a computation here, and now return instead to the comforting realm of rational droplets.

\section{Quantum entanglement}
\label{senta}

In sections \ref{seclas}--\ref{sergo} we analysed the edge modes of a 4D Hall droplet using their boundary correlator, exhibiting edge conduction channels localized on classical guiding center trajectories. It is rather natural to presume that these channels are decoupled, and therefore can be described by a collection of independent (1+1)D CFTs. We now confirm this intuition thanks to the ground state entanglement of various spatial subregions. Indeed, entanglement entropy is a well-established probe of quantum matter and of gapless edge modes in particular, as was shown \eg in 2D droplets \cite{PhysRevB.88.155314,Estienne2019hmd}, at interfaces between fractional quantum Hall states \cite{Crepel_2019a,Crepel_2019b,PhysRevLett.123.126804}, or in 3D topological insulators with hinge modes \cite{hackenbroich2020fractional,crepel2021universal}. Since we are dealing here with non-interacting fermions, calculations simplify considerably  \cite{Chung_2001,Peschel_2003,Cheong_2004,islam2015measuring,Peschel_2009}: one can relate the entanglement spectrum to that of a subregion-restricted correlation matrix, reducing the computation of many-body entanglement to a one-body problem.

We mostly focus on isotropic droplets ($\cR =1$) from now on, being understood that our conclusions extend to rational anisotropic droplets. Edge modes then propagate along Hopf fibers in $S^3$, as in fig.\ \ref{Hopf}. Two kinds of simple entangling regions will be considered, where analytical calculations can be carried out to the end. The first are devised so as to avoid crossing any boundary fiber; their entanglement entropy will turn out to receive no contribution from edge modes, confirming that the latter are indeed decoupled. By contrast, the entanglement entropy of regions that deliberately cut through fibers will include logarithmic terms, characteristic of gapless 1D systems \cite{Holzhey:1994we,Vidal:2002rm,Calabrese:2004eu}, multiplied by a CFT central charge. These considerations are 4D analogues of what has been done in \cite{Estienne2019hmd} for 2D droplets, and extend the recent results of \cite{Charles_2019,Karabali:2020zap} by including boundary effects.

\phantomsection
\addcontentsline{toc}{subsection}{Spherical cap and area law}%
\paragraph*{Spherical cap and area law.}
\label{segap}
Here we consider a subregion of $\RR^4$ whose boundary crosses none of the $S^1$ fibers supporting edge modes. For simplicity, we choose a highly symmetric region, such that we will in fact be able to evaluate explicitly the whole entanglement spectrum. An asymptotic calculation (detailed in appendix \ref{appc}) then produces the thermodynamic limit of entanglement entropy, exhibiting the area law of gapped systems. The first subleading correction of this result, normally containing the contribution of edge modes, turns out to vanish. In this sense, periodic edge modes truly are independent: each propagates on its own fiber, free of correlations with its neighbors.

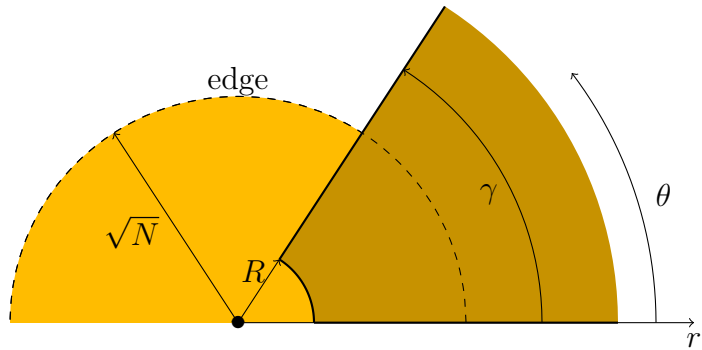
\begin{figure}[t]
\centering
\begin{tikzpicture}
\begin{scope}
\clip (-5,0)--(5,0)--(5,5)--(-5,5)--cycle;
\draw [dashed,fill=LightBrown] (0,0) circle (3);
\end{scope}
\begin{scope}
\clip[domain=0:57] plot ({5*cos(\x)}, {5*sin(\x)}) -- plot ({cos(57-\x)}, {sin(57-\x)}) -- cycle;
\fill[fill=MiddleBrown] (0,0)--(5,0)--(5,5)--(0,5)--cycle;
\end{scope}
\draw [thick,domain=0:1] plot ({1+4*\x},{0});
\draw [thick,domain=0:1] plot ({(1+4*\x)*cos(57)},{(1+4*\x)*sin(57)});
\draw [thick,domain=0:57] plot ({cos(\x)}, {sin(\x)});
\draw [smooth,dashed,domain=0:180] plot ({3*cos(\x)}, {3*sin(\x)});
\draw [domain=0:57,->] plot ({4*cos(\x)}, {4*sin(\x)});
\node at (3.3,1.7) {$\gamma$};
\draw [domain=0:37,->] plot ({5.5*cos(\x)}, {5.5*sin(\x)});
\node at (5.6,1.7) {$\theta$};
\node at (0,0) {$\bullet$};
\node at (0,3.2) {edge};
\draw [->] (0,0)--(6,0);
\node[below] at (6,0) {$r$};
\draw [<->,domain=0:1] plot ({\x*cos(57)},{\x*sin(57)});
\node[above] at (0.2,0.4) {$R$};
\draw [<->,domain=0:1] plot ({-3*\x*cos(57)},{3*\x*sin(57)});
\node at (-1.4,1.2) {$\sqrt{\cN}$};
\end{tikzpicture}
\caption{The 4D Hall droplet (yellow) and the spherical cap \eqref{secap} (brown) projected on the $(r,\theta)$ plane. In terms of the full 4D system, each point having $\theta\not\in\{0,\pi\}$ is a torus spanned by $(\alpha,\beta)$; when $\theta\in\{0,\pi\}$, only a circle with coordinate $\phii\propto\alpha+\beta$ survives (and the origin $r=0$ is just one point). The droplet is a ball with (dimensionless) radius $\sqrt{\cN}$, and the spherical cap covers radii $r\in[R,+\infty)$ and azimuths $\theta\in[0,\gamma]$.
\label{secafig}}
\end{figure}

Consider an isotropic droplet and label points in $\RR^4\cong\mathbb{C}^2$ by coordinates \eqref{hopfco}. In these terms, a `spherical cap' in $\RR^4$ is a region where (see fig.\ \ref{secafig})
\be
r\in[R,+\infty),
\qquad
\theta\in[0,\gamma],
\qquad
\alpha\in[0,2\pi],
\qquad
\beta\in[0,2\pi],
\label{secap}
\ee
with $R$ an `inner radius' and $\gamma\in(0,\pi)$ an azimuth. The coordinates $(\theta,\alpha-\beta)$ thus cover a spherical cap with opening $\gamma$ on $S^2$ (hence the terminology); $S^1$ fibers with coordinate $(\alpha+\beta)/2$ are covered entirely. Equivalently, the region \eqref{secap} is the product of a semi-infinite radial interval with a solid torus whose surface is spanned by $(\alpha,\beta)$.

Sections \ref{seclas}--\ref{secor} ensure that edge mode do not propagate across the boundary of the cap \eqref{secap}. Entanglement entropy should therefore satisfy both a bulk area law {\it and} an edge area law. Accordingly, let us evaluate, for future reference, the volume of the cap's boundary and the area of its intersection with the edge of the droplet. The boundary consists of two pieces ($r=R$ and $\theta=\gamma$) with volume forms
\be
\omega_{r=\text{cst}}
=
\frac{r^3}{4}\sin\theta\,\dd\theta\,\dd\alpha\,\dd\beta,
\qquad
\omega_{\theta=\text{cst}}
=
\frac{r^2}{2}\sin\theta\,\dd r\,\dd\alpha\,\dd\beta,
\label{omegom}
\ee
so its total volume, given some outer radius $R'$, is
\be
\text{Area(cap)}
=
2\pi^2
\bigg[
(R'^3+R^3)\sin^2(\gamma/2)
+
\frac{R'^3-R^3}{3}\sin\gamma
\bigg].
\label{arecap}
\ee
The first term is due to the angular integral of $\omega_{r=R}$ and $\omega_{r=R'}$, while the second stems from the radial integral of $\omega_{\theta=\gamma}$. (In particular, when $\gamma=\pi$, eq.\ \eqref{arecap} reduces to the sum of volumes of two $S^3$'s.) It also follows from \eqref{omegom} that the area of the torus at the intersection between the spherical cap and a three-sphere is $2\pi^2r^2\sin\gamma$. In principle, the latter measures the contribution of edge modes to entanglement entropy, but its coefficient will eventually turn out to vanish: see eq.\ \eqref{maintwo} below.

\begin{figure}[htbp]
\centering
\begin{tikzpicture}[rotate=45]
\fill[fill=LightYellow] (0,0)--(7,0)--(0,7)--cycle;
\fill[fill=MiddleBrown] (4,0)--(7,0)--(35/17,84/17)--(20/17,48/17)--cycle;
\begin{scope}
\clip (0,0)--(7,0)--(0,7)--cycle;
\draw[line width=6pt,smooth,color=LightBrown,rounded corners] (5,-1)--(20/17,48/17)--(40/17,96/17);
\end{scope}
\draw[->] (0,0) -- (8,0);
\draw (8,0) node[above] {$m$};
\node at (7,0) {\textbullet};
\draw (7,0) node[right] {$\cN$};
\draw[->] (0,0) -- (0,8);
\draw (0,8) node[above] {$n$};
\node at (0,7) {\textbullet};
\draw (0,7) node[left] {$\cN$};
\draw (7,0)--(0,7);
\draw (4,3) node[above] {$m+n=\cN$};
\draw[dashed] (-1,5)--(5,-1);
\draw (2.5,1.5) node[below] {$m+n=R^2$};
\draw[dashed] (0,0)--(2.5,6) node [near end,below,sloped,rotate=225] (TextNode) {$n/m=\tan^2(\gamma/2)$};
\draw (3.25,2.25) node {$A(m,n)\simeq1$};
\draw (1,1) node {$A(m,n)\simeq0$};
\end{tikzpicture}
\caption{A rough density plot of the overlap eigenvalues \eqref{amoval} in the triangle $0\leq m+n<\cN\gg1$. Throughout the yellow region, $A(m,n)$ nearly vanishes. Conversely, in the brown patch, the overlap is almost complete. In the transition region (thick orange line), $A(m,n)$ interpolates from zero to one; this region gives the main contribution to entanglement entropy \eqref{entrosum}. Note the similarity with figs.\ \ref{fispectrum} and \ref{secafig}: the spatial location of low-energy states \eqref{lola} is specified by their quantum numbers $(m,n)$, and states for which $m+n$ and $m$ are both `large enough' are entirely contained in the spherical cap \eqref{secap}, while those that do not satisfy this criterion are well outside the cap. The one-particle states that actually contribute to entanglement are localized on the cap's boundary.}
\label{figoverlap}
\end{figure}
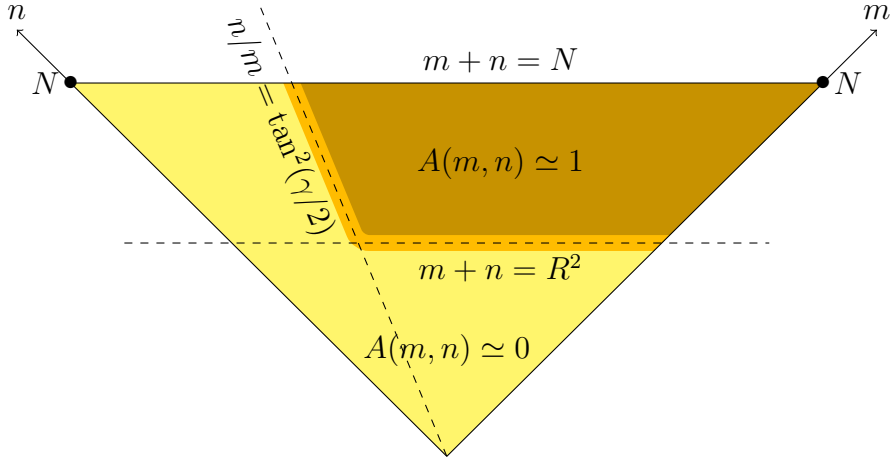

In order to evaluate entanglement entropy in a spherical cap, we use the free fermion methods of \cite{Chung_2001,Peschel_2003,Cheong_2004,klich2006lower,islam2015measuring,Peschel_2009,Rodriguez_2009,Rodriguez_2010,CMV2011}, according to which the spectrum of the reduced density matrix in a subregion is determined by the overlap of one-body wavefunctions in that subregion. Accordingly, our starting point is the overlap matrix of one-particle states \eqref{lola} in \eqref{secap}:
\be
\label{overlamm}
\AA_{mn,m'n'}
=
\int_{\text{Cap}(R,\gamma)}
\!\dd^4\bx\;
\phi_{mn}^*(\bx)\,\phi_{m'n'}(\bx).
\ee
We restrict attention to indices that contribute to the ground state \eqref{omegan}, so $m+n<\cN$ and the matrix \eqref{overlamm} has $[\cN(\cN+1)/2]^2$ entries. Most of these vanish thanks to the $\text{U}(1)\times\text{U}(1)$ symmetry of the cap \eqref{secap} under independent $\alpha,\beta$ rotations: using eq.\ \eqref{lola} for LLL states in 4D, one finds a diagonal overlap $\AA_{mn,m'n'}=A(m,n)\delta_{m,m'}\delta_{n,n'}$ with
\be
A(m,n)
=
\frac{\Gamma(m+n+2,R^2)}{\Gamma(m+n+2)}\,
\bigg[1-\frac{B\big(\cos^2(\gamma/2);m+1,n+1\big)}{B(m+1,n+1)}\bigg],
\label{amoval}
\ee
which involves incomplete gamma and beta functions. The eigenvalues $A(m,n)$ all belong to the (open) interval $(0,1)$, and are depicted, roughly, in fig.\ \ref{figoverlap}. Note the two patches where $A(m,n)\simeq0$ and $A(m,n)\simeq1$, separated by a thin `transition region'.

Since electronic interactions are neglected, the system consists of free fermions and ground state entanglement entropy can be expressed in terms of the overlap matrix:
\be
S
=
-\sum_{\substack{m,n\in\NN\\m+n<\cN}}
\Big[A(m,n)\log A(m,n)+\big(1-A(m,n)\big)\log\big(1-A(m,n)\big)\Big].
\label{entrosum}
\ee
We wish to evaluate this sum in the thermodynamic limit $\cN\gg1$. This can be done by zooming in on the `transition region' where $A(m,n)$ jumps from $0$ to $1$, since it is only those values of $A(m,n)$ that contribute substantially to \eqref{entrosum}. The details of this asymptotic analysis are exposed in appendix \ref{appc}. Their result yields an entanglement entropy whose large $\cN$ expansion takes the form stated around eq.\ \eqref{arealaw}:
\be
S\sim
s
\left[R^3\,\sin^2(\gamma/2)+\frac{\cN^{3/2}-R^3}{3}\sin\gamma\right]
+0\cdot\cN
+\cO(\cN^{1/2}),
\label{maintwo}
\ee
where $s\simeq1.80639...$ is an unimportant numerical coefficient. The leading $\cO(\cN^{3/2})$ term is the expected bulk area law: it contains a contribution $\propto R^3\,\sin^2(\gamma/2)$ due to the volume of the inner boundary of the spherical cap, while the piece $\propto(\cN^{3/2}-R^3)$ is proportional to the length $\sin(\gamma)$ of the cap's boundary. The only difference with \eqref{arecap} is the absence of the outer radius $R'$; this is because the (infinite) outer radius of the entangling region \eqref{secap} is well outside of the droplet, so its contribution to entanglement entropy vanishes. Perhaps more surprisingly, eq.\ \eqref{maintwo} exhibits a vanishing $\cO(\cN)$ term, stating that edge modes do {\it not} contribute to the area law. We now show that non-zero edge terms only occur for entangling regions that cut through classical trajectories. A slightly different region exhibiting the same effect is also studied in appendix \ref{appex}.

\phantomsection
\addcontentsline{toc}{subsection}{Gapless entanglement on an interval}%
\paragraph*{Gapless entanglement on an interval.}
\label{segapless}
As before, we consider a 4D droplet of $\sim\cN^2/2$ electrons in an isotropic harmonic trap. However, we now study an entangling region
\be
r\in[R,+\infty),
\qquad
\theta\in[0,\pi],
\qquad
\alpha\in[0,\phi],
\qquad
\beta\in[0,2\pi],
\label{otherregion}
\ee
that cuts through all $S^1$ fibers regardless of the value of $\phi\in(0,\pi)$. The $U(1)$ symmetry corresponding to the quantum number $n$ is still preserved in this geometry, resulting in a tremendous simplification. Indeed, the overlap matrix of LLL states \eqref{lola} in the interval \eqref{otherregion} takes the following block-diagonal form (up to an irrelevant phase):
\be
\mathbb{A}_{mn,m'n'}
=
\frac{\Gamma(1+\frac{m+m'}{2})}
{\sqrt{m!m'!}}
\,
\frac{\Gamma(\frac{m+m'}{2}+n+2,R^2)}
{\Gamma(\frac{m+m'}{2}+n+2)}
f_{m-m'}(\phi)
\delta_{nn'},
\label{overlap_simplified}
\ee
where $f_p(\phi)=\frac{\phi}{2\pi}\textrm{sinc}\, \frac{p\phi}{2}$. Setting $R=0$ in this expression results in a striking simplification: each block of $\mathbb{A}$ then reduces to the overlap matrix studied in \cite{Estienne2019hmd} for the entanglement entropy of 2D quantum Hall states. More precisely, in an isotropic droplet,
eq.\ \eqref{entrosum} applied to the overlap \eqref{overlap_simplified} with $R=0$ yields the exact dimensional reduction
\be
S
=
\sum_{k=1}^{N-1}S_{\text{2D}}^{(k)}(\phi),
\label{dimreduction}
\ee
where $S_{\text{2D}}^{(k)}(\phi)$ is the entanglement entropy of a 2D circular quantum Hall droplet with $k$ particles, one-body wavefunctions $\phi_m(z)=\frac{z^m}{\sqrt{\pi m!}}e^{-|z|^2/2}$, and an entangling region chosen to be a "piece of pie"  $\{z\in \mathbb{C},0<\arg z<\phi\}$. The scaling of this 2D entropy was studied in depth in \cite{Estienne2019hmd}, where the crucial result\footnote{The numerical factor $s\simeq1.80639...$ in \eqref{2d_asymptotics} is the same as in \eqref{maintwo}.}
\be
S_{\text{2D}}^{(k)}(\phi)=s\sqrt{k}+\frac{1}{6}\log\left(\sqrt{k}\sin \frac{\phi}{2}\right)+a(\phi)+O(1/\sqrt{k})
\label{2d_asymptotics}
\ee
was established at large $k$. Its logarithmic term is the contribution of the single chiral edge CFT with central charge $c=1$ present in the disk geometry. (The residual constant $a(\phi)$ stems from the corner with opening $\phi$ near $z=0$; it is not known analytically, but has been studied numerically to high precision in Refs. \cite{Rodriguez_2010,Siroisetal}.) It is then straightforward to deduce the scaling of 4D entropy from eqs.\ \eqref{dimreduction}--\eqref{2d_asymptotics}: for $R=0$, one finds
\be
\label{feny}
S
=
\frac{2s}{3}\cN^{3/2}
+
\frac{\cN}{6}\log\left(\sqrt{\cN}\sin\frac{\phi}{2}\right)
+
\left(a(\phi)-\frac{1}{12}\right)\cN
+
\cO(\cN^{1/2}).
\ee
The leading piece here is an area law, similar to the one appearing in the entropy \eqref{maintwo} of a spherical cap. The third term is a corner contribution which actually becomes $\phi$-independent when $R\neq0$. For our purposes, the most interesting bit is the middle term, due to gapless edge modes:  
\begin{empheq}[box=\othermathbox]{equation}
\label{santorini}
S_{\text{edge}}
\sim
\frac{\cN}{6}\,
\log\left(\sqrt{\cN}\sin\frac{\phi}{2}\right).
\end{empheq}
This is a crucial result: similarly to the 2D QHE (where the same formula holds without factor $\cN$, see (\ref{2d_asymptotics})), it indicates that edge modes satisfy the standard entanglement entropy formula of a 1D CFT \cite{Holzhey:1994we,Vidal:2002rm,Calabrese:2004eu} with central charges $(c,\bar c)$ such that
\be
c+\bar c
\sim
\cN.
\label{ccbar}
\ee
One should contrast this with the vanishing $\cO(\cN)$ term in the entropy \eqref{maintwo} of a spherical cap: the anisotropic nature of edge modes in the 4D QHE implies sharp `jumps' (of order $\cN$) in entanglement entropy when the entangling region is `rotated' so as to cut through the fibers that support edge modes. Such jumps measure the number $c+\bar{c}$ of independent conduction channels on the boundary. (Note that one expects $\bar c=0$ since edge modes are chiral --- at least provided a global choice of orientation has been made thanks to the symplectic gradient of the confining potential.) This should be contrasted with 2D non-interacting droplets, where $c=1$ at very strong magnetic fields. The distinction was to be expected: in 4D, edge modes propagating on $S^1$ fibers are labelled by the points of an $S^2$ whose area\footnote{The droplet's boundary $S^3\subset\RR^4$ carries a contact structure induced by the magnetic field (seen as a symplectic form) of $\RR^4$. The corresponding two-form on $S^3$ vanishes along classical trajectories, thus producing a well-defined area form --- proportional to $\cN$ --- on the quotient $S^2\cong S^3/S^1$.} is proportional to $\cN$, so their total number should be $\propto\cN$.

We restricted our attention to isotropic droplets ($\cR=1$) when writing the entropy \eqref{dimreduction}, but the generalization to arbitrary rational droplets ($\cR=p/q\in\mathbb{Q}$) is straightforward. Indeed, owing to the bounds \eqref{moc} on LLL labels in the ground state, the overlap matrix \eqref{overlap_simplified} has $\cN/\cR$ blocks and the resulting entanglement entropy takes the form \eqref{feny}--\eqref{santorini} with $\cN$ replaced by $\cN/\cR$. This readily provides a counting of edge modes in rational anisotropic droplets: keeping in mind that the entangling region \eqref{otherregion} now cuts each guiding center trajectory in $q$ intervals, the central charges of edge modes satisfy $c+\bar c\sim\cN/p$, generalizing eq.\ \eqref{ccbar}. This seems to suggest that irrational droplets, with ergodic boundary excitations, have finitely many conduction channels; we will not dwell on this intriguing possibility.

Let us conclude with a comment on higher dimensional generalizations. Consider a droplet placed on a simple product geometry $M\times\mathbb{R}^2$ for some $(d-2)$-dimensional manifold $M$, with a block-diagonal magnetic field $\bB_{d-2}+B \dd x\wedge\dd y$ (in terms of usual Cartesian coordinates $(x,y)$ in $\mathbb{R}^2$). With a confining potential $V = y$, the 
$(d-1)$-dimensional boundary of the droplet is $M\times \mathbb{R} \times \{0\}$ and the manifold $M$ plays no role for edge dynamics, as it merely labels different 1D conduction channels that propagate along $\mathbb{R}$. The number of edge modes then is nothing but the LLL degeneracy on $M$, endowed with the magnetic field $\bB_{d-2}$. In that sense, the prediction that the number of 4D edge modes is extensive in a $(d-2)$-dimensional volume is generic.

\section{Conclusion and outlook}

\paragraph*{A look back.} This work has been devoted to the properties of edge modes in a 4D quantum Hall droplet. Our point of view was deliberately `microscopic' --- based on the known energy spectrum of suitable harmonic traps --- and complements in that sense the effective field theory arguments of \cite{Polchinski2002,Karabali2002,Karabali2003,Karabali2004,Karabali2016}. As we have seen, 4D edge modes are more complex than their 2D peers. First, their chirality means that they propagate in only one direction along the edge (rather than spreading on the whole edge as one might have naively expected); this direction is given by the symplectic gradient of the confining potential, as ensured by the classical intuition of section \ref{seclas}. Second, because of this localization, edge modes wind on boundary tori in a way that is sensitive to the confining potential: isotropic traps give rise to edge modes on Hopf fibers in an $S^3$ boundary, but this is not the case in anisotropic traps. In particular, generic traps produce ergodic guiding center motion. To our knowledge, this is the first time in the literature that the existence of such ergodic edge modes is pointed out.

The purpose of this paper has been to confirm these intuitions with sharp quantitative estimates of edge correlations and entanglement. In particular, our asymptotic formula \eqref{main} for correlations in rational droplets was useful both as a check of localization on guiding centers in the quantum theory, and as a motivation for the fractal edge correlations built in section \ref{sergo}. We have also seen in section \ref{senta} that entanglement entropy in regions whose boundary is parallel to edge modes satisfies a strict area law \eqref{maintwo}, while regions that cut through $S^1$ fibers supporting edge modes exhibit a logarithmic term \eqref{santorini} typical of 1D CFTs. As a corollary, eq.\ \eqref{santorini} also provided the total central charge of edge modes, namely $c+\bar c\sim\cN$, which is related to the Fermi energy.

\paragraph*{A look forward.} Our hope is that this work will be relevant for experiments involving synthetic dimensions. Numerous proposals for such simulations have indeed been put forward \cite{Price2015,Ozawa2016,Yuan,PriceOzawa,OzawaCarusotto,Ozawa2019} and quasiperiodic photonic waveguides have in fact been successfully used to directly simulate the lattice version of the model we discussed \cite{Kraus2012,Kraus2013,Zilberberg2017,Lohse2018}. However, most of these setups appear to rely on a simple factorization between spatial directions, in which case the subtle edge effects described in the present work do not arise. (When the boundary is flat, edge modes simply propagate on 1D straight lines \cite{Polchinski2002}.) It thus seems important to understand how more complicated traps --- such as the harmonic traps studied here --- could be simulated. What seems to be crucial, especially in view of the ergodic considerations in section \ref{sergo}, is that droplet's boundary be {\it compact}, allowing edge modes to propagate in a (quasi-)periodic manner. Assuming this is feasible, a natural second step will be to deform the trap continuously, so as to generate bifurcations of edge mode windings such as depicted in figs.\ \ref{fiMainPlot}--\ref{fig:irr1}--\ref{fig:irr2}. To some extent, this is related to recent studies of the geometry of the 4D QHE \cite{Karabali2016}, including \eg the higher-dimensional analogue of Hall viscosity \cite{AvronSeiler,Levay}.

Somewhat more remotely, two key extensions of the present work need to be addressed. First, the `integrable' setup considered here did not allow us to address what is presumably the most interesting situation, namely chaotic edge dynamics. Second, it would be interesting to take interactions into account and see how they affect ergodicity.

\section*{Acknowledgements}

We are grateful to Andrei Bernevig, Laurent Charles, Viet Dang, Beno\^it Dou\c{c}ot, Nathan Goldman, Nicolas Regnault and C\'ecile Repellin for illuminating discussions on the QHE, its higher-dimensional generalizations, and ergodicity. In particular, we thank Laurent Charles for pointing out the potential chaotic behavior of boundary dynamics in a 4D Hall droplet. B.E. also 
thanks Andrei Bernevig, Nicolas Regnault and C\'ecile Repellin for their collaboration at an early stage of this work. The work of B.E.\ and B.O.\ is supported by the ANR grant {\it TopO} No.\ ANR-17-CE30-0013-01. B.O.\ is also supported by the European Union's Horizon 2020 research and innovation programme under the Marie Sk\l odowska-Curie grant agreement No.\ 846244.

\appendix

\newpage
\section{One-body dynamics in a harmonic trap}
\label{app_quadratic}
\setcounter{equation}{0}
\renewcommand{\theequation}{\thesection.\arabic{equation}}

This appendix displays the exact solution, both classical and quantum, of the dynamics of a charged particle in a magnetic field $\bB=B(\dd x\wedge\dd y+\dd u\wedge\dd v)$ trapped by the quadratic potential \eqref{trap}. The classical solution illustrates the emergence of the guiding center approximation at large magnetic fields, thus confirming eq.\ \eqref{gc} in section \ref{seclas}. The quantum setup similarly accompanies section \ref{secor}, exhibiting small corrections to the approximate spectrum obtained in \eqref{lolspec}--\eqref{lola} thanks to a projection to the lowest Landau level.

\phantomsection
\addcontentsline{toc}{subsection}{Classical dynamics; guiding center approximation}%
\paragraph*{Classical dynamics; guiding center approximation.}
\label{app_quadratic_classical}
Given the harmonic trap \eqref{trap}, the unprojected one-body Hamiltonian \eqref{hav}
(with unit mass and charge) reads 
\be
H
=
\frac{1}{2}(\bp-\bA)^2
+
\frac{k}{2}(x^2+y^2)
+
\frac{k'}{2}(u^2+v^2).
\label{hamil}
\ee
In symmetric gauge $\bA=\frac{B}{2}(x\,\dd y-y\,\dd x+u\,\dd v-v\,\dd u)$, this is a quadratic function on phase space, so its flow is linear and trivially integrable. Furthermore, the $(x,y)$ and $(u,v)$ degrees of freedom decouple, with respective eigenmodes $(a,b)$ and $(c,d)$ given by
\be
\begin{split}
\sqrt{2g}\,a
&=
i(p_x+ip_y)+g \cyclotron(x+iy)/2,\\
\sqrt{2g}\,b
&=
i(p_x-ip_y)+g \cyclotron(x-i y)/2,
\end{split}
\qquad
\begin{split}
\sqrt{2g'}\,c &= i(p_u+ip_v)+g' \cyclotron(u+iv)/2,\\
\sqrt{2g'}\,d &= i(p_u-ip_v)+g' \cyclotron(u-i v)/2
 \end{split}
 \label{eigenmodes}
\ee
where $g\equiv\sqrt{1+4k/B^2}$, $g'\equiv\sqrt{1+4k'/B^2}$. The normalization is chosen for later convenience: we will soon relate these modes to the lowering operators of eqs.\ \eqref{annihop}. In particular, they diagonalize the Hamiltonian \eqref{hamil}:
\be
H
=
\frac{g+1}{2}|a|^2
+
\frac{g-1}{2}|b|^2
+
\frac{g'+1}{2}|c|^2
+
\frac{g'-1}{2}|d|^2,
\label{diagoh}
\ee
and their time evolution reads
\be
 a(t) = e^{-iB \frac{g+1}{2}t} a(0) ,
 \;\;\;
 b(t) = e^{-i B \frac{g-1}{2}t} b(0),
 \;\;\;
  c(t) = e^{-iB \frac{g'+1}{2}t} c(0),
  \;\;\;
  d(t) = e^{-i B \frac{g'-1}{2}t} d(0).
  \nn
\ee
This exhibits a sharp separation of time scales at large $B$: the modes $(b,d)$ oscillate with low frequencies $B(g-1)/2\sim k/B\equiv\omega$ and $B(g'-1)/2\sim k'/B\equiv\omega'$, while $(a,c)$ evolve at cyclotron frequencies $B(g+1)/2\sim B(g'+1)/2\sim B$. These fast oscillations imply that low energy configurations have vanishingly small $(a,c)$ amplitudes in the limit $B\to\infty$, effectively enforcing the constraint $\bp-\bA\approx0$ initially encountered in \eqref{pix}. The slow dynamics of $(b,d)$ then boils down to the expected guiding center equations \eqref{gc}.

\phantomsection
\addcontentsline{toc}{subsection}{Quantum dynamics; LLL projection}%
\paragraph*{Quantum dynamics; LLL projection.}
\label{app_quadratic_quantum}
In symmetric gauge, the problem of finding the spectrum of the 4D Hamiltonian \eqref{hamil} reduces to two 2D Landau problems in isotropic harmonic traps. Accordingly, introduce dimensionless complex coordinates 
\be
z
\equiv
\sqrt{\frac{g}{2}}\,
\frac{x+iy}{\ell_B},
\qquad
w
\equiv
\sqrt{\frac{g' }{2}}\,
\frac{u+iv}{\ell_B},
\label{malenafo}
\ee
where $\ell_B = \sqrt{\hbar/B}$ is the magnetic length. At large $B$, one has $g\sim g'\sim1$ and the definitions \eqref{malenafo} become those introduced above eqs.\ \eqref{annihop}. Similarly, the eigenmodes \eqref{eigenmodes} expressed in complex coordinates (and divided by $\sqrt{\hbar B}$) become lowering operators
\be
a=\frac{z}{2}+\der_{\bar z},
\qquad
b=\frac{\bar z}{2}+\der_z,
\qquad
c=\frac{w}{2}+\der_{\bar w},
\qquad
d=\frac{\bar w}{2}+\der_w
\ee
that reduce to \eqref{annihop} in the limit $B\to\infty$. The quantized Hamiltonian \eqref{hamil} then coincides with $\hbar B\times$\eqref{diagoh} up to an irrelevant zero-point energy (being understood that $|a|^2$, $|b|^2$, etc.\ are replaced by $a^{\dagger}a$, $b^{\dagger}b$, etc.). In particular, the integer eigenvalues $(\mu,m,\nu,n)$ of $a^{\dagger}a$, $b^{\dagger}b$, $c^{\dagger}c$ and $d^{\dagger}d$ label uniquely the eigenstates of the Hamiltonian, whose energies
\be
E_{\mu,\nu;m,n}
=
\hbar\cyclotron
\left(\frac{g+1}{2}\mu
+
\frac{g-1}{2}m
+
\frac{g'+1}{2}\nu
+
\frac{g'-1}{2} n\right)
\label{egevala}
\ee
reproduce the aforementioned time/energy scale separation between the high frequency $(g+1)\cyclotron/2\sim\cyclotron$ and the low frequency $(g-1)\cyclotron/2  \sim\omega\ll\cyclotron$ (and similarly for $g\to g'$). The indices $\mu,\nu$ thus label Landau-like energy levels, while $m,n$ lift degeneracies in each level. For example, the available energies at $\mu=\nu=0$ read $E_{0,0;m,n} =\hbar B[(g-1)m+(g'-1)n]/2$ and reduce to \eqref{lolspec} when $B^2/k\to\infty$. The corresponding eigenfunctions coincide with the LLL-projected expressions \eqref{lola} up to slightly different length scales in the definition \eqref{malenafo} of $(z,w)$.

\section{Asymptotics of the incomplete gamma function}
\label{appa}
\setcounter{equation}{0}
\renewcommand{\theequation}{\thesection.\arabic{equation}}

This appendix accompanies section \ref{sesotrap}. It is devoted to the proof of the asymptotic formula \eqref{mare} for the incomplete gamma function
\be
\Gamma(\cN,x)
\equiv
\int_x^{+\infty}\dd t\,t^{\cN-1}\,e^{-t},
\label{gamma}
\ee
where $x\in\mathbb{C}$ and the integral runs along any path from $x$ to positive real infinity. Specifically, we wish to obtain an approximate expression of this function in the large $\cN$ limit, keeping $|x|/\cN$ finite and assuming $\arg x\neq0$. To achieve this, start by changing variables: letting $x\equiv\cN\lambda\,e^{i\phii}$, define $s\in\mathbb{C}$ by $t\equiv \cN\lambda\,e^s$ and rewrite \eqref{gamma} as
\be
\Gamma(\cN,\cN\lambda e^{i\phii})
=
(\cN\lambda)^{\cN}\int_{i\phii}^{+\infty}\dd s\,e^{\cN\left(s-\lambda e^s\right)}.
\label{gamin}
\ee
Here the integrand is a holomorphic function of $s\in\mathbb{C}$, so the integral can be evaluated along any path that connects the starting point $s=i\phii$ to the endpoint $s=+\infty$. Accordingly, the plan of action is as follows: (i) find stationary phase paths; (ii) choose a convenient integration contour (as we shall see, depending on whether $\lambda\leq1$ or $\lambda>1$); (iii) compute the resulting integrals in the large $\cN$ limit.

\phantomsection
\addcontentsline{toc}{subsection}{(i)~~Stationary phase paths}%
\paragraph*{(i)~~Stationary phase paths.} We think of the integrand of \eqref{gamin} as a holomorphic function of $s\in\mathbb{C}$. To perform the integral, any path connecting $i\phii$ to real positive infinity does the job; we will choose (a concatenation of) steepest-descent paths, that is, paths along which the phase of the integrand is constant.\footnote{The fact that {\it steepest-descent} paths are also {\it stationary phase} paths is standard but non-trivial, and follows from holomorphicity of the integrand. See \eg \cite[chap.\ 6]{Bender} for details.} We now describe such paths.

\begin{figure}[t]
\centering
\includegraphics[width=0.60\textwidth]{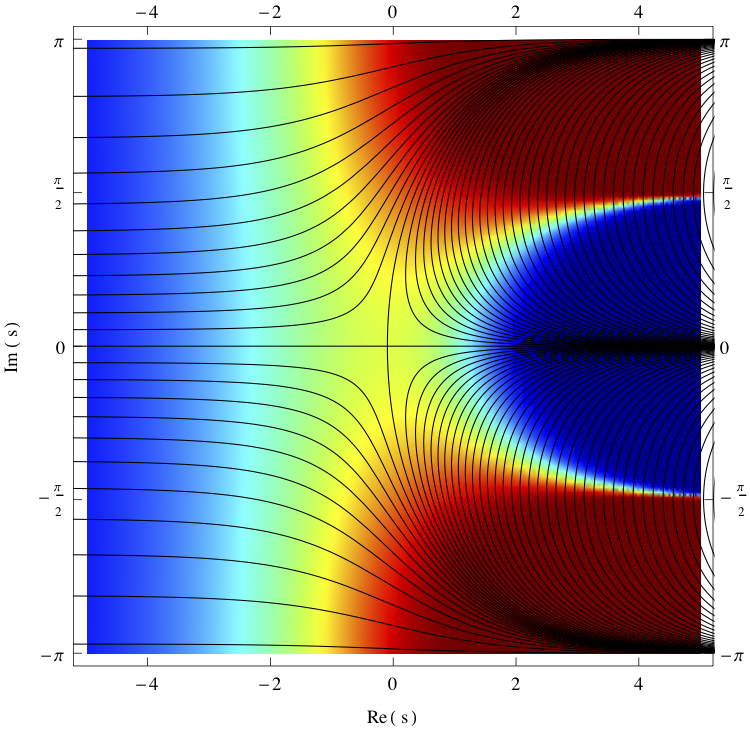}
\caption{The steepest descent contours \eqref{stepc} at $\lambda=1.1$, with $\log c$ ranging from roughly $-8$ to $8$. The picture is $2\pi$-periodic along the imaginary axis, so we let $\text{Im}(s)$ range from $-\pi$ to $\pi$. Note the crossing of paths at the saddle point $s=-\log\lambda$. Background colors reflect the norm of the function $\exp[\cN(s-\lambda e^s)]$: blue means it approaches zero, while red means it blows up. Accordingly, all stationary phase paths that escape to minus real infinity are steepest {\it descent} curves when traced from right to left, while those trapped in the region $\text{Re}(s)\geq-\log\lambda$ can be steepest {\it descent} or steepest {\it ascent} when traced from left to right, respectively depending on whether they end up in a blue zone or a red zone.
\label{stepa}}
\end{figure}

Consider the function $f(s)\equiv s-\lambda e^s$ such that the integrand of \eqref{gamin} reads $e^{\cN f(s)}$. Steepest descent ($=$ stationary phase) paths are such that the imaginary part of $f(s)$ is constant. Writing $s=a+ib$ with $a,b\in\RR$, one has $\text{Im}(f(s))=b-\lambda e^a\sin b$, so a path with constant imaginary phase $c$ is a set of points $a+ib$ in the plane such that
\be
b-\lambda e^a\sin b=c
\qquad\Leftrightarrow\qquad
a=\log\left[\frac{b-c}{\lambda\sin b}\right].
\label{stepc}
\ee
A family of such paths, labelled by $c$, is depicted in fig.\ \ref{stepa}. In particular, if the path is to pass through the point $s=i\phii$, then $c=\phii-\lambda\sin\phii$ and
\be
a
=
\log\left[\frac{b-\phii+\lambda\sin\phii}{\lambda\sin b}\right]
\qquad
\text{(path going through $s=i\phii$).}
\label{stephi}
\ee
Note that these paths generally do not cross each other, since the right equation in \eqref{stepc} is the unique solution of the equation $b-\lambda e^a\sin b=c$ when $c\neq0$. The only exception occurs when $c=0$, in which case a {\it second} solution is simply $b=0$, that is, the real axis in the complex $s$ plane; this is manifest in fig.\ \ref{stepa}. Thus, for $c=0$, a crossing between two steepest-descent paths occurs at the saddle point $s=-\log\lambda$.

It is also manifest from the left equation in \eqref{stepc} that any increase of $b$ by an integer multiple of $2\pi$ can be absorbed by a corresponding increase in $c$. Stationary phase paths are therefore repeated with $2\pi$ periodicity along the imaginary axis in the complex $s$ plane, and the integrand of \eqref{gamin} has infinitely many saddle points at $s^*_n=-\log\lambda+2\pi in$, $n\in\ZZ$. One is thus free to restrict attention to the strip $|\text{Im}(s)|\leq\pi$ in the complex $s$ plane when carrying out the integral --- a restriction we shall use implicitly from now on. Within that strip, the integrand has a unique saddle point at $s^*_0\equiv s^*=-\log\lambda$.

\begin{figure}[t]
\centering
\includegraphics[width=0.45\textwidth]{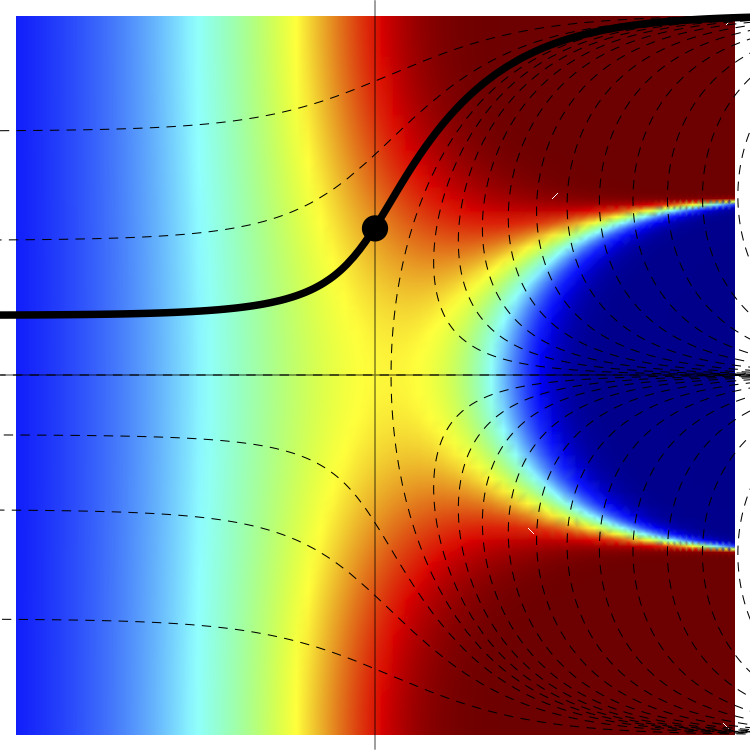}
\hfill
\includegraphics[width=0.45\textwidth]{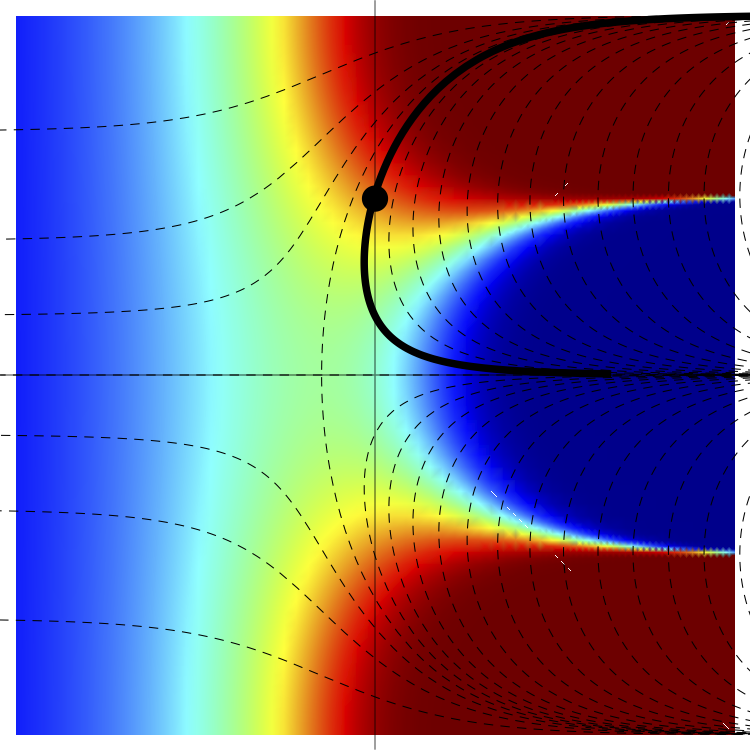}
\caption{The steepest descent path passing through the point $s=i\phii$ can be of one of two types: it connects $+\infty\pm i\pi$ either to $-\infty+i\phii-i\lambda\sin\phii$ (left panel), or to $+\infty$ (right panel). In both cases, the steepest {\it descent} contour is traced in the direction of monotonously decreasing $|\text{Im}(s)|$. However, when $\lambda<1$ as in the left panel, no steepest descent path connects $i\phii$ to $+\infty$. Conversely, when $\lambda>1$ as in the right panel, there exists a steepest descent path connecting $i\phii$ to $+\infty$ provided $|\phii|$ is smaller than (or equal to) the critical value $\phii_c>0$ given by $\phii_c=\lambda\sin\phii_c$. When $\lambda>1$ but $|\phii|>\phii_c$, the steepest descent path going through $i\phii$ escapes again to $-\infty+i\phii-i\lambda\sin\phii$.
\label{codet}}
\end{figure}

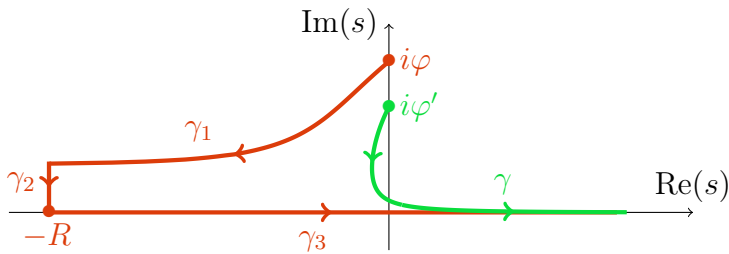
\begin{figure}[t]
\centering
\begin{tikzpicture}
\draw[smooth,->] (-5,0) -- (4,0);
\draw (4,0) node[above] {$\text{Re}(s)$};
\draw[smooth,->] (0,-0.5) -- (0,2.5);
\draw (0,2.5) node[left] {$\text{Im}(s)$};
\tikzset{->-/.style={decoration={
  markings,
  mark=at position .5 with {\arrow{>}}},postaction={decorate}}}
\draw[->-,ultra thick, MyRed, samples=100, domain=2:0.009+2-1.5*sin(360*2/(2*pi)),variable=\b] plot ({ln((\b-(2-1.5*sin(360*2/(2*pi))))/(1.5*sin(360*\b/(2*pi))))},{\b});
\draw[->-,ultra thick, MyRed, samples=100, domain=0.038+2-1.5*sin(360*2/(2*pi)):0,variable=\b] plot ({0.028+ln((0.01+2-1.5*sin(360*2/(2*pi))-(2-1.5*sin(360*2/(2*pi))))/(1.5*sin(360*(0.01+2-1.5*sin(360*2/(2*pi)))/(2*pi))))},{\b});
\draw[->-,ultra thick, MyRed, samples=100, domain=ln((0.01+2-1.5*sin(360*2/(2*pi))-(2-1.5*sin(360*2/(2*pi))))/(1.5*sin(360*(0.01+2-1.5*sin(360*2/(2*pi)))/(2*pi)))):3,variable=\a] plot ({\a},{0});
\draw[->-,ultra thick, MyGreen, samples=100, domain=1.4:0.1,variable=\b] plot ({ln((\b-(1.4-1.5*sin(360*1.4/(2*pi))))/(1.5*sin(360*\b/(2*pi))))},{\b});
\draw[->-,ultra thick, MyGreen, samples=100, domain=0.1:0.002,variable=\b] plot ({ln((\b-(1.4-1.5*sin(360*1.4/(2*pi))))/(1.5*sin(360*\b/(2*pi))))},{\b});
\node[MyRed] at (0,2) {\textbullet};
\draw[MyRed] (0,2) node[right] {$i\phii$};
\node[MyGreen] at (0,1.4) {\textbullet};
\draw[MyGreen] (0,1.4) node[right] {$i\phii'$};
\node[MyRed] at (-4.473,0) {\textbullet};
\draw[MyRed] (-4.5,0) node[below] {$-R$};
\draw[MyRed] (-2.5,0.8) node[above] {$\gamma_1$};
\draw[MyRed] (-4.5,0.4) node[left] {$\gamma_2$};
\draw[MyRed] (-1,-0.1) node[below] {$\gamma_3$};
\draw[MyGreen] (1.5,0.1) node[above] {$\gamma$};
\end{tikzpicture}
\caption{Integration curves for \eqref{gamin} at $\lambda>1$. When $\phii>\phii_c$, the path (red) contains three pieces $\gamma_1,\gamma_2,\gamma_3$, where $\gamma_1$ and $\gamma_3$ are steepest descents while $\gamma_2$ is not (but its contribution to the integral vanishes in the limit $R\to+\infty$). The same kind of piecewise-smooth path is taken when $\lambda\leq1$. Conversely, when $\lambda>1$ and $\phii=\phii'\leq\phii_c$, a single steepest descent path $\gamma$ (green) suffices to evaluate the integral \eqref{gamin}.
\label{papic}}
\end{figure}

\phantomsection
\addcontentsline{toc}{subsection}{(ii)~~Integration contours}%
\paragraph*{(ii)~~Integration contours.} To evaluate the integral \eqref{gamin} at large $\cN$, connect the point $s=i\phii$ to positive real infinity by stationary phase curves. Let us first assume that $\lambda\leq1$, so that the saddle point of the integrand of \eqref{gamin} lies at $s^*=-\log\lambda\geq0$. Then, for any $\phii\neq0$, there is no steepest-descent contour that connects the point $i\phii$ to positive real infinity (see the left panel in fig.\ \ref{codet}), and one is forced to compute the integral by taking a sequence of several stationary phase paths. Specifically:
\begin{enumerate}[leftmargin=*]\setlength\itemsep{0em}
\item[(i)] Starting at $s=i\phii$, follow the steepest descent path that moves towards the negative real axis, decreasing $|\text{Im}(s)|$, until $\text{Re}(s)$ reaches some value $-R$ (the eventual intent being to let $R\to+\infty$). Call that portion of path $\gamma_1$.
\item[(ii)] Connect the endpoint of $\gamma_1$ to the real axis by following a path at constant real part $-R$. Call that (short) portion of path $\gamma_2$; its endpoint is $s=-R$.
\item[(iii)] Starting at $s=-R$, follow the real axis in the positive direction, all the way to $+\infty$. This is a stationary phase path; call it $\gamma_3$.
\end{enumerate}
This concatenation of curves is depicted in red in fig.\ \ref{papic}. By construction, $\gamma_1$ and $\gamma_3$ are steepest descent paths. Furthermore, since the norm of the integrand of \eqref{gamin} decreases exponentially when the real part of $s$ goes to $-\infty$, the contribution of $\gamma_2$ to the integral vanishes in the limit $R\to+\infty$. Accordingly, we will soon compute the large $\cN$ asymptotics of the integral \eqref{gamin} using the paths $\gamma_1$ and $\gamma_3$, with $R=+\infty$. Note that, in that limit, the imaginary part of the endpoint of $\gamma_1$ converges to its asymptotic value $\phii-\lambda\sin\phii$, obtained by setting $a=-\infty$ in the steepest-descent equation \eqref{stephi}.

The situation is somewhat different when $\lambda>1$. Depending on the value of $\phii$, two possibilities may occur (see the right panel in fig.\ \ref{codet}):
\begin{enumerate}[leftmargin=*]\setlength\itemsep{0em}
\item[(i)] If $|\phii|$ is `large' (in a sense to be defined shortly), it is not enclosed by the zero-phase curve that connects $+\infty+i\pi$ to $+\infty-i\pi$ and contains the saddle point $s=-\log\lambda$. Then the integration path used to evaluate \eqref{gamin} is essentially the same as for $\lambda\leq1$, and consists of the concatenation of $\gamma_1$, $\gamma_2$ and $\gamma_3$ (red curve in fig.\ \ref{papic}).
\item[(ii)] If, on the other hand, $|\phii|$ is small enough, then it is enclosed by the zero-phase path that goes through the saddle point. This allows us to take a {\it single} steepest descent contour $\gamma$ joining $s=i\phii$ to $s=+\infty$ (green curve in fig.\ \ref{papic}).
\end{enumerate}
Thus, for $\lambda>1$, a Stokes phenomenon \cite[sec.\ 6.6]{Bender} occurs at a critical value of $\phii$ given by the points $s=\pm i\phii_c$ that belong to the {\it vanishing-phase} steepest descent contour connecting $+\infty+i\pi$ to $+\infty-i\pi$. This curve also passes through the saddle point at $s=-\log\lambda$, so $\phii_c$ can be found by setting $a=-\log\lambda$ and $b=0$ in \eqref{stephi}, giving the transcendental equation $\phii_c=\lambda\sin\phii_c$. The resulting function $\phii_c(\lambda)$ increases monotonously with $\lambda$, satisfying $\phii_c\sim\sqrt{6(\lambda-1)}$ as $\lambda\to1^+$ and $\phii_c\sim\pi-\pi/\lambda$ as $\lambda\to+\infty$. In practice, this phenomenon has no effect on the correlation \eqref{cofaz}, so we will not dwell on it.

\phantomsection
\addcontentsline{toc}{subsection}{(iii)~~Evaluating asymptotics}%
\paragraph*{(iii)~~Evaluating asymptotics.} We now compute the integral \eqref{gamin} in the large $\cN$ limit using the (concatenations of) steepest-descent paths described above. To begin, assume that either $\lambda\leq1$, or $\lambda>1$ and $\phii>\phii_c$. In that case, \eqref{gamin} can be estimated by following the piecewise-smooth curve depicted in red in fig.\ \ref{papic}. In the limit $R\to+\infty$, the contribution of $\gamma_2$ vanishes (exponentially in $R$), so the integral reads
\be
\int_{i\phii}^{+\infty}\dd s\,e^{\cN\left(s-\lambda e^s\right)}
=
\int_{\gamma_1^{\infty}}\dd s\,e^{\cN\left(s-\lambda e^s\right)}
+
\int_{-\infty}^{+\infty}\dd s\,e^{\cN\left(s-\lambda e^s\right)},
\ee
where $\gamma_1^{\infty}$ is the limit of $\gamma_1$ as its left endpoint goes to $-\infty+i\phii-i\lambda\sin\phii$, while $\gamma_3$ yields the second term. The latter is a complete gamma function, so \eqref{gamin} can be recast as
\be
\Gamma(\cN,\cN\lambda e^{i\phii})
=
\Gamma(\cN)
+
(\cN\lambda)^{\cN}\int_{\gamma_1^{\infty}}\dd s\,e^{\cN\left(s-\lambda e^s\right)}.
\label{gamora}
\ee
This can be made more explicit by labelling the points of the path $\gamma_1^{\infty}$ by their imaginary part $b$, ranging from $\phii$ to $\phii-\lambda\sin\phii$, using the fact that their real part $a$ is given by eq.\ \eqref{stephi}. Including the Jacobian of the change of variable $s\to b$, one thus finds\vspace{-.5em}
\be
\begin{split}
\Gamma(\cN,\cN\lambda e^{i\phii})
&=
\Gamma(\cN)
-
\left[\cN\lambda e^{i(\phii-\lambda\sin\phii)}\right]^{\cN}
\!\!\!\!\!\!\int
\limits_{-\lambda\sin\phii}^{0}\!\!\!\!\!\!\dd b\,
\left(i+
\frac{1}{b+\lambda\sin\phii}-\cot(b+\phii)
\right)
\times\\
&
~~~~~~~~~\times\exp\left[\cN\left(%
\log\left[\frac{b+\lambda\sin\phii}{\lambda\sin(b+\phii)}\right]-(b+\lambda\sin\phii)\cot(b+\phii)
\right)\right]
\end{split}
\label{hint}
\ee
where we have renamed $b$ into $b+\phii$. This is as far as one can get without approximations. The integrand involves an exponential that increases monotonously on the interval $[-\lambda\sin\phii,0]$, with a slope at $b=0$ that grows with $\cN$. Accordingly, the large $\cN$ asymptotics of the integral \eqref{hint} is governed by the behavior of the integrand near $b=0$. Let us therefore assume $\phii\in\,(0,\pi)$ so that $\sin\phii>0$, and Taylor-expand the exponent as
\be
\log\left[\frac{b+\lambda\sin\phii}{\lambda\sin(b+\phii)}\right]-(b+\lambda\sin\phii)\cot(b+\phii)
\sim
-\lambda\cos\phii
+b\frac{\lambda+1/\lambda-2\cos\phii}{\sin\phii}
+\cO(b^2),
\label{apro}
\ee
where we note that $\lambda+1/\lambda-2\cos\phii>0$ for all $\phii\in\,(0,\pi)$ and any $\lambda>0$. One can then derive the asymptotics of the integral \eqref{hint} as $\cN\to+\infty$ using Laplace's method \cite[sec.\ 6.4]{Bender} --- namely expanding the integrand around its endpoint (here $b=0$) and computing the remaining exponential integral while neglecting higher-order corrections. Up to $\cO(1/\cN)$ corrections and provided either $\lambda\leq1$, or $\lambda>1$ and $\phii>\phii_c$, one has
\be
\Gamma(\cN,\cN\lambda e^{i\phii})
\stackrel{\cN\to+\infty}{\sim}
\Gamma(\cN)
-\frac{1}{\cN}
\left[\cN\lambda e^{i\phii-\lambda e^{i\phii}}\right]^{\cN}
\frac{1-\lambda e^{-i\phii}}{\lambda^2+1-2\lambda\cos\phii},
\label{marebis}
\ee
which is nothing but the aforementioned result \eqref{mare}.

The computation is nearly identical when $\lambda>1$ and $\phii<\phii_c$: the integration contour then is the green steepest-descent path of fig.\ \ref{papic}, connecting $s=i\phii$ to $s=+\infty$. The large $\cN$ asymptotics of the integral follow from the same considerations as those just described for $\phii>\phii_c$ (or $\lambda\leq1$), so the second term of eq.\ \eqref{marebis} remains valid. What changes is that the $\Gamma(\cN)$ on the right-hand side of \eqref{marebis} disappears. As far as the asymptotics of the incomplete gamma function is concerned, this is a huge difference: a term of order $\Gamma(\cN)\sim (\cN/e)^{\cN}$ disappears abruptly when $|\phii|$ goes below $\phii_c$. However, as mentioned in footnote \ref{funnyfoo}, this term essentially makes no difference for the correlation function \eqref{cofaz}.

\section{Edge correlations in anisotropic droplets}
\label{appb}
\setcounter{equation}{0}
\renewcommand{\theequation}{\thesection.\arabic{equation}}

In this appendix, we derive eq.\ \eqref{main} for the correlation function of edge modes in an anisotropic trap with rational ratio \eqref{rr}. As mentioned in section \ref{senisotro}, we proceed in three steps: first, a `method of images' simplifies the summation bounds in \eqref{t3}; then we establish an asymptotic formula for each of the individual terms in the sum; finally, the large $\cN$ limit allows us to replace the sum \eqref{t3} by a series, which we then evaluate. The result will include the earlier isotropic formula \eqref{asycofo} as a special case, without ever relying on the detailed gamma function asymptotics of appendix \ref{appa}. It will also imply, as an immediate corollary, the irrational correlations \eqref{conjec} conjectured in section \ref{sergo}.

\phantomsection
\addcontentsline{toc}{subsection}{(i)~~Method of images}%
\paragraph*{(i)~~Method of images.} We wish to rewrite the sum \eqref{t3} --- call it $S$ --- with simpler bounds on indices. Accordingly, let $x\equiv\bar zz'$, $y\equiv\bar ww'$ and define $m'\equiv qm$, $n'\equiv pn$ to get
\be
S
\equiv
\sum_{\substack{m,n\in\NN \\ m+\cR n<\cN}}
\frac{x^m\,y^n}{m!\,n!}
=
\sum_{\substack{m',n'\in\NN \\ m'+n'<q \cN \\ q|m',\;p|n'}}
\frac{x^{m'/q}\,y^{n'/p}}{(m'/q)!\,(n'/p)!}.
\label{sums}
\ee
Now rename $(m',n')$ into $(m,n)$ and use the fact that
\be
\frac{1}{q}
\sum_{\mu=0}^{q-1}e^{-2\pi i\mu m/q}
=
\begin{cases}
1 & \text{if }q|m,\\
0 & \text{otherwise}
\end{cases}
\qquad\forall\text{ integer }m
\ee
to recast the sum \eqref{sums} as
\be
S
=
\frac{1}{pq}
\sum_{\mu=0}^{q-1}
\sum_{\nu=0}^{p-1}
\sum_{\substack{m,n\in\NN \\ m+n< q \cN}}
e^{-2\pi i\mu m/q}
e^{-2\pi i\nu n/p}
\frac{x^{m/q}\,y^{n/p}}{(m/q)!\,(n/p)!}.
\label{slums}
\ee
The sum over $m,n$ now runs over a `symmetric' domain, as in an isotropic trap. One can think of this trick as a method of images, since the term $e^{-2\pi i\mu/q}x^{1/q}$ is the $\mu^{\text{th}}$ `image' of $x^{1/q}=(\bar zz')^{1/q}$ under the $\ZZ_q$ action $z^{1/q}\mapsto e^{-2\pi i\mu/q}z^{1/q}$. The sum over $(\mu,\nu)$ in \eqref{slums} thus runs over elements of the group $\ZZ_q\times\ZZ_p$. See also fig.\ \ref{fimeth} in the main text.

\phantomsection
\addcontentsline{toc}{subsection}{(ii)~~Individual asymptotics}%
\paragraph*{(ii)~~Individual asymptotics.} Our goal is to compute the asymptotics of the correlation function \eqref{t3} when $\bx$ and $\bx'$ are on the edge. Using the method of images to rewrite the sum $S$ in the form \eqref{slums}, the summand of the correlation function \eqref{t3} is a term
\be
t_{mn}
\equiv
e^{-\frac{1}{2}(|z|^2+|w|^2+|z'|^2+|w'|^2)}\,
e^{-2\pi i\mu m/q}
e^{-2\pi i\nu n/p}
\frac{(\bar zz')^{m/q}\,(\bar ww')^{n/p}}{(m/q)!\,(n/p)!}.
\label{summand}
\ee
Now write the complex coordinates of $\bx$ and $\bx'$ as in \eqref{zz}, with $\theta\in(0,\pi)$ for simplicity. In order to `zoom in' on the azimuth $\theta$ on the edge, define new indices $(j,k)$ by\footnote{For \eqref{nonoc} to make sense in terms of integer $(m,n)$, the angle $\theta$ must belong to a suitable discrete set; at sufficiently large $\cN$, one can approximate any $\theta$ arbitrarily closely in this way, so this is not worrisome.}
\be
m\equiv q \cN\cos^2(\theta/2)-j-k,
\qquad
n\equiv q \cN\sin^2(\theta/2)+k,
\label{nonoc}
\ee
and ask how the summand \eqref{summand} behaves in the limit $\cN\to+\infty$ with finite $j,k$. The answer turns out to be
\be
t_{mn}
\sim
\frac{\sqrt{\cR}}{2\pi\cN sc}\,
\exp\left[-\frac{(j+k)^2}{2q^2\cN c^2}-\frac{k^2}{2p q \cN s^2}\right]\,
e^{-i(\alpha+2\pi\mu)(q \cN c^2-j-k)/q}\,
e^{-i(\beta+2\pi\nu)(q \cN s^2+k)/p}
\label{tom}
\ee
where we write $s\equiv\sin(\theta/2)>0$ and $c\equiv\cos(\theta/2)>0$ to lighten notations. It remains to plug this into the sum \eqref{t3} giving the correlation function.

\phantomsection
\addcontentsline{toc}{subsection}{(iii)~~Series approximation}%
\paragraph*{(iii)~~Series approximation.} The method of images \eqref{slums}, along with the redefinition \eqref{nonoc} and the approximate summand \eqref{tom}, can be used to rewrite the correlator \eqref{t3} as
\be
\begin{split}
&\CC(\bx,\bx')
\sim
\frac{1}{2\pi^3\cN q^2 \sqrt{\cR}\,sc}
\sum_{\mu=0}^{q-1}
\sum_{\nu=0}^{p-1}
e^{-i(\alpha+2\pi\mu)\cN c^2}
e^{-i(\beta+2\pi\nu)\cN s^2/\cR}\\
&\quad\times
\sum_{j=1}^{+\infty}e^{ij(\alpha+2\pi\mu)/q}\,e^{-j^2/(2q^2\cN c^2)}
\sum_{k=-\infty}^{+\infty}
e^{ik\left[\frac{\alpha+2\pi\mu}{q}-\frac{\beta+2\pi\nu}{p}+\frac{ij}{q^2\cN c^2}\right]}\,
e^{-\frac{k^2}{2q \cN}\left(\frac{1}{ps^2}+\frac{1}{qc^2}\right)},
\end{split}
\label{cco}
\ee
where we let all summation bounds $\propto\cN$ go to infinity. The series in $k$ is essentially a periodic delta function setting $q\beta=p\alpha$, which already exhibits the expected localization on guiding center trajectories \eqref{gc}. More precisely, keeping $\cN$ large but finite, one may recognize in the last line a Jacobi theta function $\vartheta(z\,;\tau)\equiv\sum_{n\in\ZZ}e^{i\pi n^2\tau+2\pi i nz}$. Using also the modular property $\vartheta(z\,;\,\tau)\-=\frac{1}{\sqrt{-i\tau}}\-e^{-i\pi z^2/\tau}$ $\vartheta(z/\tau\,;\,-1/\tau)$, eq.\ \eqref{cco} can be recast as
\be
\begin{split}
&\CC(\bx,\bx')
\sim
\frac{1}{\pi^2\sqrt{2\pi\cN}\,q}\,
\frac{e^{-i(\alpha\cN c^2+\beta\cN s^2/\cR)}}
{\sqrt{c^2+\cR s^2}}\,
\sum_{\mu=0}^{q-1}
\sum_{\nu=0}^{p-1}
e^{-2\pi i(\mu\cN c^2+\nu\cN s^2/\cR)}\\
&
\times
\exp\left[-\frac{\cN s^2c^2}{2p}\frac{[p(\alpha+2\pi\mu)-q(\beta+2\pi\nu)]^2}{qc^2+ps^2}\right]\,
\sum_{j=1}^{+\infty}e^{ij(\alpha+2\pi\mu)/q}\,e^{-j^2/(2q^2\cN c^2)}.
\end{split}
\ee
To simplify this further, we assume that $\cN c^2$ is an integer and that $\cN s^2$ is an integer multiple of $\cR$ so that $e^{-2\pi i(\mu\cN c^2+\nu\cN s^2/\cR)}=1$; as before, this can be achieved with arbitrary accuracy in $\theta$ when $\cN$ is large enough (at fixed $p,q$). Finally, the series over $j$ yields a periodic power-law decay of correlations along classical trajectories:
\be
\begin{split}
&\CC(\bx,\bx')
\sim
-
\frac{1}{\pi^2}\,
\frac{1}{i\sqrt{8\pi\cN}\,q}\,
\frac{1}{\sqrt{c^2+\cR s^2}}\,
e^{-i\alpha\cN c^2}
e^{-i\beta\cN s^2/\cR}\\
&\times
\sum_{\mu=0}^{q-1}
\sum_{\nu=0}^{p-1}
\exp\left[-\frac{\cN s^2c^2}{2p}\frac{[p(\alpha+2\pi\mu)-q(\beta+2\pi\nu)]^2}{qc^2+ps^2}\right]\,
\frac{e^{i(\alpha+2\pi\mu)/2q}}{\sin(\frac{\alpha+2\pi\mu}{2q})}.
\end{split}
\label{mainbis}
\ee
This is the detailed form of the aforementioned result \eqref{main}. Note that it reproduces \eqref{asycofo} with $a=b=\delta\theta=0$, including numerical factors, in the special case $p=q=1$.

The strict $\cN=\infty$ limit of \eqref{mainbis} is used in section \ref{sergo} to infer the correlation function of ergodic edge modes. In particular, in that limit, the Gaussian dressing becomes an indicator function for classical trajectories:
\be
\lim_{\cN\to\infty}
e^{-\frac{\cN}{8}(\alpha+2\pi\mu-\frac{\beta+2\pi\nu}{\cR})^2\sin^2\theta\,F(\theta)}
=
\begin{cases}
1 & \text{if }\alpha=\frac{\beta+2\pi\nu}{\Delta}-2\pi\mu,\\
0 & \text{otherwise}.
\end{cases}
\ee
As a result, the definition \eqref{limico} of $\cK(\alpha,\beta)$ applied to the asymptotics \eqref{mainbis} yields
\be
\cK(\alpha,\beta)
=
\frac{1}{\sqrt{8\pi^5}}
\frac{1}{\sqrt{c^2+\Delta s^2}}
\begin{cases}
\frac{\ds1}{\ds q|\sin(\tfrac{\beta+2\pi\nu}{2p})|} & \text{if }\alpha=\frac{\beta+2\pi\nu}{\Delta}\!\!\!\!\mod{2\pi}\text{ for $\nu\in\ZZ$},\\
0 & \text{otherwise},
\end{cases}
\label{limicobis}
\ee
which reproduces the announced formula \eqref{limain} up to normalization. The irrational limit of this expression is well defined: letting $p\to\infty$ and $q\to\infty$ while keeping $p/q$ finite and converging to some irrational ratio $\Delta$, eq.\ \eqref{limicobis} yields
\be
\cK_{\cR,\theta,\beta}(\alpha)
=
\frac{1}{\sqrt{2\pi^5}}
\frac{\Delta}{\sqrt{c^2+\Delta s^2}}
\begin{cases}
\frac{\ds1}{\ds|\beta+2\pi\nu|} & \text{if }\alpha=\frac{\beta+2\pi\nu}{\Delta}\!\!\!\!\mod{2\pi}\text{ for $\nu\in\ZZ$},\\
0 & \text{otherwise}.
\end{cases}
\label{cirrbis}
\ee
Up to normalization, this is the result announced in eq.\ \eqref{conjec}.

\section{Entanglement in a spherical cap}
\label{appc}
\setcounter{equation}{0}
\renewcommand{\theequation}{\thesection.\arabic{equation}}

This appendix spells out the computations needed in section \ref{segap} to evaluate the ground state entanglement spectrum of a spherical cap \eqref{secap} in the thermodynamic limit. Recall that the $\text{U}(1)\times\text{U}(1)$ symmetry of the cap leads to a diagonal overlap matrix with eigenvalues \eqref{amoval}. Our goal is to approximate these eigenvalues in the `transition regions' of fig.\ \ref{figoverlap}, where $A(m,n)$ interpolates from $0$ to $1$. Those are indeed the only regions that contribute to the sum \eqref{entrosum} for entanglement entropy in the thermodynamic limit.

\phantomsection
\addcontentsline{toc}{subsection}{(i)~~Overlap asymptotics}%
\paragraph*{(i)~~Overlap asymptotics.} Keeping parameters $(R,\gamma)$ fixed, both factors of the eigenvalue \eqref{amoval} are monotonous functions of $m$ and $n$. Furthermore, $A(m,n)$ is nearly constant ($\simeq$ $0$ or $1$) in most of the $(m,n)$ plane (see fig.\ \ref{figoverlap}), where the summand of \eqref{entrosum} almost vanishes. The `jump' of $A(m,n)$ from $0$ to $1$ is concentrated in two transition regions, reflecting the factorized form of eq.\ \eqref{amoval} and the two kinds of boundaries of the spherical cap: the first is the line $m+n\sim R^2$, where the gamma function jumps; the second is the line $n/m=\tan^2(\gamma/2)$, where the beta function jumps. In both cases, the jump can be described by standard asymptotic methods, as follows.

Consider first the normalized incomplete gamma function, $\Gamma(m+n+2,R^2)/\Gamma(m+n+2)$, in eq.\ \eqref{amoval}. We focus on its transition region by writing $m+n\equiv R^2+aR\sqrt{2}$, where $a$ is an $\cO(1)$ variable. The asymptotics of $\Gamma(R^2+aR\sqrt{2},R^2)/\Gamma(R^2+aR\sqrt{2})$ as $R\to+\infty$ then follow from a variant of the saddle-point approximation, and read
\be
\frac{\Gamma(R^2+aR\sqrt{2},R^2)}{\Gamma(R^2+aR\sqrt{2})}
\,\stackrel{R\to+\infty}{\sim}\,
\frac{1}{2}\text{erfc}(-a)
-\frac{1}{3\sqrt{2\pi}R}e^{-a^2}(1+a^2)
+\cO(1/R^2)
\label{gamas}
\ee
in terms of a complementary error function. Note that the subleading term on the right-hand side of \eqref{gamas} is an {\it even} function of $a$; this will have important consequences for entanglement entropy, whose subleading correction will consequently vanish.

We now turn to the normalized incomplete beta function in eq.\ \eqref{amoval}. In contrast to the gamma function that only depended on the sum $m+n$, {\it two} variables are now needed to describe the transition region of the $(m,n)$ plane, occurring this time at $n/m\sim\tan^2(\gamma/2)$. Accordingly, define a `slow' coordinate $\tau$ and a `fast' coordinate $b$ by
\be
\begin{split}
m
&\equiv
\cos^2(\gamma/2)\,\tau\cN-b\sin(\gamma/2)\cos(\gamma/2)\sqrt{2\cN},\\
n
&\equiv
\sin^2(\gamma/2)\,\tau\cN+b\sin(\gamma/2)\cos(\gamma/2)\sqrt{2\cN},
\end{split}
\label{monotop}
\ee
where $\cN\gg1$. Eqs.\ \eqref{monotop} ensure that $m+n=\tau\cN$ holds exactly, without any $\cO(\sqrt{\cN})$ correction, while $b$ roughly measures the difference $m-n$. Using these variables in the integral definition of the beta function, one can compute the resulting asymptotic behavior using a minor variation of the saddle-point method and find
\be
1-\frac{B\big(\cos^2(\gamma/2);m+1,n+1\big)}{B(m+1,n+1)}
\sim
\frac{1}{2}\text{erfc}(b)
-
\frac{\sqrt{2}\cot\gamma}{3\sqrt{\pi\tau\cN}}\,
e^{-b^2}
(2-b^2)
+\cO(1/\cN).
\label{betas}
\ee
Note that the leading order of this asymptotic series is a function of $b$ alone; the $\tau$ variable only appears in subleading terms. Also note, similarly to eq.\ \eqref{gamas}, that the $\cO(\cN^{-1/2})$ correction is an even function of $b$. We will return to this below.

The asymptotics \eqref{gamas}--\eqref{betas} describe the overlap matrix \eqref{amoval} all along the transition region of fig.\ \ref{figoverlap}. To express compactly the resulting formulas, it is most convenient to define labels $j\equiv m+n$ and $k\equiv\cot(\gamma/2)n-\tan(\gamma/2)m$, in terms of which the transition region consists of the two orthogonal intervals shown in fig.\ \ref{figoverlabis}:
\be
\begin{split}
\text{Transition region I} &=\big\{(j,k)\big|j\simeq R^2,\;k\in[-R^2\tan(\gamma/2),0]\big\},\\
\text{Transition region II} &=\big\{(j,k)\big|j\in[R^2,\cN],\;k\simeq0\big\}.
\end{split}
\label{tati}
\ee
Overlap eigenvalues satisfy simple asymptotics in these two regions. First, in region I, where $j=m+n\sim R^2$, the beta factor of \eqref{amoval} essentially equals one and $A(m,n)$ reduces to a (normalized) incomplete gamma function, given in the transition region by eq.\ \eqref{gamas}:
\be
A(m,n)\Big|_{\text{I}}
\sim
\frac{1}{2}\text{erfc}\left(\frac{R^2-j}{R\sqrt{2}}\right)
-\frac{1}{3\sqrt{2\pi}R}e^{-\frac{(j-R^2)^2}{2R^2}}\left[1+\frac{(j-R^2)^2}{2R^2}\right]
+\cO(1/R^2).
\label{amore}
\ee
Note that this is independent of $k$. Similarly, in region II, the gamma factor of \eqref{amoval} essentially equals one, and the beta asymptotics \eqref{betas} suffices to deduce
\be
A(m,n)\Big|_{\text{II}}
\sim
\frac{1}{2}\text{erfc}\left(\frac{k}{\sqrt{2j}}\right)
-
\frac{\sqrt{2}\cot\gamma}{3\sqrt{\pi j}}\,
e^{-k^2/(2j)}
\left[2-\frac{k^2}{2j}\right]
+\cO(1/\cN),
\label{amoreto}
\ee
which emphatically depends on {\it both} $j$ and $k$. We now use eqs.\ \eqref{amore}--\eqref{amoreto} to evaluate entanglement entropy on the spherical cap in the thermodynamic limit.

\begin{SCfigure}[2][t]
\centering
\begin{tikzpicture}
\draw[dashed] (0,0)--(7,2.5) node [near end,above,sloped] (TextNode) {$k=\cot(\gamma/2)j$};
\draw[dashed] (0,0)--(7,-5.25) node [near end,below,sloped] (TextNode) {$k=-\tan(\gamma/2)j$};
\draw[dashed] (7,2.5)--(7,-5.25);
\fill[fill=LightYellow] (0,0)--(7,-5.25)--(7,2.5)--cycle;
\fill[fill=MiddleBrown] (4,0)--(7,0)--(7,-5.25)--(4,-3)--cycle;
\begin{scope}
\clip (0,0)--(7,-5.25)--(7,2.5)--cycle;
\draw[line width=6pt,smooth,color=LightBrown] (4,-4)--(4,0)--(7,0);
\end{scope}
\draw[->] (-.4,0) -- (8.5,0);
\draw (8.5,0) node[above] {$j$};
\node at (4,0) {\textbullet};
\draw (4,0) node[above] {$R^2$};
\draw[dashed] (4,2)--(4,-4);
\draw (5.5,0) node[below] {region II};
\node at (7,0) {\textbullet};
\draw (7,0) node[above] {$\cN$};
\draw[->] (0,-5.25) -- (0,2.5);
\draw (0,2.5) node[left] {$k$};
\draw (4,-1.5) node[above,rotate=90] {region I};
\draw (5.5,-2.25) node {$A(m,n)\simeq1$};
\draw (2,-.3) node {$A(m,n)\simeq0$};
\end{tikzpicture}
\caption{The heat map of fig.\ \ref{figoverlap} in terms of $j\equiv m+n$ and $k\equiv n/t-tm$, with $t\equiv\tan(\gamma/2)$. The triangle over which one integrates in order to evaluate entanglement entropy is represented in yellow and brown, but, as usual, only the (orange) transition region contributes substantially to the integral \eqref{entrint}. Owing to the choice of coordinates, this transition region now consists of two segments at nearly constant $j\simeq R^2$ and at nearly constant $k\simeq0$.}
\label{figoverlabis}
\end{SCfigure}

\phantomsection
\addcontentsline{toc}{subsection}{(ii)~~Leading entanglement}%
\paragraph*{(ii)~~Leading entanglement.} Entanglement entropy is given by the sum \eqref{entrosum}. In practice, the dependence of $A(m,n)$ on $(m,n)$ is `slow' in all regions that contribute to entropy because, as shown in eqs.\ \eqref{gamas}--\eqref{betas}, the overlap eigenvalues in the transition region are asymptotic to functions with $\cO(1)$ derivatives whose arguments take the form $j/\sqrt{\cN}$ or $k/\sqrt{\cN}$. Accordingly, at large $\cN$, the sum \eqref{entrosum} may be replaced by an integral:
\be
S
\sim
\int
_{\substack{m,n>0\\m+n<\cN}}\dd m\,\dd n\,F\big[A(m,n)\big],
\qquad
F[\lambda]\equiv
-\lambda\log\lambda-(1-\lambda)\log(1-\lambda).
\label{efdef}
\ee
This quantity is most easily evaluated by changing the integration variables $(m,n)$ into $(j,k)$ as defined above \eqref{tati}, with a Jacobian $\sin(\gamma)/2$:
\be
\frac{2S}{\sin\gamma}
\sim
\!\!\int\limits_{\substack{\text{triangle}\\ \text{in fig.\ \ref{figoverlabis}}}}\!\!
\dd j\,\dd k\,
F[A(m,n)]
\sim
\int\limits_{\text{region I}}
\!\!\dd j\,\dd k\,
F[A(m,n)]\,\,\,
+\!\!\!
\int\limits_{\text{region II}}
\!\!\dd j\,\dd k\,
F[A(m,n)].
\label{entrint}
\ee
Using the leading terms of eqs.\ \eqref{amore}--\eqref{amoreto} in regions I and II, respectively, one can express both integrals on the right-hand side of \eqref{entrint} in terms of a common parameter $s\equiv\sqrt{2}\int_{-\infty}^{+\infty}\dd a\,F\Big[\frac{1}{2}\text{erfc}(a)\Big]\simeq1.80639...$, namely as $\int_{\text{I}}\sim s\,R^3\,\tan(\gamma/2)$ and $\int_{\text{II}}\sim\frac{2s}{3}\,(\cN^{3/2}-R^3)$. The leading asymptotic value of the entanglement entropy \eqref{entrint} follows:
\be
S
\sim
s
\left[R^3\,\sin^2(\gamma/2)+\frac{\cN^{3/2}-R^3}{3}\sin\gamma\right].
\label{arealaw}
\ee
This is nothing but the leading term of \eqref{maintwo}, exhibiting the expected area law that involves the volume \eqref{arecap} of the boundary of the cap.

\phantomsection
\addcontentsline{toc}{subsection}{(iii)~~Subleading entanglement}%
\paragraph*{(iii)~~Subleading entanglement.} Eq.\ \eqref{arealaw} states that {\it bulk} entanglement entropy satisfies an area law \cite{Karabali:2020zap}. To quantify the contribution of {\it edge} modes to entanglement, one needs to include subleading $\cO(\cN)$ corrections of eq.\ \eqref{arealaw}, which follow from the subleading terms of eqs.\ \eqref{amore}--\eqref{amoreto}. Both of these subleading terms actually give {\it vanishing} contributions to entanglement entropy. Indeed, consider for definiteness the $1/R$ correction in eq.\ \eqref{amore}, which we write schematically as $A(m,n)|_{\text{I}}\sim\text{leading}+\frac{1}{R}E(j)+\cO(1/R^2)$, where $E(j)$ is an even function of $j$. We can plug this into the function $F$ defined in \eqref{efdef}, and expand the function in powers of $1/R$. The result is a correction to the integral over region I that involves the quantity $\int_{-\infty}^{+\infty}\dd a\,F'\big[\tfrac{1}{2}\text{erfc}(a)\big]\,E(a)$. The latter vanishes because $F[\tfrac{1}{2}\text{erfc}(a)]$ is even in $a$, so that $F'[\tfrac{1}{2}\text{erfc}(a)]$ is odd, while $E(a)$ is even.

There is thus no correction of order $1/R=\cO(\cN^{-1/2})$ to the entanglement entropy due to transition region I. A similar argument applies to region II, since the subleading $\cO(\cN^{-1/2})$ correction of \eqref{amoreto} is even in $k$. This allows us to conclude that the expansion \eqref{arealaw} holds at least up to (and including) order $\cO(\cN)$, confirming the result \eqref{maintwo}.

\newpage
\section{Entanglement in a rectangle}
\label{appex}

This appendix completes section \ref{senta}. It is devoted to the ground state entanglement of a finite rectangle in a boundary torus of an isotropic droplet. The rectangle's sides are chosen either to be parallel to guiding centre trajectories, or to cut them orthogonally. The resulting entropy \eqref{finalent} confirms the logarithmic formula \eqref{santorini} while exhibiting a uniform density of edge modes along the direction transverse to their motion.

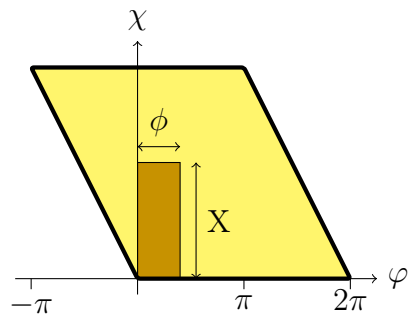
\begin{SCfigure}[2][b]
\begin{tikzpicture}[scale=.7]
\draw[fill = LightYellow] (0,0)--(4,0)--(2,4)--(-2,4)--cycle;
\draw[fill = MiddleBrown] (0,0)--(0.8,0)--(0.8,2.2)--(0,2.2)--cycle;
\draw[->] (-2.3,0) -- (4.5,0);
\draw (4.5,0) node[right] {$\phii$};
\draw[->] (0,-.3) -- (0,4.5);
\draw (0,4.5) node[above] {$\chi$};
\draw[<->] (1.1,0)--(1.1,2.2);
\draw (1.1,1.1) node[right] {$\upchi$};
\draw[<->] (0,2.5)--(0.8,2.5);
\draw (0.4,2.5) node[above] {$\phi$};
\draw[ultra thick,rounded corners=1] (0,0)--(4,0)--(2,4)--(-2,4)--cycle;
\draw[-] (-2,0)--(-2,-.2);
\draw (-2,-.1) node[below] {$-\pi$};
\draw[-] (2,0)--(2,-.2);
\draw (2,-.1) node[below] {$\pi$};
\draw[-] (4,0)--(4,-.2);
\draw (4,0) node[below] {$2\pi$};
\end{tikzpicture}
\caption{The torus defined by identifications \eqref{phichi} in the $(\phii,\chi)$ plane, along with the (brown) rectangle \eqref{secan}. Without loss of generality, we choose the coordinates $(\phii,\chi)$ to cover the yellow patch shown here. Edge modes propagate horizontally. Note that the periods \eqref{phichi} imply that the entangling region closes when $\upchi=4\pi$, in which case the rectangle becomes an annular strip that intersects each classical trajectory {\it twice}.}
\label{fitoto}
\end{SCfigure}

Given the Hopf coordinates \eqref{hopfco}, define new angles $\phii\equiv(\alpha+\beta)/2$, $\chi\equiv\alpha-\beta$ on a torus at fixed $r,\theta$. The identifications $(\alpha,\beta)\sim(\alpha+2\pi,\beta)\sim(\alpha,\beta+2\pi)$ then entail the following periodicity conditions on $(\phii,\chi)$:
\be
\label{phichi}
(\phii,\chi)\sim(\phii+\pi,\chi+2\pi)\sim(\phii+\pi,\chi-2\pi).
\ee
In these terms, edge modes propagate along circles at fixed $(r,\theta,\chi)$; each such circle, whose points are labelled by the angle $\phii$, supports a 1D CFT. It is therefore natural to define a `rectangle' as an entangling region where
\be
r\in[R,+\infty),
\qquad
\phii\in[0,\phi],
\qquad
\theta\in[0,\pi],
\qquad
\chi\in[0,\upchi]
\label{secan}
\ee
for some bounds $\phi\in(0,\pi)$ and $\upchi\in[0,4\pi]$. This region cuts a certain number ($\propto\upchi$) of edge fibers, and contains an interval of the same length $\phi$ in each such fiber. It is displayed in fig.\ \ref{fitoto}. Note that it does {\it not} reduce to a cylinder (of finite width $\phi$) when $\upchi=2\pi$; instead, the rectangle only closes when $\upchi=4\pi$, in which case it cuts each edge mode trajectory in {\it two} disjoint intervals of identical length $\phi$. 

In contrast to the cap \eqref{secap} or the interval \eqref{otherregion}, the rectangle \eqref{secan} breaks both factors of the U$(1) \times$ U$(1)$ symmetry whose quantum numbers $(m,n)$ label energy eigenstates \eqref{lola}. It is therefore challenging to obtain the corresponding entanglement spectrum exactly. What makes the problem nevertheless tractable is that, for non-interacting fermions, entanglement entropy is related to the statistics of charge fluctuations in the subregion of interest 
\cite{Klich_2006,Klich_2009,Calabrese_2012}. In particular, when fluctuations are Gaussian, or nearly so (in that higher cumulants are suppressed relatively to the particle variance $\kappa$), one gets 
\begin{align}
\label{eq:relationC2S}
S \sim \frac{\pi^2}{3} \kappa\,.
\end{align}
In the gapped bulk of a quantum Hall droplet (of any dimension), charge fluctuations are known not be Gaussian \cite{Charles_2019}. By contrast, charge fluctuations of edge modes are indeed expected to follow a normal distribution. This was established in \cite{Estienne2019hmd} in 2D, but it remains true in higher dimensions. Indeed, consider a droplet placed on the product manifold $M\times\mathbb{R}^2$ considered at the very end of section \ref{senta}, with the same boundary $M\times\mathbb{R}\times\{0\}$ as before. Then $M$ labels independent conduction channels, each of which is strictly identical to that of a 2D droplet, so its charge fluctuations are Gaussian.

With this motivation in mind, we now compute the second cumulant and evaluate the ensuing entropy, which will turn out to follow the logarithmic law characteristic of 1D gapless theories. Our starting point is full counting statistics in the rectangle \eqref{secan}, that is, the probability distribution of the number of particles contained in it. Its variance, \ie its {\it second cumulant}, can be expressed in terms of the correlation function as
\be
\kappa
=
\text{tr}\Big[
\CC\big|_{\cI}-\big(\CC\big|_{\cI}\big)^2
\Big]
=
\int_{\text{rectangle}}\dd^4\bx\int_{\text{rectangle}^c}\dd^4\by\;
\CC(\bx,\by)\CC(\by,\bx),
\ee
where the superscript $c$ denotes the complement. This variance contains two contributions: one is a pure bulk area law, the other is an edge contribution. We are interested in the latter, which originates from the asymptotics \eqref{asycofo} of the correlation function near the edge.  Accordingly, following \cite[sec.\ IV]{Estienne2019hmd}, instead of computing the variance directly, we evaluate its second derivative with respect to the parameter $\phi$:
\be
\der_{\phi}^2\kappa_{\text{edge}}
=
-\frac{1}{8}
\int\limits_R^{+\infty}r^3\dd r
\int\limits_0^{\pi}\sin\theta\,\dd\theta
\int\limits_0^{\upchi}\dd\chi
\int\limits_0^{+\infty}r'^3\dd r'
\int\limits_0^{\pi}\sin\theta\,\dd\theta'
\int\limits_0^{\upchi}\dd\chi'
\big|\CC_{\text{edge}}(\bx,\bx')_{\alpha=\phi}\big|^2
\label{gaussian}
\ee
where $\CC_{\text{edge}}$ is the edge correlator on the right-hand side of \eqref{asycofo}. At large $\cN$, the integral \eqref{gaussian} is readily evaluated since the integrand is a product of Gaussians. The result,
\be
\der_{\phi}^2\kappa_{\text{edge}}
\sim
-\frac{\cN\upchi}{16\pi^3\,\sin^2(\phi/2)},
\label{decum}
\ee
is just $\cN\upchi/(2\pi)$ times what one finds in a 2D droplet for a `piece of pie' 
(see \cite{Estienne2019hmd}, eq.~(28)).

As argued above, edge modes have Gaussian charge fluctuations, so their entanglement entropy is given by eq.\ \eqref{eq:relationC2S}. One can then integrate the second derivative \eqref{decum} of the variance of the number of edge modes to write their entropy
\be
S_{\text{edge}}
\sim
\frac{\cN\upchi}{12\pi}\,
\log\left(\sqrt{\cN}\sin\frac{\phi}{2}\right),
\label{finalent}
\ee
where we added a factor $\sqrt{\cN}$ in the logarithm by hand. (From the current perspective, this factor is merely an integration constant.) This is manifestly extensive in $\upchi$, confirming that the density of edge modes in the transverse direction is uniform. It also reduces to \eqref{santorini} when $\upchi=2\pi$, which was bound to be the case since the rectangle \eqref{secan} then intersects each edge trajectory exactly once. Finally, note that the rectangle closes for $\upchi=4\pi$, in which case entanglement entropy is {\it twice} the single interval value \eqref{santorini}, consistently with the fact that the corresponding cylindrical subregion cuts each trajectory in {\it two} subintervals of the same length.

\newpage
\addcontentsline{toc}{section}{References}
\providecommand{\href}[2]{#2}

\end{document}